\documentclass[a4paper,fleqn]{cas-dc}

\usepackage[authoryear]{natbib}
\usepackage{array}
\usepackage{booktabs}
\usepackage{siunitx}
\usepackage{tikz}
\usetikzlibrary{arrows.meta,positioning,calc,fit,backgrounds,decorations.markings,decorations.pathmorphing}
\usepackage{float}

\def\sectionfont{\rmfamily\fontsize{11pt}{13pt}\bfseries}
\def\ssectionfont{\rmfamily\fontsize{10pt}{12pt}\itshape\selectfont}
\def\sssectionfont{\rmfamily\fontsize{10pt}{12pt}\itshape\selectfont}

\newcommand{\nSites}{10}         %
\newcommand{\nSitesHeld}{2}     %
\newcommand{\nLayouts}{10}       %
\newcommand{\nStationsMin}{4}   %
\newcommand{\nStationsMax}{21}   %
\newcommand{\nSeaStates}{3000}     %
\newcommand{\depthMin}{14}       %
\newcommand{\depthMax}{45}       %
\newcommand{\tpMin}{2.8}          %
\newcommand{\tpMax}{17.8}          %
\newcommand{\windowSeconds}{100}  %

\newcommand{\hsMin}{0.4}           %
\newcommand{\hsMax}{8.4}           %
\newcommand{\lineShort}{75}       %
\newcommand{\lineLong}{105}        %
\newcommand{\simSeconds}{1380}      %
\newcommand{\simMinutes}{23}      %
\newcommand{\recordSamples}{13801}   %
\newcommand{\solverStep}{0.1}      %
\newcommand{\rampSeconds}{30}     %
\newcommand{\elemSize}{0.75}        %
\newcommand{\repeatSeconds}{1206.4}   %
\newcommand{\solverHoursMed}{37}  %
\newcommand{\solverHoursMin}{26}  %
\newcommand{\solverHoursMax}{60}  %

\newcommand{\seaRhoHTLo}{0.48}  %
\newcommand{\seaRhoHTHi}{0.70}  %
\newcommand{\seaRhoHCLo}{$-$0.05}  %
\newcommand{\seaRhoHCHi}{0.30}  %

\newcommand{\seaErrRhoSone}{$-$0.04}      %
\newcommand{\seaErrRhoStwo}{$-$0.02}      %
\newcommand{\seaErrSRise}{0.02}        %
\newcommand{\seaErrRhoWithheld}{0.13}  %

\newcommand{\axialSpeed}{2871}       %
\newcommand{\axialCrossLong}{0.037}   %
\newcommand{\dissipFreq}{1}       %
\newcommand{\nyquistFreq}{5}      %

\newcommand{\sOneMAE}{42.9}   %

\newcommand{\sOneGroundedRsq}{0.990}   %
\newcommand{\sOneTouchdownRsq}{0.991}  %
\newcommand{\sOneSuspendedRsq}{0.945}  %
\newcommand{\sOneThrNinety}{12\,630}              %
\newcommand{\sOneSnapShareNinety}{7}        %
\newcommand{\sOneNonSnapUpNinety}{7.9}        %
\newcommand{\sOneNonSnapBiasNinety}{+132}      %
\newcommand{\sOneSnapUpNinety}{32.2}           %
\newcommand{\sOneThrNinetyNine}{15\,291}          %
\newcommand{\sOneSnapShareNinetyNine}{32}    %
\newcommand{\sOneNonSnapUpNinetyNine}{10.7}    %
\newcommand{\sOneNonSnapBiasNinetyNine}{+227}  %
\newcommand{\sOneSnapUpNinetyNine}{49.7}       %

\newcommand{\sOneFullMAE}{38.0}       %
\newcommand{\sOneFullMAETable}{38.4}  %
\newcommand{\sOneFullGapPct}{1.2}    %
\newcommand{\oodTestShare}{49}      %

\newcommand{\sTwoRsq}{0.9994}   %
\newcommand{\sTwoMAE}{32.3}   %

\newcommand{\srMAEfour}{34.5}      %
\newcommand{\srMAEfull}{30.9}      %
\newcommand{\srPenTest}{1.12}      %
\newcommand{\srPenBase}{5.41}      %
\newcommand{\srRsqFour}{0.9994}      %
\newcommand{\srRsqBaseFour}{0.9803}  %

\newcommand{\srTouchdownRsq}{0.991}      %
\newcommand{\srTouchdownRsqBase}{0.172}  %
\newcommand{\srGroundedRsq}{0.995}       %
\newcommand{\srGroundedRsqBase}{0.988}   %
\newcommand{\srSuspendedRsq}{0.989}      %
\newcommand{\srSuspendedRsqBase}{0.898}  %

\newcommand{\snapEvents}{902}       %
\newcommand{\snapReachBefore}{24}  %
\newcommand{\snapReachAfter}{73}   %

\newcommand{\maskSamples}{1643}       %
\newcommand{\maskShare}{0.0040}         %
\newcommand{\maskSims}{109}          %
\newcommand{\ceilingRefSims}{2874}    %
\newcommand{\ceilingCleanMax}{25.3}   %
\newcommand{\ceilingRatio}{2.0}      %
\newcommand{\ceilingCases}{36}      %
\newcommand{\ceilingSamples}{638}    %
\newcommand{\discardNonfinite}{11}  %
\newcommand{\discardShort}{2}      %
\newcommand{\nInUse}{2987}            %

\newcommand{\wSub}{240.5}               %
\newcommand{\catCleanSims}{2891}       %
\newcommand{\catCandBelow}{65}       %
\newcommand{\catCandAbove}{1}       %
\newcommand{\catMomBelow}{1.5}        %
\newcommand{\catMomAbove}{15}        %
\newcommand{\catPoolShare}{0.37}       %
\newcommand{\catDropSone}{487}        %
\newcommand{\catDropSoneTen}{354}     %
\newcommand{\catDropStwo}{47}        %
\newcommand{\catDropStwoTen}{35}     %

\newcommand{\thrSegWeight}{1.26}         %
\newcommand{\thrSegDiffMed}{1\,275}        %
\newcommand{\thrSegResOrd}{6}         %
\newcommand{\thrCleanSamplesM}{12.3}     %
\newcommand{\thrCohCleanMax}{1.21}       %
\newcommand{\thrCohHiN}{2\,699}            %
\newcommand{\thrCohResLo}{10.9}          %
\newcommand{\thrSegResSnap}{62}        %
\newcommand{\thrCohEvMed}{1.006}          %
\newcommand{\thrCohMargin}{2.5}         %
\newcommand{\thrCohMaskMax}{41}        %
\newcommand{\thrCohCatMax}{1.6}         %
\newcommand{\thrFairOnlySamples}{8}   %
\newcommand{\thrFairOnlySims}{5}      %
\newcommand{\thrCrestM}{8.25}            %
\newcommand{\thrCrestPct}{1.73}          %
\newcommand{\thrCrestAbove}{21}        %
\newcommand{\thrEvIsoTenPct}{17}       %
\newcommand{\thrNlSust}{274}            %
\newcommand{\thrNlIso}{1770}             %
\newcommand{\thrNlMid}{162}             %
\newcommand{\thrFloorAdd}{21}          %
\newcommand{\thrFloorEval}{19}         %
\newcommand{\thrSlackZeroPct}{79}      %
\newcommand{\thrSlackCatMax}{3}       %
\newcommand{\thrIsoMaskMax}{14}        %
\newcommand{\thrPrevMed}{11}           %
\newcommand{\thrNextMed}{21}           %
\newcommand{\thrSettleMed}{38}         %
\newcommand{\thrTabUpLo}{7.7}           %
\newcommand{\thrTabUpHi}{9.2}           %
\newcommand{\thrTabReachLo}{71.5}        %
\newcommand{\thrTabReachHi}{72.5}        %
\newcommand{\thrTabNinetyNineLo}{9.5}   %
\newcommand{\thrTabNinetyNineHi}{11.7}   %
\newcommand{\thrTabNinetyNineTen}{20}  %
\newcommand{\thrCapSoneLo}{0.85}         %
\newcommand{\thrCapSoneLoN}{42}        %
\newcommand{\thrCapSoneHi}{0.73}         %
\newcommand{\thrCapSoneHiN}{55}        %
\newcommand{\thrCapStwoLo}{0.99}         %
\newcommand{\thrCapStwoLoN}{18}        %
\newcommand{\thrCapStwoHi}{0.91}         %
\newcommand{\thrCapStwoHiN}{27}        %

\newcommand{\repeatShare}{55}         %
\newcommand{\repeatSoneMAEDrop}{5.2}   %
\newcommand{\repeatStwoMAEDrop}{8.8}   %
\newcommand{\repeatActSiteMin}{8}    %
\newcommand{\repeatActSiteMax}{13}    %
\newcommand{\repeatActPooled}{11.2}     %
\newcommand{\repeatActPooledLo}{9.7}   %
\newcommand{\repeatActPooledHi}{12.5}   %
\newcommand{\repeatSpanStart}{9679}     %
\newcommand{\repeatMatchMax}{0.7}      %
\newcommand{\repeatEquivBound}{2.5}    %

\newcommand{\pinShareSix}{0.03}          %
\newcommand{\pinShareTen}{0.11}          %
\newcommand{\pinShareTrainLo}{11}      %
\newcommand{\pinShareTrainHi}{24}      %

\newcommand{\nStatNodeFeat}{12}         %
\newcommand{\nDynNodeFeat}{22}          %
\newcommand{\nNodeFeat}{34}             %
\newcommand{\nStatEdgeFeat}{10}         %
\newcommand{\nDynEdgeFeat}{4}          %
\newcommand{\nEdgeFeat}{14}             %
\newcommand{\nEnvChannels}{13}          %
\newcommand{\windowSamples}{1000}         %
\newcommand{\gatHeads}{4}              %
\newcommand{\gatHidden}{128}             %
\newcommand{\gatConcat}{512}             %
\newcommand{\gatOut}{128}                %
\newcommand{\gatDropout}{0.1}            %
\newcommand{\lstmLayers}{2}            %
\newcommand{\lstmHidden}{384}            %
\newcommand{\lstmOut}{768}               %
\newcommand{\lstmDropout}{0.1}           %
\newcommand{\headHidden}{128}            %
\newcommand{\headDropout}{0.1}           %
\newcommand{\decDim}{128}                %
\newcommand{\decHeads}{4}              %
\newcommand{\decFFN}{256}                %
\newcommand{\decBands}{8}              %
\newcommand{\decDropout}{0.1}            %
\newcommand{\optLr}{1 \times 10^{-3}}                 %
\newcommand{\optWd}{1 \times 10^{-4}}                 %
\newcommand{\optClip}{1}               %
\newcommand{\optBatch}{32}              %
\newcommand{\schedFactor}{0.5}           %
\newcommand{\schedThreshold}{10^{-4}}        %
\newcommand{\schedDecays}{5}           %
\newcommand{\schedPatience}{5}         %
\newcommand{\schedFloor}{5 \times 10^{-5}}            %
\newcommand{\mapeThrN}{1000}              %
\newcommand{\nLossTerms}{8}            %
\newcommand{\pinballTau}{0.9}            %
\newcommand{\pinballThr}{5.75}            %
\newcommand{\pinballPct}{85}            %
\newcommand{\tensionMean}{2.79}           %
\newcommand{\tensionStd}{2.73}            %
\newcommand{\selRsqDelta}{0.0010}           %
\newcommand{\selUpCap}{10}              %
\newcommand{\sOneEpochsMax}{100}         %
\newcommand{\sOneEpochsRun}{62}         %
\newcommand{\sOneSelEpoch}{32}          %
\newcommand{\sOneAdmissibleN}{3}       %
\newcommand{\sTwoAdmissibleN}{76}       %
\newcommand{\sTwoEpochs}{78}            %
\newcommand{\sTwoSelEpoch}{76}          %
\newcommand{\selEarlyStopPatience}{30}  %

\newcommand{\searchPool}{60000}     %
\newcommand{\searchLayouts}{6}  %
\newcommand{\searchSUni}{0.515}     %
\newcommand{\searchSBi}{0.479}      %
\newcommand{\searchNarrow}{256}   %
\newcommand{\searchArchPct}{1}  %
\newcommand{\searchSeedPct}{3}  %
\newcommand{\searchSeeds}{three}    %
\newcommand{\searchLRPct}{3}    %

\newcommand{\sOnePostSelEpochs}{30}  %
\newcommand{\sOnePostSelHours}{3}   %
\newcommand{\sOneValMAESel}{38}      %
\newcommand{\sOneValMAEBest}{18}     %
\newcommand{\sTwoLateEpochs}{23}     %
\newcommand{\sTwoLateHours}{10}      %
\newcommand{\sTwoLateSkillDrop}{0.002}  %
\newcommand{\recordReconSec}{0.2}     %
\newcommand{\sOneTrainHours}{7}     %
\newcommand{\sTwoTrainHours}{34}     %
\newcommand{\sOneTestMin}{10}        %
\newcommand{\sTwoTestMin}{13}        %
\newcommand{\sOneTestMs}{12}         %
\newcommand{\sTwoTestMs}{16}         %

\newcommand{\resRsqMin}{0.9992}                   %
\newcommand{\resRsqMax}{0.9994}                   %
\newcommand{\resSkillMin}{0.36}                 %
\newcommand{\resSkillMax}{0.39}                 %
\newcommand{\resMAEFour}{57.9}                  %
\newcommand{\resMAEFine}{38.1}                  %
\newcommand{\resMAEFineHi}{39.1}                %
\newcommand{\resUpNNFour}{41.1}                 %
\newcommand{\resUMNNFour}{110}                 %
\newcommand{\resUpNNFifteen}{9.8}              %
\newcommand{\resUMNNFifteen}{116}              %
\newcommand{\siteSixUpSone}{16.0}               %
\newcommand{\siteDevRatioSone}{1.008}            %
\newcommand{\siteDevErrSone}{0.86}              %
\newcommand{\siteDevErrStwo}{0.31}              %
\newcommand{\siteTenRatioSone}{0.974}            %
\newcommand{\siteTenErrSone}{2.70}              %
\newcommand{\siteTenUpSone}{94.0}               %
\newcommand{\siteTenErrStwo}{4.62}              %
\newcommand{\siteElevenRatioSone}{1.013}         %
\newcommand{\siteElevenErrSone}{1.34}           %
\newcommand{\siteElevenUpSone}{2.8}            %
\newcommand{\siteElevenErrStwo}{1.01}           %
\newcommand{\siteDevUpLo}{1.7}                 %
\newcommand{\siteDevUpHi}{6.9}                 %
\newcommand{\siteSixHL}{0.193}                   %
\newcommand{\siteTenHL}{0.200}                   %
\newcommand{\siteElevenHL}{0.368}                %
\newcommand{\siteNextHL}{0.373}                  %
\newcommand{\sOneGroundedNeg}{7.9}             %
\newcommand{\sOneTouchdownNeg}{0.0}            %
\newcommand{\srTouchdownNegBase}{47.8}          %
\newcommand{\sOneSuspendedNeg}{4.4}            %
\newcommand{\gcOffGrounded}{-0.03}               %
\newcommand{\gcOnGrounded}{0.97}                %
\newcommand{\sOneNonSnapUMNinety}{10}         %
\newcommand{\sOneSnapUMNinety}{296}            %
\newcommand{\sOneNonSnapUMNinetyNine}{36}     %
\newcommand{\sOneSnapUMNinetyNine}{537}        %
\newcommand{\snapNEvents}{154}                 %
\newcommand{\snapCapDev}{0.938}                  %
\newcommand{\snapNTen}{4}                    %
\newcommand{\snapCapEleven}{0.943}               %
\newcommand{\snapStationAvailSone}{97}        %
\newcommand{\snapStationNASone}{57}           %
\newcommand{\snapStationN}{six}                %
\newcommand{\snapStationLo}{0.43}               %
\newcommand{\snapStationHi}{0.98}               %
\newcommand{\snapStationTwoN}{six}             %
\newcommand{\snapStationTwoMedian}{1.02}        %
\newcommand{\snapStationTwoLo}{0.90}            %
\newcommand{\snapStationTwoHi}{1.18}            %
\newcommand{\snapStationSone}{0.789}             %
\newcommand{\snapStationSonePct}{79}          %
\newcommand{\snapStationStwoFull}{0.920}         %
\newcommand{\snapStationStwoFour}{0.938}         %
\newcommand{\snapStationStwoFourPct}{94}      %
\newcommand{\snapStationCommonN}{25}          %
\newcommand{\snapStationCommonDiff}{+0.021}       %
\newcommand{\snapStationCommonPct}{60}        %
\newcommand{\snapStationSiteDevFull}{0.938}      %
\newcommand{\snapStationSiteDevFour}{0.926}      %
\newcommand{\snapStationSiteNDevSone}{43}     %
\newcommand{\snapStationSiteNDevStwo}{19}     %
\newcommand{\snapStationSiteTenFull}{0.816}      %
\newcommand{\snapStationSiteTenFour}{0.302}      %
\newcommand{\snapStationSiteNTenSone}{1}     %
\newcommand{\snapStationSiteNTenStwo}{1}     %
\newcommand{\snapStationSiteElevenFull}{0.915}   %
\newcommand{\snapStationSiteElevenFour}{0.954}   %
\newcommand{\snapStationSiteNElevenSone}{53}  %
\newcommand{\snapStationSiteNElevenStwo}{25}  %

\newcommand{\alongTenBias}{76}     %
\newcommand{\alongTenBiasPct}{2.3}  %
\newcommand{\alongTenBiasLo}{69}   %
\newcommand{\alongTenBiasHi}{83}   %
\newcommand{\alongTenDrift}{9}    %

\newcommand{\srMAELo}{30.4}                    %
\newcommand{\srMAEHi}{34.5}                    %
\newcommand{\srPenBaseSix}{2.42}               %
\newcommand{\srPenBaseFive}{4.31}              %
\newcommand{\srSkillFour}{0.342}                %
\newcommand{\srSkillBaseFour}{0.414}            %
\newcommand{\srSkillFull}{0.326}                %
\newcommand{\sOneFullSkill}{0.366}              %
\newcommand{\srSkillPenBase}{1.13}             %
\newcommand{\srFullGainPct}{19}              %
\newcommand{\srMAPECross}{15}                %
\newcommand{\figFourSite}{loc10}                %
\newcommand{\figFourHs}{4.0}                  %
\newcommand{\figFourTdRsq}{0.974}               %
\newcommand{\figFourTdMAE}{44}               %
\newcommand{\figFourTdRsqInterp}{-7.6}         %
\newcommand{\figFourTdMAEInterp}{1101}         %
\newcommand{\figFourTdArc}{0.80}               %
\newcommand{\figFourBSite}{loc11}               %
\newcommand{\figFourBTdMAE}{31}              %
\newcommand{\figFourBTdMAEInterp}{426}        %
\newcommand{\figFourBTdArc}{0.60}              %
\newcommand{\sTwoThrNinety}{12\,622}              %
\newcommand{\sTwoNonSnapUpNinety}{64.5}        %
\newcommand{\sTwoNonSnapBiasNinety}{-8}      %
\newcommand{\sTwoNonSnapUMNinety}{40}        %
\newcommand{\sTwoSnapUMNinety}{570}           %
\newcommand{\sTwoThrNinetyNine}{15\,278}          %
\newcommand{\sTwoSnapUMNinetyNine}{1\,043}       %
\newcommand{\snapNStwo}{45}                  %
\newcommand{\noiseFairX}{0.35}            %
\newcommand{\noiseFairZ}{0.33}            %
\newcommand{\noiseSnrInstr}{16.6}         %
\newcommand{\noiseVelFactor}{7.1}        %
\newcommand{\noiseSigMin}{0.01}           %
\newcommand{\noiseSigMax}{0.25}           %
\newcommand{\noiseRsqFourInstr}{0.983}     %
\newcommand{\noiseMAEFourInstr}{228.3}     %
\newcommand{\noiseMAPEFourInstr}{14.6}    %
\newcommand{\noiseMAPEFourClean}{2.6}    %
\newcommand{\noiseMAEFourInstrB}{297.0}    %
\newcommand{\noiseKeepSigma}{0.10}        %
\newcommand{\noiseLoseSigma}{0.25}        %
\newcommand{\noiseSnapFourClean}{0.955}    %
\newcommand{\noiseSnapFourInstr}{0.979}    %
\newcommand{\noiseMAEFourClean}{34.5}     %
\newcommand{\noiseMAEFourDiff}{428.4}      %
\newcommand{\noiseRsqFourDiff}{0.934}      %
\newcommand{\noiseMAEFourOff}{247.8}       %
\newcommand{\noiseSkillFourClean}{0.34}   %
\newcommand{\noiseSkillFourInstr}{6.42}   %
\newcommand{\noiseSkillFourOff}{0.76}     %
\newcommand{\resSSShareOne}{50}    %
\newcommand{\resSSShareTwo}{56}    %
\newcommand{\lrSubsetPct}{10}     %
\newcommand{\lrStride}{5}        %
\newcommand{\lrTestWindows}{49\,513}   %
\newcommand{\lrInOne}{34}         %
\newcommand{\lrInTwo}{70}         %
\newcommand{\lrInThree}{286}       %
\newcommand{\lrInFour}{1618}        %
\newcommand{\lrLagMin}{0.1}        %
\newcommand{\lrLagShort}{5}      %
\newcommand{\lrLagLong}{40}       %
\newcommand{\lrRsqConst}{0.973}      %
\newcommand{\lrWinMAE}{95.8}        %
\newcommand{\lrMemGain}{2.5}       %
\newcommand{\lrMAEFour}{106.5}       %
\newcommand{\lrSFour}{0.81}         %
\newcommand{\lrSOne}{0.37}          %
\newcommand{\lrMAERatio}{2.5}      %

\begin{document}
\let\WriteBookmarks\relax
\def\floatpagepagefraction{1}
\def\textpagefraction{.001}

\shorttitle{Full-line mooring tension from sparse motion sensing}
\shortauthors{A. Konstantaras et~al.}

\title [mode = title]{Spatial Super-Resolution of Mooring Tension using a Topology-General, Physics-Informed Graph Attention Network}

\author[1]{Athanasios Konstantaras}[orcid=0009-0007-2856-646X]
\cormark[1]
\ead{thanoskonstantaras@gmail.com}
\credit{Conceptualization, Methodology, Software, Validation, Formal analysis,
  Investigation, Data curation, Visualization, Writing -- original draft}

\affiliation[1]{organization={Offshore and Dredging Engineering, Delft
    University of Technology},
  addressline={Postbus 5},
  postcode={2600 AA},
  city={Delft},
  country={The Netherlands}}

\author[2]{Juan P. {Aguilar-L\'{o}pez}}[orcid=0000-0003-2930-5812]
\ead{J.P.AguilarLopez@tudelft.nl}
\credit{Conceptualization, Supervision, Writing -- review \& editing}

\author[2]{Oriol {Colom\'{e}s}}[orcid=0000-0002-5552-9695]
\ead{J.O.ColomesGene@tudelft.nl}
\credit{Conceptualization, Supervision, Writing -- review \& editing}

\affiliation[2]{organization={Faculty of Civil Engineering and Geosciences,
    Delft University of Technology},
  addressline={Stevinweg 1},
  postcode={2628 CN},
  city={Delft},
  country={The Netherlands}}

\cortext[1]{Corresponding author}

\begin{abstract}
Digital twins for monitoring, maintenance and fatigue assessment of floating
structures require the tension along the whole mooring line, including its
touchdown zone. Many data-driven surrogates, however, predict it
at one point, usually the fairlead, and
lose accuracy outside their training conditions and at peaks. Using a graph attention
network, bidirectional recurrence and a query decoder, this study reconstructs
the full tension field from motion at as few as four stations: the fixed anchor,
the fairlead and two sensors dividing the line into thirds. Across
\nSeaStates{} high-fidelity finite-element simulations based on tangential
differential calculus, one set of weights serves \nLayouts{} sensor layouts
(\nStationsMin{}--\nStationsMax{} stations) and \nSites{} North Sea sites
\depthMin{}--\depthMax{}~m deep, two withheld from training. Snap loads are
separated from solver artifacts and retained. From four stations, the model
reconstructs the tension at all \nStationsMax{} with an $R^2$ of
\srRsqFour{} and recovers a median \snapStationStwoFourPct{}\% of the tension
where each of \snapNStwo{} distinct snap events occurs. The robustness of the
model was tested against white motion noise of \noiseSigMin{}--\noiseSigMax{}~m.
Reconstructing a \simMinutes{}-minute record takes
under a second, against the solver's median of \solverHoursMed{}~h.
\end{abstract}

\begin{keywords}
Mooring tension reconstruction \sep Digital twin \sep  Graph neural network \sep Touchdown zone \sep Snap loads \sep
Spatial super-resolution 
\end{keywords}

\begingroup
\ExplSyntaxOn
\cs_set_protected:Npn \hbox_to_wd:nn #1#2 { \hbox to #1 { #2 \hss } }
\ExplSyntaxOff
\maketitle
\endgroup

\section{Introduction}\label{sec:intro}

Floating offshore wind power is moving from demonstration projects to commercial deployment, with global installed capacity expected to reach 16.5 GW by 2030 \citep{GWEC2025,IRENA2024}, about sixty times the 278 MW installed at the end of 2024 \citep{GWEC2025}. A floating platform is held in place by its mooring system, lines of chain, steel wire or synthetic rope that connect the floater to anchors on the seabed, and this system
plays a crucial role as one of the main drivers of cost optimisation of such structures
\citep{kikuchi2019upscaling,martinez2022mapping}. A mooring-line failure can have severe economic
and safety consequences \citep{ma2020mooringbreakage}. To prevent it, the tension in the mooring
line must be quantified, as it is the governing quantity in both the ultimate and the fatigue limit
states \citep{thies2014mooring, benassai2014ultimate}. Many innovative mooring designs for floating
offshore wind turbines (FOWTs) have recently been proposed, with new configurations and materials
such as synthetic fibre ropes, taut rather than catenary lines and lines with integrated damping
devices \citep{xu2021design,tian2026innovative,garcia2026review}, all aimed
at reducing the maximum tension or the cyclic tension amplitude.

Despite these design efforts, mooring-line tension remains among the hardest quantities to monitor
on a floating structure. It is typically measured by load cells or in-line load pins,
whose accuracy and long-term reliability offshore are limited, particularly where submerged access
for maintenance is impractical \citep{elman2013reducing,gordon2014mooring,hashmani2024review}.
Motion, by contrast, is cheaper and more robust to monitor. Low-cost global navigation satellite
system (GNSS) receivers, combined with micro-electromechanical system (MEMS) sensors on the
platform, have recently been used successfully to monitor mooring lines
\citep{revertcalabuig2023gnss,jayasinghe2022model,minnebo2014mooring,dealteriis2022positioning}.

Motion measurements, however, do not provide the tension directly, so a model that maps the
measured motion to the tension of the line is needed. Physics-based models impose the motion of the
fairlead, the point at which the line is attached to the floater, as a boundary condition on a
numerical model of the line, as proposed for monitoring the
stress along the mooring lines of FOWTs \citep{chen2023realtime}. They range from linear and quasi-static approximations, which neglect the inertia of the line and the
drag acting on it, to fully dynamic lumped-mass and finite-element formulations
\citep{walker2021digitaltwin}. Lumped-mass models, validated against model-scale tests
\citep{hall2015moordyn}, discretise the line into point masses connected by elastic segments.
Finite-element models describe the strain, and with it the tension, as a continuous field along
the line and avoid the
non-physical oscillations that lumped-mass models develop when a slack line becomes taut again, but
in a direct comparison they needed about \SI{460}{\second} where the lumped-mass model needed
\SI{5}{\second} \citep{agarwal2026finitestrain}. With the finite-element model used in this
study, a \simMinutes{}-minute simulation takes a median of \solverHoursMed{}~h
(Section~\ref{subsec:fe_data}), far too slow to follow a monitoring campaign. Lowering the fidelity
reduces the cost only at the expense of accuracy, and physics-based modelling has been found unable
to combine low computational cost with low error, which data-driven surrogates may achieve
\citep{glasspool2025balancing}.

Data-driven surrogate models, called surrogates hereafter, learn the relation between an input, here
motion, and an output, here tension, from data. Early work replaced most
of a finite-element simulation with a neural network that estimated the fairlead tension from the
floater motion, 20 times faster \citep{guarize2007ann}. Surrogates have
since been trained on the motion of floating production, storage and offloading (FPSO) vessels
\citep{qiao2021lstm,li2022spm}, a
semi-submersible platform \citep{wang2022lstm} and semi-submersible FOWTs
\citep{payenda2024rnn,tang2025cnnlstm,grafe2024virtual}, and, for design, on wave elevation, mean
wind speed and current across water depths and the design variables of a taut mooring with load
reduction devices \citep{festa2025lrd}. Three
limitations recur in these models. First, they return the tension at one point of each line, usually the fairlead, and an
asymmetric mooring layout would require a separate model for each fairlead
\citep{grafe2024virtual}. Second, their accuracy drops outside the conditions seen in training
\citep{xie2024domain,yan2026dacla,kammerling2026robust}. Third, their largest errors occur at the
tension peaks \citep{chen2022skewness,kammerling2026robust}. Snap loads are the extreme case of such
peaks, sharp spikes in tension when a slack line becomes taut again, which in basin tests of a FOWT
exceeded the cyclic tension maxima by 37--68\,\% and can break the line or shorten its life
\citep{hsu2017snap,qiao2020snaptaut}.

The first limitation has been addressed by \citet{xie2026sparse}, who forecast, from sparse
measurements of the platform motion of one FOWT, the tension of 50 segments per line over the next
30 s, but report errors averaged over each line. Such an error says little about the touchdown
zone, where the line meets the seabed, which for risers, the pipes that carry fluids between the
seabed and a floating platform, is the hardest part of the structure to predict because of its
nonlinear contact with the seabed \citep{dasilva2022slender}. This zone also deserves attention in its
own right. The dynamics of a mooring line matter especially where it meets the seabed
\citep{agarwal2026finitestrain}, and the chain--seabed interaction is one of the main
uncertainties in mooring analysis \citep{wu2026review}. Predicting the tension along the whole line, including the touchdown zone, for mooring lines of different designs, calls for a model whose outputs are not fixed to
the segments of one line, such as a graph neural network.

Graph neural networks operate directly on the nodes and segments of a structure. Each node is
updated with messages exchanged with its neighbours through functions shared by all nodes
\citep{scarselli2009gnn,gilmer2017mpnn}, so that a trained network also applies to discretisations
of other sizes than those seen in training \citep{sanchezgonzalez2020gns}. Graph
attention networks (GATs) weight the message of each neighbour with a learned coefficient
\citep{velickovic2018gat}, and their GATv2 variant makes this weighting depend on the receiving
node, increasing expressiveness at the same cost \citep{brody2022gatv2}. In offshore
engineering, graph attention has been used to detect a break in a mooring line from the horizontal
motion of an FPSO \citep{costa2024spectral}, and a GATv2 encoder combined with a recurrent
one has predicted the tower loads of a semi-submersible FOWT under a single sea state, on a graph of
its tower, hull and fairleads \citep{yung2025omae,yung2025structural}. Graph networks have also
reconstructed full fields from sparse measurements. The stress over the whole finite-element mesh
of a semi-submersible floater was recovered from simulated gauges on \SI{5}{\percent} of its
elements, with a mean absolute error about nine to twelve times smaller than that of dense,
convolutional and recurrent networks that did not use the mesh connectivity
\citep{jiang2025gtwin}. In none of
these studies does the graph follow the topology of the mooring line, although its discretisation
is already a chain of nodes connected by segments.

In the present study, the tension along the entire length of a catenary mooring line is
reconstructed from the motion measured at a few fixed points on the line itself, here called
stations. Only a few stations are used, because powering sensors underwater and communicating with
them is difficult, and fitting many along a line is costly \citep{jin2024inclinometers}. In
wave-flume tests on a model mooring line, inertial units attached at two points have already been
used to recover the displacement at those points \citep{you2026imu}. The model places a graph on
the stations of the line, encodes the interaction between neighbouring stations by graph attention,
derives from them a representation of every station at which the tension is required, according to
its position along the line, and follows each station in time with a bidirectional recurrent
network. Its training loss adds physical constraints, the material law and the equilibrium of the
line, to the data-fitting terms. The study addresses the three limitations above. First, one set
of weights returns the tension at all \nStationsMax{} stations of the line, including the touchdown
region, for \nLayouts{} sensor layouts with \nStationsMin{} to \nStationsMax{} sensed stations. From
four of them, the fixed anchor, the fairlead and two line-mounted sensors, it is evaluated against
an interpolated matched-resolution baseline. Second, the same weights serve \nSites{} sites in the
southern North Sea, \nSitesHeld{} of which are withheld from training. Third, snap loads are
separated from solver artifacts and kept rather than removed, so the extremes are assessed and
their calibration is quantified. The sensitivity of the reconstruction to motion-measurement noise
is also evaluated.

The remainder of the manuscript is structured as follows. Section~\ref{sec:data} describes the
finite-element simulations, the sites and sea states, the data splits and the separation of snap
loads from solver artifacts. Section~\ref{sec:method} explains the surrogate model, its training
objective and its evaluation. Section~\ref{sec:results} reports the results when every output
station is sensed, when only a few stations are sensed and when the motion measurements are noisy,
and discusses their practical implications and limitations. Section~\ref{sec:conclusions}
concludes.

\section{Dataset and data preparation}\label{sec:data}

This section covers everything done to the simulation archive before any model is involved, and
Section~\ref{sec:method} then explains the surrogate model. The dataset comprises \nSeaStates{}
finite-element simulations of catenary mooring lines at \nSites{} sites in the southern North Sea,
with two unstretched line lengths and a range of water depths (Section~\ref{subsec:sites}). This
section describes the solver, the environmental conditions and the archived fields, defines the
reconstruction task and the data splits, and separates snap loads from solver artifacts.

\begin{figure}[pos=H]
\centering
\resizebox{\linewidth}{!}{%
\begin{tikzpicture}[
    x=0.72cm, y=1.05cm,
    >=Latex,
    sensed/.style={circle, draw=cAccent, fill=cAccent, minimum size=4.4pt, inner sep=0pt,
                   line width=0.5pt},
    unsensed/.style={circle, draw=cLine!70!black, fill=white, minimum size=2.6pt,
                     inner sep=0pt, line width=0.4pt},
    force/.style={-{Latex[length=1.5mm,width=1.1mm]}, line width=0.55pt, color=black!85},
    dragf/.style={-{Latex[length=1.5mm,width=1.1mm]}, line width=0.55pt, color=red!65!black},
  ]
  \colorlet{cWater}{blue!18!white}
  \colorlet{cBed}{brown!35!black}
  \colorlet{cLine}{blue!55!black}
  \colorlet{cAccent}{red!65!black}

  \fill[cWater] (-0.6,0) rectangle (12.6,3.05);
  \draw[decorate, decoration={snake, amplitude=0.7pt, segment length=6pt}, cLine!60!white,
        line width=0.5pt] (-0.6,3.05) -- (12.6,3.05);
  \draw[->, cLine!70!white, line width=0.5pt] (1.0,3.45) -- (2.4,3.45);
  \draw[->, cLine!70!white, line width=0.5pt] (1.0,1.55) -- (2.4,1.55);
  \node[anchor=west, font=\tiny, color=cLine!70!black] at (2.55,3.45) {waves};
  \node[anchor=west, font=\tiny, color=cLine!70!black] at (2.55,1.55) {current};

  \fill[cBed!25!white] (-0.6,-0.55) rectangle (12.6,0);
  \draw[cBed, line width=0.6pt] (-0.6,0) -- (12.6,0);

  \coordinate (A) at (0,0);
  \coordinate (TD) at (3.2,0);
  \coordinate (F) at (10.6,2.55);
  \draw[cLine, line width=1.1pt]
    (A) -- (TD) .. controls (5.6,0.05) and (8.0,1.15) .. (F);

  \draw[cLine, line width=0pt,
        postaction={decorate, decoration={markings,
          mark=between positions 0 and 1 step 1/20 with {\node[unsensed]{};}}}]
    (A) -- (TD) .. controls (5.6,0.05) and (8.0,1.15) .. (F);

  \draw[cLine, line width=0pt,
        postaction={decorate, decoration={markings,
          mark=between positions 0 and 1 step 1/3 with {\node[sensed]{};}}}]
    (A) -- (TD) .. controls (5.6,0.05) and (8.0,1.15) .. (F);

  \fill[cBed!70!white] (-0.10,-0.22) -- (0.10,-0.22) -- (0,0) -- cycle;
  \node[anchor=north, font=\tiny] at (0,-0.30) {anchor};
  \node[anchor=south west, font=\tiny] at (10.75,2.60) {fairlead};

  \draw[<->, line width=0.4pt] (-0.35,0) -- (-0.35,3.05);
  \node[rotate=90, anchor=south, font=\tiny] at (-0.42,1.5) {water depth $h_0$};

  \draw[->, line width=0.4pt] (3.2,0.78) -- (3.2,0.10);
  \node[anchor=south, font=\tiny] at (3.2,0.78) {touchdown point};
  \node[anchor=south, font=\tiny, color=black!75] at (1.6,0.07) {grounded};
  \node[anchor=north west, font=\tiny, color=black!75] at (7.9,0.62) {suspended};

  \node[anchor=west, font=\scriptsize] at (-0.6,4.15) {(a)};
  \node[sensed] at (0.9,4.15) {};
  \node[anchor=west, font=\tiny] at (1.05,4.15) {measured ($N_{\mathrm{in}}$)};
  \node[unsensed] at (5.6,4.15) {};
  \node[anchor=west, font=\tiny] at (5.75,4.15) {reconstructed ($N_{\mathrm{out}}$)};

  \node[anchor=west, font=\scriptsize] at (-0.6,-1.05) {(b)};
  \coordinate (P1) at (1.7,-2.55);
  \coordinate (P2) at (3.5,-1.65);
  \coordinate (C)  at (2.6,-2.10);
  \draw[cLine, line width=1.6pt] (P1) -- (P2);
  \draw[force] (P2) -- ++(0.9,0.45) node[anchor=west, font=\tiny] {$T(s+\mathrm{d}s)$};
  \draw[force] (P1) -- ++(-0.9,-0.45) node[anchor=east, font=\tiny] {$T(s)$};
  \draw[force] (C) -- ++(0,-0.80) node[anchor=north, font=\tiny] {$w_{s}\,\mathrm{d}s$};
  \draw[dragf] (C) -- ++(-0.696,0.654) node[anchor=south, font=\tiny] {$\mathbf{b}_{\mathrm{drag,n}}$};
  \draw[dragf] ($(C)+(0.10,-0.20)$) -- ++(0.70,0.35)
    node[anchor=north west, font=\tiny] {$\mathbf{b}_{\mathrm{drag,t}}$};
  \node[anchor=north, font=\tiny, color=black!75] at (2.6,-3.20) {suspended element};
  \begin{scope}[shift={(0,0.86)}]
  \colorlet{cSpring}{yellow!60!orange}
  \fill[cBed!25!white] (6.7,-4.02) rectangle (10.9,-3.37);
  \draw[cBed, line width=0.6pt] (6.7,-3.37) -- (10.9,-3.37);
  \coordinate (G1) at (7.8,-3.31);
  \coordinate (G2) at (9.8,-3.31);
  \foreach \xs in {7.95, 9.20} {
    \draw[cSpring!80!black, line width=0.5pt] (\xs,-3.34) -- (\xs,-3.44);
    \draw[cSpring!80!black, line width=0.5pt, decorate,
          decoration={zigzag, segment length=1.6pt, amplitude=1.3pt}] (\xs,-3.44) -- (\xs,-3.84);
    \draw[cSpring!80!black, line width=0.5pt] (\xs,-3.84) -- (\xs,-3.92);
    \draw[cSpring!80!black, line width=0.5pt] (\xs+0.30,-3.34) -- (\xs+0.30,-3.58)
      (\xs+0.25,-3.58) -- (\xs+0.35,-3.58)
      (\xs+0.22,-3.50) -- (\xs+0.22,-3.70) -- (\xs+0.38,-3.70) -- (\xs+0.38,-3.50)
      (\xs+0.30,-3.70) -- (\xs+0.30,-3.92);
    \draw[cSpring!80!black, line width=0.7pt] (\xs-0.10,-3.92) -- (\xs+0.40,-3.92);
  }
  \draw[cLine, line width=1.6pt] (G1) -- (G2);
  \draw[force] (G2) -- ++(0.9,0) node[anchor=west, font=\tiny] {$T(s+\mathrm{d}s)$};
  \draw[force] (G1) -- ++(-0.9,0) node[anchor=east, font=\tiny] {$T(s)$};
  \draw[force] (8.80,-2.68) -- (8.80,-3.26);
  \node[anchor=south, font=\tiny] at (8.80,-2.68) {$w_{s}\,\mathrm{d}s$};
  \draw[force] (8.80,-3.97) -- (8.80,-3.37);
  \node[anchor=east, font=\tiny] at (8.76,-3.80) {$\mathbf{b}_{\mathrm{bed}}$};
  \node[anchor=north, font=\tiny, color=black!75] at (8.6,-4.06) {grounded element};
  \end{scope}
\end{tikzpicture}%
}
\caption{Line configuration and forces. (a) Filled markers: measured stations (here
$N_{\mathrm{in}} = 4$, the anchor, the fairlead and two line-mounted sensors). Open markers: the
$N_{\mathrm{out}} = 21$ native stations. (b) Forces on an element of length $\mathrm{d}s$: the
tension $T$ along the tangent at both ends, the submerged weight $w_{s}\,\mathrm{d}s$, the drag
$\mathbf{b}_{\mathrm{drag,t}}$ and $\mathbf{b}_{\mathrm{drag,n}}$ along and normal to the tangent
(Eq.~\eqref{eq:drag}) and, on the grounded section only, the seabed reaction
$\mathbf{b}_{\mathrm{bed}}$ of the spring and damper of Eq.~\eqref{eq:seabed}.}
\label{fig:configuration}
\end{figure}
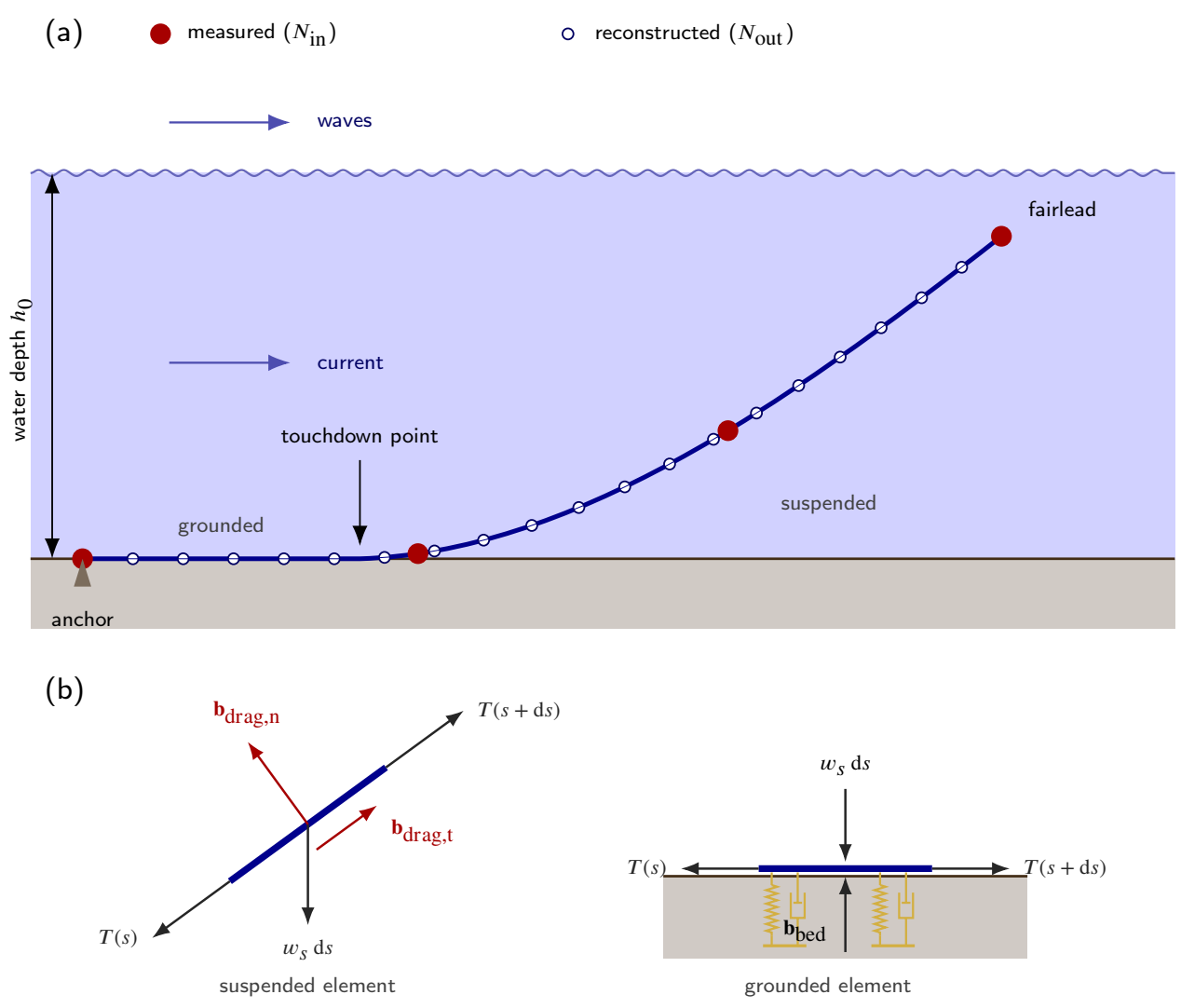

Figure~\ref{fig:configuration}a shows the geometry. The lower part of the line rests on the seabed,
the grounded section, and the upper part hangs in the water, the suspended section, which reaches
the fairlead. Motion is recorded at a few stations along the line, whereas tension is required at
every station, including the touchdown point, which migrates with the sea state and therefore
cannot be instrumented directly. Figure~\ref{fig:configuration}b shows the forces acting on an
element of the line, which Section~\ref{subsec:fe_data} formulates, and Table~\ref{tab:properties}
lists the properties of the line, the seabed and the water.

\subsection{Finite-element mooring model}\label{subsec:fe_data}

Ground truth comes from a geometrically nonlinear finite-strain cable model formulated with
tangential differential calculus \citep{agarwal2026finitestrain} and implemented in the Julia
language \citep{bezanson2017julia} with the Gridap finite-element library
\citep{badia2020gridap}. The line is treated as a one-dimensional manifold embedded in the
vertical plane, so the curvature of the catenary and its large displacements enter the kinematics
without linearisation. This matters here because the line lifts off and settles back on the seabed
and passes between slack and taut states, motions far from any static shape about which a model
could be linearised. Equilibrium is imposed on the deformed configuration $\Gamma$,
Eq.~\eqref{eq:equilibrium},
\begin{equation}
    \rho\,\ddot{\mathbf{u}} - \operatorname{div}_{\Gamma}\boldsymbol{\sigma}(\mathbf{u})
    - \mathbf{b}(\mathbf{u}) = \mathbf{0},
    \label{eq:equilibrium}
\end{equation}
where $\rho$ is the material density, $\mathbf{u}$ the displacement field,
$\operatorname{div}_{\Gamma}$ the surface divergence, $\boldsymbol{\sigma}$ the Cauchy stress
and $\mathbf{b}$ the external body force per unit volume.

The material is Saint Venant--Kirchhoff, the simplest hyperelastic law, linear between the
Green--Lagrange strain and the second Piola--Kirchhoff stress. It admits large displacements and
rotations while the strain stays small, as it does in a mooring line. For a cable of Young modulus
$E$ and Poisson ratio $\nu$ the Lam\'{e} constants reduce to $\lambda = 0$ and
$\mu = E/2(1+\nu)$, so the stress follows from the tangential Green--Lagrange strain
$\mathbf{E}_{\mathrm{t}}$ through the surface deformation gradient $\mathbf{F}$ and its
determinant $J$, Eq.~\eqref{eq:constitutive},
\begin{equation}
    \boldsymbol{\sigma}
    = \tfrac{1}{J}\,\mathbf{F}\,\bigl(2\mu\,\mathbf{E}_{\mathrm{t}}\bigr)\,\mathbf{F}^{\!\top}.
    \label{eq:constitutive}
\end{equation}
Axial tension, the quantity reconstructed here, is the cross-sectional resultant of
Eq.~\eqref{eq:constitutive}. It therefore varies continuously along the line rather than being
constant within a segment, which lets the formulation represent a stress wave travelling along the
line during a sharp transient \citep{agarwal2026finitestrain}. At the time step used here, no such
wave is carried (Section~\ref{subsec:snaps}).

The body force collects self-weight, buoyancy, seabed reaction and hydrodynamic drag
(Fig.~\ref{fig:configuration}b). The seabed is a damped spring acting only on the grounded
portion, with stiffness $k_{\mathrm{b}}$, damping $c_{\mathrm{b}}$ and penetration measured against
the bed elevation $z_{\mathrm{b}}$, Eq.~\eqref{eq:seabed},
\begin{equation}
    \mathbf{b}_{\mathrm{bed}} =
    \begin{cases}
        \bigl[k_{\mathrm{b}}\,(z_{\mathrm{b}} - z) - c_{\mathrm{b}}\,\dot{z}\bigr]\,\mathbf{e}_{z},
            & z \le z_{\mathrm{b}},\\[2pt]
        \mathbf{0}, & z > z_{\mathrm{b}},
    \end{cases}
    \label{eq:seabed}
\end{equation}
and drag is quadratic in the velocity of the line relative to the fluid,
$\mathbf{v}_{\mathrm{rel}} = \mathbf{v}_{\mathrm{wave}} + \mathbf{v}_{\mathrm{current}}
- \dot{\mathbf{u}}$, resolved into components along and normal to the local tangent with
separate coefficients, Eq.~\eqref{eq:drag},
\begin{equation}
    \mathbf{b}_{\mathrm{drag},i}
    = \tfrac{1}{2}\,\rho_{\mathrm{w}}\,C_{\mathrm{d},i}\,d\,
      \lVert \mathbf{v}_{\mathrm{rel},i}\rVert\,\mathbf{v}_{\mathrm{rel},i},
    \qquad i \in \{\mathrm{t},\mathrm{n}\}.
    \label{eq:drag}
\end{equation}
Added mass and added damping are neglected, since their effect on the overall response of a
line of this type is small \citep{hall2015moordyn}. The inertia of the line is therefore that of
its own mass, which the momentum term of the training loss also assumes
(Section~\ref{subsec:loss}). Table~\ref{tab:properties} lists the constants of the model, those of
the line and the seabed as given by \citet{santjer2025}.

\begin{table}[t]
\centering
\caption{Properties of the line, the seabed, the water and the discretisation of the
finite-element model.}
\label{tab:properties}
\footnotesize
\setlength{\tabcolsep}{3pt}
\begin{tabular}{@{}l l l@{}}
\toprule
{Property} & {Symbol} & {Value} \\
\midrule
Line diameter & $d$ & \SI{0.068}{\metre} \\
Cross-sectional area & $A$ & \SI{0.003619}{\metre\squared} \\
Mass per unit length & $\rho A$ & \SI{28.23}{\kilo\gram\per\metre} \\
Submerged weight per unit length & $w_{s}$ & \SI{240.5}{\newton\per\metre} \\
Axial stiffness & $EA$ & \SI{232.7}{\mega\newton} \\
Unstretched length & $L_{0}$ & \SI{75}{\metre} or \SI{105}{\metre} \\
Normal drag coefficient & $C_{\mathrm{d,n}}$ & 2.6 \\
Tangential drag coefficient & $C_{\mathrm{d,t}}$ & 1.4 \\
Seabed stiffness & $k_{\mathrm{b}}$ & \SI{3e6}{\pascal\per\metre} \\
Seabed damping & $c_{\mathrm{b}}$ & \SI{3e5}{\pascal\second\per\metre} \\
Sea-water density & $\rho_{\mathrm{w}}$ & \SI{1025}{\kilo\gram\per\cubic\metre} \\
Element length (quadratic) &  & \SI{0.75}{\metre} \\
Time step & $\Delta t$ & \SI{0.1}{\second} \\
\bottomrule
\end{tabular}

\end{table}

Each line is meshed uniformly with quadratic elements of \elemSize{}~m, which fixes the spatial
resolution of the solution, and the 21 archived stations lie at element boundaries
(Section~\ref{subsec:campaign}). The resulting second-order system is integrated with the
generalised-$\alpha$ method under asymptotic annihilation \citep{gobat2001generalizedalpha}, which
suppresses spurious high-frequency oscillation while leaving the wave-frequency response intact
\citep{chung1993generalizedalpha}, a property that Section~\ref{subsec:snaps} uses to separate
snap loads from solver artifacts. The formulation has been verified by mesh, element-order and time-step convergence studies and
against standard benchmarks \citep{agarwal2026finitestrain}. In a sample of 30 simulations across
the ten sites, a single \simMinutes{}-minute simulation takes between \solverHoursMin{} and
\solverHoursMax{}~h of solver time, with a median of \solverHoursMed{}~h, which puts the full
campaign at the order of $10^{5}$ hours of solver time and is the practical motivation for a
surrogate.

\subsection{Sites and sea states}\label{subsec:sites}

The environmental conditions, the waves and current that load the line, are taken from the study
of \citet{santjer2025}, which drives the same mooring model with conditions extracted from the
Dutch Continental Shelf Model, a
three-dimensional hydrodynamic model of the north-western European shelf coupled to the SWAN
spectral wave model \citep{zijl2023dcsm,booij1999swan}. Of the 12 sites in that study,
\nSites{} are used here, covering water depths from \depthMin{} to \depthMax{}~m with
unstretched line lengths of \lineShort{}~m at the six shallower sites and \lineLong{}~m at the
four deeper ones.

For each site, \citet{santjer2025} took the largest significant wave height ($H_s$) of each day
from September to April over five years of hindcast, together with its peak period ($T_p$) and the
largest current magnitude at five depths within four hours of it, fitted a Gaussian-copula
Bayesian network to their joint
distribution and drew $n = 300$ cases from it to study how these conditions drive the tension of
the line. The cases are reused here unchanged. They therefore follow the dependence observed at
each site rather than filling the space of conditions uniformly, as a space-filling design such
as a Latin hypercube would. The dependence is moderate. The rank correlation between $H_s$ and
$T_p$ is \seaRhoHTLo{} to \seaRhoHTHi{} at the ten sites, and that between $H_s$ and the current
\seaRhoHCLo{} to \seaRhoHCHi{}, so the cases still cover a broad range of conditions
(Table~\ref{tab:sites}). For a surrogate meant to monitor lines at these sites, this concentrates
the training data on the conditions the lines experience, and no simulation is spent on
combinations of wave height and period that do not occur there. The price is that rare
combinations are represented by few cases, which Section~\ref{subsec:res_limits} returns to. The
input of the model is, moreover, the measured motion of the line, and the sea-state parameters
only condition it (Section~\ref{subsec:graph}).

Each case is forced by a JONSWAP spectrum parameterised by $H_s$ and $T_p$. A wave time series is synthesised from the
spectrum by inverse Fourier transform and converted to particle kinematics by linear wave
theory, and the current profile is interpolated through the water column from the five sampled
depths. Waves and current propagate along the line, the more severe alignment. The anchor is fixed
and the fairlead follows the local wave-particle motion, so the line is excited without a floater
model, an assumption whose consequences are discussed in Section~\ref{subsec:res_limits}. These
are the boundary conditions of the model. Each simulation starts from the static configuration of
the line, and the forcing is ramped up from rest (Section~\ref{subsec:campaign}).

Table~\ref{tab:sites} lists the sites. Pooled over all \nSeaStates{} cases, $H_s$ spans
\hsMin{} to \hsMax{}~m and $T_p$ spans \tpMin{} to \tpMax{}~s. Two sites are withheld from
training and used only for testing. The withheld site loc10 (\SI{15.0}{\metre}) and the training
site loc06 (\SI{14.5}{\metre}) are the two shallowest in the campaign, and the next shallowest
training site lies at \SI{30.1}{\metre}, so loc10 has a single close neighbour in depth. The other
withheld site, loc11, lies within a cluster of four training sites spanning
\SIrange{35.7}{44.7}{\metre}. The transfer to loc10 is therefore the hardest this dataset poses, and
Section~\ref{sec:results} quantifies its cost.

\begin{table}[t]
\centering
\caption{The \nSites{} sites: line geometry and the per-site range of $H_s$ and $T_p$ over the
300 cases sampled there.}
\label{tab:sites}
\footnotesize
\setlength{\tabcolsep}{4pt}
\begin{tabular}{@{}l S[table-format=2.1] S[table-format=3.0] S[table-format=1.1] S[table-format=1.1] S[table-format=2.1] S[table-format=2.1] l@{}}
\toprule
 & {$h_0$} & {$L_0$} & \multicolumn{2}{c}{$H_s$ [\si{\metre}]} & \multicolumn{2}{c}{$T_p$ [\si{\second}]} & \\
\cmidrule(lr){4-5}\cmidrule(lr){6-7}
{Site} & {[\si{\metre}]} & {[\si{\metre}]} & {min} & {max} & {min} & {max} & {Splits} \\
\midrule
loc01 & 30.1 & 75  & 0.6 & 8.0 & 3.3 & 13.3 & train / val / test \\
loc03 & 35.7 & 75  & 0.8 & 5.7 & 3.4 & 14.9 & train / val / test \\
loc04 & 44.7 & 105 & 0.4 & 6.6 & 4.0 & 14.7 & train / val / test \\
loc05 & 39.2 & 105 & 0.7 & 8.3 & 3.3 & 14.9 & train / val / test \\
loc06 & 14.5 & 75  & 0.5 & 6.2 & 3.0 & 17.8 & train / val / test \\
loc07 & 40.7 & 105 & 0.7 & 8.4 & 3.3 & 16.6 & train / val / test \\
loc08 & 30.2 & 75  & 0.6 & 8.0 & 3.4 & 15.3 & train / val / test \\
loc09 & 33.6 & 75  & 0.4 & 8.1 & 4.0 & 15.3 & train / val / test \\
loc10 & 15.0 & 75  & 0.5 & 6.4 & 2.8 & 17.0 & test only \\
loc11 & 38.6 & 105 & 0.5 & 7.9 & 3.4 & 14.2 & test only \\
\bottomrule
\end{tabular}
\end{table}

\subsection{Archived fields and sampling}\label{subsec:campaign}

Each simulation runs for \simSeconds{}~s and is archived at every solver step, giving
\recordSamples{} samples at an interval of \solverStep{}~s. A station is a fixed point of the line,
identified by its position along the unstretched line, at which motion is measured or tension is
evaluated. Horizontal and vertical position and axial tension are recorded at 21 such stations,
the native stations, spaced equally in unstretched arc length, each at an element boundary of the
mesh. This is the finest spatial resolution of any ground truth in this work and the output
resolution used throughout.

Two properties of the archive matter for what follows. First, the archive interval is the
integration step of the solver, not a decimation of a finer solution, and each archived value is
the state of the solution at its step rather than an average over the step. A tension transient
occupying a single sample therefore lies at the limit of what the simulation resolves rather than
of what was written out. Second, the synthesised wave train is periodic. Its 1441 spectral components are
evenly spaced up to a cut-off of \SI{7.5}{\radian\per\second}, which gives a repeat period of
\repeatSeconds{}~s within a \simSeconds{}~s record, and Section~\ref{subsec:task} quantifies its
effect on the evaluation. The solver ramps the forcing up from rest over the first
\rampSeconds{}~s of each record, and the fairlead motion has reached its later amplitude by the
end of that interval. The ramp is not removed from the data used for training and testing, whereas
\citet{santjer2025} disregard the first three minutes of their records as ramp-up.

\subsection{Reconstruction task, discretisation and splits}\label{subsec:task}

The task is co-temporal reconstruction. Given the motion of the line at $N_{\mathrm{in}}$ sensed
stations, where motion is measured, over a window of \windowSeconds{}~s, the model estimates
axial tension at $N_{\mathrm{out}}$ output stations over the same window, either the sensed ones
($N_{\mathrm{out}} = N_{\mathrm{in}}$) or the \nStationsMax{} native stations
($N_{\mathrm{out}} = \nStationsMax{}$). It is not a forecast, and dynamic tension is never an
input. The static pretension profile is a node feature, but it is taken once per simulation from
the first archived sample, at the start of the ramp of Section~\ref{subsec:campaign} and before
wave or current loading acts, so it carries no information about the dynamic response.

Two models serve the two cases, and Section~\ref{sec:method} describes both. Stage~1, the
matched-resolution model, reconstructs the tension at the stations it senses. Stage~2 reconstructs
it at all \nStationsMax{} native stations from fewer sensed ones.

Sensor layouts are produced by resampling the 21-station record in unstretched arc length onto
\begin{equation}
    N \in \{4,\; 5,\; 6,\; 7,\; 8,\; 10,\; 12,\; 15,\; 18,\; 21\}
    \label{eq:layouts}
\end{equation}
stations, with the anchor and the fairlead preserved exactly in every layout. Windows are
$W = 1000$ samples (\windowSeconds{}~s) long.

Splits are built per case (Table~\ref{tab:splits}). At the eight training sites the last
\SI{30}{\percent} of each record is reserved as a temporal test set. The first \SI{70}{\percent}
is divided into non-overlapping blocks of 2000 samples, assigned at random, \SI{70}{\percent} to
training and \SI{30}{\percent} to validation, and every window must lie within a single block, so
no window appears in more than one split. The two withheld sites enter only the test set. For
Stage~2, windows are drawn paired across sensor layouts, so the same physical window is evaluated
at every $N$ and comparisons across sensor counts are not confounded by the draw. Input and target
normalisation statistics are fitted on training windows only.

The repeat of the forcing (Section~\ref{subsec:campaign}) falls inside the temporal test region,
so \repeatShare{}\,\si{\percent} of the temporal test windows at the training sites cover forcing
that the same site's training windows have already seen, raising the concern that the model
recognises a repeated forcing realisation instead of reconstructing tension from it. The task
limits how much such recognition could help. The model is given the very motion the forcing
produced and only has to return the tension consistent with it, so having encountered the same
forcing before adds little, the input already contains what a memorised realisation would supply.
A forecaster, which must instead infer the coming motion from what came before rather than receive
it, would gain a real advantage from recognising a repeated pattern, since it could recall the
continuation instead of predicting it.

Appendix~\ref{app:repeat} tests this directly: windows that cover the repeat score no better than
unaffected windows of matched activity, and any advantage of the repeat to either model is bounded
below \repeatEquivBound{}\,\si{\percent}. The withheld sites, which carry the generalisation claim,
are unaffected in any case.

\begin{table}[t]
\centering
\caption{Windows by split. Training and validation: budgets shared by both stages. Test: windows
evaluated after the screening of Section~\ref{subsec:snaps}, out of \num{50000} drawn per stage;
Stage~2 counts a window once per sensor layout.}
\label{tab:splits}
\footnotesize
\setlength{\tabcolsep}{3pt}
\begin{tabular}{@{}l l l r r@{}}
\toprule
{Split} & {Sites} & {Time span} & {Stage~1} & {Stage~2} \\
\midrule
Training & 8 training & first \SI{70}{\percent}, blocked & \multicolumn{2}{c}{1\,000\,000} \\
Validation & 8 training & first \SI{70}{\percent}, blocked & \multicolumn{2}{c}{10\,000} \\
Test, temporal & 8 training & last \SI{30}{\percent} & 24\,883 & 25\,080 \\
Test, withheld & loc10, loc11 & whole record & 24\,630 & 24\,450 \\
\bottomrule
\end{tabular}

\end{table}

\subsection{Snap loads versus solver artifacts}\label{subsec:snaps}

The archived records contain many tension spikes that occupy a single sample, and two kinds must be
told apart. A snap load is physical. A slack section of the line extends until it is taut, its
motion is arrested and the tension rises sharply for a short time. Snap loads are the extreme
loads a study of mooring tension must characterise, so they are identified and kept rather than
treated as noise. A solver artifact is a value that the solver writes without a physical cause,
such as a single station jumping to a very large tension and back, or a run whose tension
diverges. Artifacts are fewer, but they must be removed. Duration cannot separate the two, since
both occupy a single sample. They differ instead in how the tension is distributed along the line
and in time, and these differences follow from how the simulated line carries tension within one
solver step, which the next paragraphs derive.

With $EA = \SI{232.7}{\mega\newton}$ and a mass per unit length of
\SI{28.23}{\kilo\gram\per\metre}, the axial wave speed is
$c = \sqrt{EA/\rho A} = \SI{\axialSpeed}{\metre\per\second}$, which crosses the longest line in
\axialCrossLong{}~s, well within one solver step of \solverStep{}~s. The generalised-$\alpha$
scheme under asymptotic annihilation removes response above $1/(10\Delta t)$, or
\dissipFreq{}~\si{\hertz} at this step, from the computed solution \citep{agarwal2026finitestrain},
well below the Nyquist frequency of the record, \nyquistFreq{}~\si{\hertz}. A snap therefore
appears in the archive with the value that the solver computes at the step in which the arrest
falls, not as an average over the step, and a transient shorter than one step is not resolved.
Axial waves are absent, and at every archived sample the line is in axial equilibrium
with the loads acting at that instant. The seabed of Eq.~\eqref{eq:seabed} acts normal to the bed
only, so friction holds no tension in the grounded section. The axial force balance between a
station $i$ and the fairlead then gives
\begin{equation}
    T_{N-1} - T_{i} = w_{s}\,(z_{N-1} - z_{i}) + \sum_{j} m_{j}\,a_{\mathrm{t},j} + D_{i},
    \label{eq:axial_balance}
\end{equation}
where $w_{s} = \SI{\wSub}{\newton\per\metre}$ is the submerged weight per unit length, $z$ the
elevation, $m_{j}$ and $a_{\mathrm{t},j}$ the lumped mass and tangential acceleration of the
stations between the two, and $D_{i}$ the tangential drag over the same stretch. Three
consequences follow for a genuine record. First, when a slack section becomes taut its motion is
arrested. Because the line equilibrates axially within a sample, the arrest is complete within a
step or two, so a snap appears as a short peak that follows a period of slack, after which the
line typically remains taut at a lower tension. The wave response, in contrast, changes smoothly
over many samples. Second,
neighbouring stations read nearly the same value during an arrest, and a grounded stretch, which
does not rise, transmits its tension unchanged, so identical readings at adjacent stations are
expected. Third, at wave frequency the inertia and drag terms are small compared with the weight
term, and the fairlead is the highest point of the line, so no station can carry more than the
fairlead at the same sample. This holds for the simulated line only. In a physical line an axial
wave reflected at the anchor can raise the anchor tension above that of the fairlead for a few
hundredths of a second, but the solver filters out such high-frequency waves, as noted above.

Writing $T[t,i]$ for the archived tension at time $t$ and station $i$, with station $N-1$ the
fairlead, four ratios are formed,
\begin{align}
    \mathrm{iso}  &= \frac{T[t,i]}
        {\max\!\bigl(\tfrac{1}{2}(T[t-1,i] + T[t+1,i]),\, \SI{1}{\newton}\bigr)},
        \label{eq:iso}\\[2pt]
    \mathrm{coh}  &= \frac{T[t,i]}{\max\bigl(T[t,i-1],\, T[t,i+1]\bigr)},
        \label{eq:coh}\\[2pt]
    \mathrm{cat}  &= \frac{T[t,i]}{T[t,N-1]},
    \;\;
    \mathrm{fair}  = \frac{T[t,i]}{\max_{t'} T[t',N-1]}.
        \label{eq:cat}
\end{align}
where $t'$ ranges over every sample of the record. Temporal isolation, Eq.~\eqref{eq:iso},
measures whether the value is large relative to the preceding and following samples of the same
station, and spatial coherence, Eq.~\eqref{eq:coh}, whether it is confined to a single station.
The two catenary ratios of Eq.~\eqref{eq:cat} compare the station with the fairlead at the same
instant and with the largest fairlead tension of the record. A sample is classified as a solver
artifact when either of two rules fires:
\begin{itemize}
    \item The shape rule requires $T > \SI{10}{\kilo\newton}$ and $\mathrm{iso} > 10$ and
    either $\mathrm{coh} > 3$ or $\mathrm{fair} > 3$.
    \item The magnitude ceiling requires $T > \SI{50}{\kilo\newton}$.
\end{itemize}
A sample that neither rule flags is classified as a snap load when all of the following hold:
\begin{itemize}
    \item $T > \SI{3}{\kilo\newton}$, $\mathrm{iso} > 3$ and $\mathrm{coh} < 3$.
    \item The fairlead condition holds, that is, no station other than the fairlead carries more
    than \SI{1}{\kilo\newton} with $\mathrm{cat} > 1$. The floor keeps the condition silent on a
    slack line, where both tensions are near zero.
    \item The tension at the station fell below $0.25\,T$ within the 10 preceding samples.
\end{itemize}
The shape rule catches spikes confined to one or two stations. The magnitude ceiling catches
sustained divergence, a run in which the tension grows to unphysical values and stays there over
many samples. The samples around such a value are as large as the value itself, so $\mathrm{iso}$
stays near 1 and the shape rule, which requires $\mathrm{iso} > 10$, cannot flag it. A
single-sample jump of the whole line in which tension falls towards the fairlead escapes both
artifact rules. It is spatially smooth, which silences the coherence ratio of Eq.~\eqref{eq:coh},
and the record maximum
of the fairlead is often itself inflated by a blowup elsewhere in the run, which silences the
$\mathrm{fair}$ test. The fairlead condition prevents such a sample from being classified as a
snap. It is not masked, so it remains in the training data, and the test windows covering it are
screened out as described below. Appendix~\ref{app:thresholds} justifies the thresholds of both
rules from Eq.~\eqref{eq:axial_balance} and the archived records.

Exclusion is applied per window rather than per simulation. A simulation is discarded outright
only if it contains non-finite values (\discardNonfinite{} simulations) or ended too early to
yield one window (\discardShort{}), which leaves \num{\nInUse} of the \num{\nSeaStates}
simulations. Otherwise the flagged
samples are masked and every window covering one is dropped. The mask covers \num{\maskSamples}
samples in \maskSims{} simulations, \maskShare{}\,\si{\percent} of the archive, of which the
ceiling flags \num{\ceilingSamples}. Excluding per window preserves the extremes. Of
\snapEvents{} catalogued snap events, discarding every simulation the mask touches would leave
\snapReachBefore{}\,\si{\percent} reachable, whereas per-window exclusion leaves
\snapReachAfter{}\,\si{\percent} (Table~\ref{tab:snapaccounting}).

The fairlead condition was applied to the test windows only after both models had been trained.
Training used the mask above, and \catPoolShare{}\,\si{\percent} of the windows available for
training and validation contain a sample that violates the condition. In the test draws, every
window covering such a sample is removed, \catDropSone{} of the \num{50000} Stage-1 windows and
\catDropStwo{} of the \num{5000} Stage-2 windows, most of them at loc10 (\catDropSoneTen{} and
\catDropStwoTen{}). All test results in Section~\ref{sec:results} are computed on the remaining
windows.

\begin{table}[t]
\centering
\caption{Snap events reachable under each exclusion policy, by split. Whole-case: every
simulation the mask touches is discarded. Reachable: at least one window covering the event lies
in the split and contains no masked sample.}
\label{tab:snapaccounting}
\footnotesize
\setlength{\tabcolsep}{3pt}
\begin{tabular}{@{}l r r r r@{}}
\toprule
{Split} & {Present} & {Whole-case} & {Per-window} & {Gain} \\
\midrule
Training and validation & 517 & 124 & 381 & $+257$ \\
Test, temporal & 194 & 46 & 152 & $+106$ \\
Test, withheld & 191 & 43 & 121 & $+78$ \\
\midrule
Total & 902 & 213 & 654 & $+441$ \\
\bottomrule
\end{tabular}

\end{table}

\section{Surrogate model}\label{sec:method}

\begin{figure*}[pos=t]
\centering
\resizebox{\textwidth}{!}{%
\begin{tikzpicture}[
    >=Latex,
    font=\normalsize,
  ]
  \colorlet{cLine}{blue!55!black}
  \colorlet{cAccent}{red!65!black}
  \colorlet{cFill}{blue!7!white}
  \colorlet{cFillB}{red!6!white}

  \tikzset{
    blk/.style={rounded corners=2.5pt, draw=cLine!75!black, fill=cFill,
                line width=0.7pt, text width=3.05cm, align=center,
                inner sep=4pt, minimum height=2.15cm},
    blkq/.style={blk, draw=cAccent!75!black, fill=cFillB, dashed},
    blkout/.style={blk, draw=cLine, fill=white, line width=1.1pt},
    dims/.style={font=\scriptsize, align=center, text width=3.4cm,
                 color=black!72},
    flow/.style={-{Latex[length=2.2mm,width=1.6mm]}, line width=0.9pt,
                 color=cLine!75!black},
    gate/.style={rounded corners=1.2pt, draw=cLine!75!black, fill=cFill,
                 line width=0.5pt, minimum width=0.62cm, minimum height=0.42cm,
                 inner sep=0pt, font=\scriptsize},
    op/.style={circle, draw=cLine!75!black, fill=white, line width=0.5pt,
               inner sep=0pt, minimum size=0.36cm, font=\scriptsize},
    cellflow/.style={-{Latex[length=1.6mm,width=1.2mm]}, line width=0.6pt,
                     color=cLine!75!black},
  }

  \node[blk] (in) at (0,0)
    {\textbf{Input}\\[2pt]
     {\footnotesize\itshape graph on the line,\\ motion over \SI{100}{\second}}};
  \node[blk] (gat) at (3.75,0)
    {\textbf{Spatial encoder}\\[1pt]{\footnotesize GATv2 $\times 2$}\\[2pt]
     {\footnotesize\itshape message passing along\\ the line, global readout}};
  \node[blkq] (dec) at (7.5,0)
    {\textbf{Query decoder}\\[1pt]{\footnotesize cross-attention}\\[2pt]
     {\footnotesize\itshape $N_{\mathrm{in}}$ measured\\ $\rightarrow$ \nStationsMax{} queried}};
  \node[blk] (lstm) at (11.25,0)
    {\textbf{Temporal encoder}\\[1pt]{\footnotesize bidirectional LSTM $\times 2$}\\[2pt]
     {\footnotesize\itshape per station, over the\\ whole window}};
  \node[blk] (head) at (15.0,0)
    {\textbf{Prediction head}\\[1pt]{\footnotesize MLP}\\[2pt]
     {\footnotesize\itshape per station and sample,\\ shared weights}};
  \node[blkout] (out) at (18.75,0)
    {\textbf{Tension field}\\[3pt]
     $\widehat{\mathbf{T}} \in \mathbb{R}^{W \times N_{\mathrm{out}}}$\\[2pt]
     {\footnotesize\itshape the same \SI{100}{\second}}};

  \foreach \a/\b in {in/gat, gat/dec, dec/lstm, lstm/head, head/out}
    \draw[flow] (\a) -- (\b);

  \node[dims, below=3pt of in.south]
    {static node $[N_{\mathrm{in}}\times\nStatNodeFeat{}]$,
     dynamic node $[W\times N_{\mathrm{in}}\times\nDynNodeFeat{}]$\\
     static edge $[E\times\nStatEdgeFeat{}]$,
     dynamic edge $[W\times E\times\nDynEdgeFeat{}]$\\[2pt]
     $W=\windowSamples{}$ samples};
  \node[dims, below=3pt of gat.south]
    {$\nNodeFeat{} \rightarrow \gatConcat{}\,(\gatHeads{}\times\gatHidden{})
      \rightarrow \gatOut{}$\\[2pt]
     one embedding per station\\ at every sample};
  \node[dims, below=3pt of dec.south]
    {$\gatOut{} \rightarrow \decDim{}$, \decHeads{} heads\\[2pt]
     interpolation skip is\\ the identity when
     $N_{\mathrm{in}} = \nStationsMax{}$};
  \node[dims, below=3pt of lstm.south]
    {hidden \lstmHidden{} per direction,
     \lstmLayers{} layers $\rightarrow \lstmOut{}$\\[2pt]
     full sequence, not a summary};
  \node[dims, below=3pt of head.south]
    {$\lstmOut{} \rightarrow \headHidden{} \rightarrow 1$\\[2pt]
     one evaluation per window};
  \node[dims, below=3pt of out.south]
    {axial tension at every\\ queried station};

  \draw[cAccent!70!black, line width=0.5pt, dashed, rounded corners=3pt]
    ($(dec.north west)+(-0.16,0.42)$) rectangle ($(dec.south east)+(0.16,-0.16)$);
  \node[font=\footnotesize\itshape, color=cAccent!80!black, anchor=south]
    at ($(dec.north)+(0,0.50)$) {added for sparse sensing};

\end{tikzpicture}%
}
\caption{Model architecture, end to end. Every block is defined per station or per segment,
so no width depends on $N_{\mathrm{in}}$ and one set of weights serves all \nLayouts{}
sensor layouts.}
\label{fig:architecture}
\end{figure*}
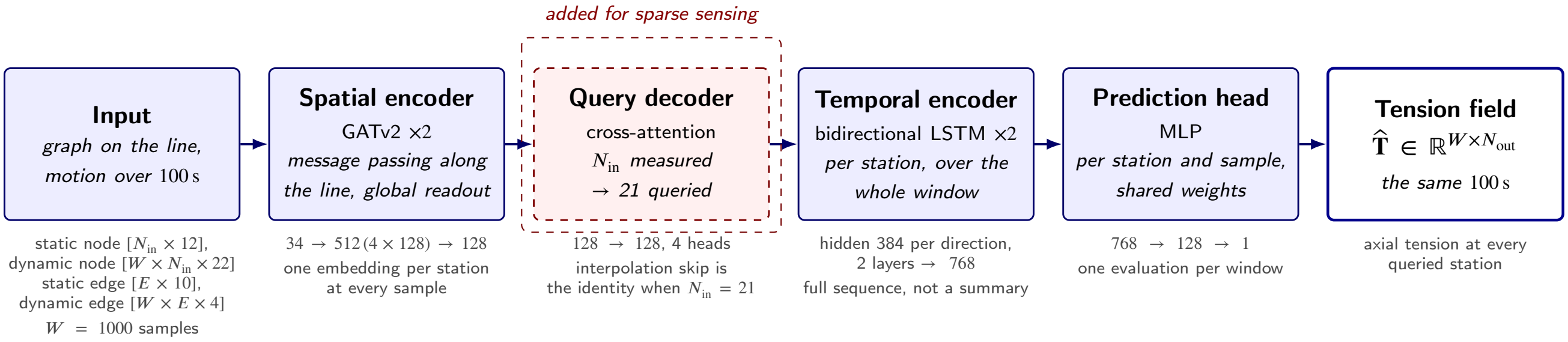

This section explains the surrogate model for the map of Section~\ref{subsec:task}, from motion at
$N_{\mathrm{in}}$ stations over a \windowSeconds{}~s window to axial tension at $N_{\mathrm{out}}$
stations over the same window. One window of motion enters the model and the tension over that
window leaves it in a single evaluation. No prediction is fed back as an input, so the model is not
autoregressive. Figure~\ref{fig:pipeline} shows how a window of the archive becomes the input of
the model and its target, and Fig.~\ref{fig:architecture} the path through the model and the width
of every representation along it. The subsections follow that path, through the graph and its
features (Section~\ref{subsec:graph}), the spatial encoder (Section~\ref{subsec:gat}), the query
decoder that Stage~2 adds (Section~\ref{subsec:decoders}), the temporal encoder
(Section~\ref{subsec:lstm}) and the prediction head (Section~\ref{subsec:head}), and end with the
training loss (Section~\ref{subsec:loss}) and the evaluation, training and selection procedure
(Section~\ref{subsec:selection}).

\begin{figure*}[pos=t]
\centering
\resizebox{\textwidth}{!}{%
\begin{tikzpicture}[
    >=Latex,
    font=\normalsize,
  ]
  \colorlet{cLine}{blue!55!black}
  \colorlet{cAccent}{red!65!black}
  \colorlet{cFill}{blue!7!white}
  \colorlet{cFillT}{red!6!white}

  \tikzset{
    pblk/.style={rounded corners=2.5pt, draw=cLine!75!black, fill=cFill,
                 line width=0.7pt, text width=3.55cm, align=center,
                 inner sep=4pt, minimum height=3.10cm},
    pblkt/.style={rounded corners=2.5pt, draw=cAccent!75!black, fill=cFillT,
                  line width=0.7pt, align=center, inner sep=4pt, minimum height=1.25cm},
    flow/.style={-{Latex[length=2.2mm,width=1.6mm]}, line width=0.9pt,
                 color=cLine!75!black},
    flowt/.style={-{Latex[length=2.2mm,width=1.6mm]}, line width=0.9pt,
                  color=cAccent!75!black},
    sensedp/.style={circle, draw=cAccent, fill=cAccent, minimum size=5.5pt, inner sep=0pt},
  }

  \node[pblk] (arc) at (0,0)
    {\textbf{Simulation archive}\\[4pt]
     {\footnotesize \textbullet\ \nStationsMax{} native stations\\[1pt]
      \textbullet\ \recordSamples{} samples at \solverStep{}~s\\[1pt]
      \textbullet\ $x$, $z$ and $T$}};
  \node[pblk] (lay) at (4.15,0)
    {\textbf{Sensor layout and window}\\[4pt]
     {\footnotesize \textbullet\ $x, z$ resampled in arc length $\rightarrow N_{\mathrm{in}}$\\[1pt]
      \textbullet\ windows $W = \windowSamples{}$ samples (\windowSeconds{}~s)}};
  \node[pblk] (der) at (8.30,0)
    {\textbf{Derived channels}\\[4pt]
     {\footnotesize \textbullet\ $\mathrm{d}x/\mathrm{d}t,\ \mathrm{d}z/\mathrm{d}t \rightarrow
      v_{\mathrm{t}}, v_{\mathrm{n}} \rightarrow F_{\mathrm{t}}, F_{\mathrm{n}}$\\[1pt]
      \textbullet\ seabed contact, penetration and reaction\\[3pt]
      {\itshape Appendix~\ref{app:features}}}};
  \node[pblk] (gr) at (12.45,0) {};
  \node[pblk] (inp) at (16.60,0)
    {\textbf{Model input}\\[4pt]
     {\footnotesize static node $[N_{\mathrm{in}}\times\nStatNodeFeat{}]$\\
      dynamic node $[W\times N_{\mathrm{in}}\times\nDynNodeFeat{}]$\\
      static edge $[E\times\nStatEdgeFeat{}]$\\
      dynamic edge $[W\times E\times\nDynEdgeFeat{}]$\\[3pt]
      {\itshape Appendix~\ref{app:features}}}};

  \node[font=\normalsize\bfseries, anchor=north] at ($(gr.north)+(0,-0.10)$)
    {Graph at each sample};
  \foreach \i/\dx in {0/-1.20, 1/-0.40, 2/0.40, 3/1.20}
    \node[sensedp] (g\i) at ($(gr.center)+(\dx,0.18)$) {};
  \foreach \a/\b in {0/1, 1/2, 2/3} {
    \draw[-{Latex[length=1.3mm,width=1.0mm]}, cLine, line width=0.55pt]
      ($(g\a.east)+(0.04,0.06)$) -- ($(g\b.west)+(-0.04,0.06)$);
    \draw[-{Latex[length=1.3mm,width=1.0mm]}, cLine, line width=0.55pt]
      ($(g\b.west)+(-0.04,-0.06)$) -- ($(g\a.east)+(0.04,-0.06)$);
  }
  \node[anchor=north, align=left, font=\footnotesize] at ($(gr.center)+(0,-0.15)$)
    {\textbullet\ station: \nNodeFeat{} features\\[1pt]
     \textbullet\ segment: \nEdgeFeat{} features\\[1pt]
     \textbullet\ both directions};

  \foreach \a/\b in {arc/lay, lay/der, der/gr, gr/inp}
    \draw[flow] (\a) -- (\b);

  \node[pblkt, text width=11.25cm] (tgt) at (8.30,-2.85)
    {\textbf{Target}\quad {\footnotesize $T$ resampled in arc length $\rightarrow N_{\mathrm{out}}$
      ($N_{\mathrm{in}}$ in Stage~1, \nStationsMax{} in Stage~2) $\rightarrow$ standardised
      $\rightarrow [W\times N_{\mathrm{out}}]$; never an input}};
  \node[pblkt, text width=3.55cm] (loss) at (16.60,-2.85)
    {\textbf{Loss and metrics}\\[1pt]
     {\footnotesize against the prediction $\widehat{\mathbf{T}}$}};
  \draw[flowt] (arc.south) |- (tgt.west);
  \draw[flowt] (tgt) -- (loss);
  \draw[flowt, dashed] (inp.south) -- (loss.north);
\end{tikzpicture}%
}
\caption{From the simulation archive to the model. Motion at the sensed stations becomes the
input (top row) and tension at the output stations of the same window the target (bottom row).
$E = 2(N_{\mathrm{in}} - 1)$ directed edges.}
\label{fig:pipeline}
\end{figure*}
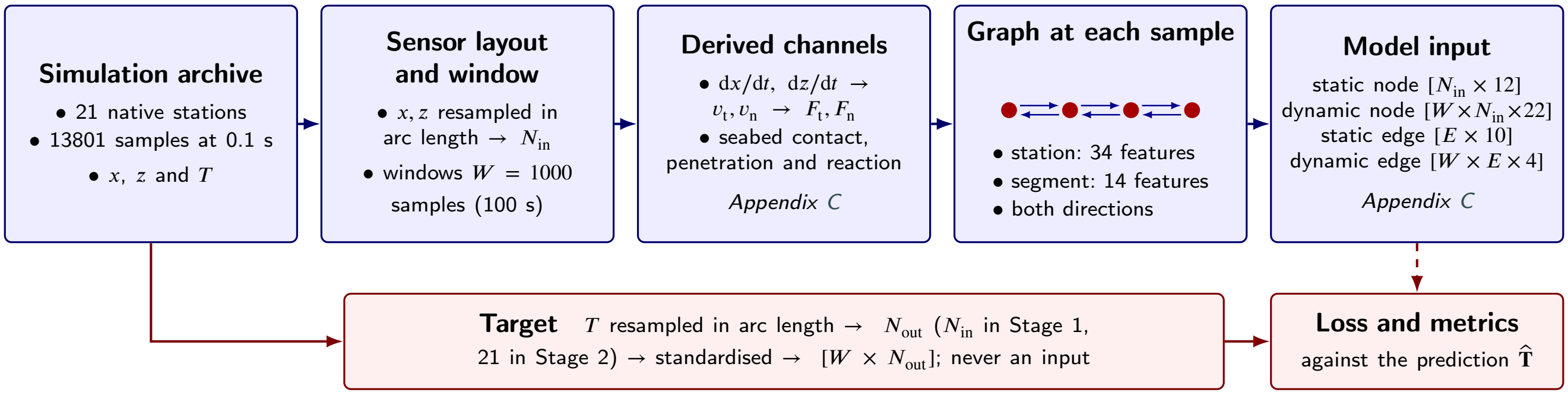

\subsection{Graph representation and features}\label{subsec:graph}

The stations of a sensor layout lie along the line in a chain, each joined to the next by a
segment of the line, so they map onto a graph that follows the physical topology of the line.
Stations become vertices, ordered from the anchor at $i = 0$ to the fairlead at
$i = N_{\mathrm{in}} - 1$, and each segment becomes a pair of directed edges, one pointing towards
the anchor and one towards the fairlead, so that information travels both ways along the line
during message passing. No self-loops are added, so the edges are exactly the physical segments.

Four feature matrices describe the line (Fig.~\ref{fig:pipeline}, defined channel by channel in
Appendix~\ref{app:features}). Two are static, fixed for a simulation, and two carry one row per
sample of the window. Their concatenated widths are \nNodeFeat{} per station and \nEdgeFeat{} per
segment.

Most dynamic channels are computed from the measured positions rather than read from the archive,
which holds only position and tension. Known physics is thus supplied to the network as additional
input channels, rather than left for it to rediscover from coordinates. Velocities are first
differences of position in time, central inside the window and
one-sided at its edges. Differencing position along the line instead gives the instantaneous unit
tangent, which rotates each velocity into the local frame of the line. The rotation matters because
drag on a slender cylinder is strongly anisotropic, so a velocity in global coordinates alone
cannot express the drag law. The axial and normal components then give approximate quadratic drag
forces, computed from the velocity of the line rather than its velocity relative to the water
(Appendix~\ref{app:features}). Seabed contact contributes a binary flag, the penetration below the bed and a single
reaction that combines submerged weight with a distributed spring and damper term, in newtons. The
\nEnvChannels{} environmental channels, constant within a simulation, are attached to every station
so that the encoder can condition on the sea state.

Continuous channels are z-scored with statistics fitted on training windows only. The two binary
channels, the nodal contact flag and the touchdown indicator, are passed through unchanged, since
they are already bounded and a rarely active column has a standard deviation close to zero.

\subsection{Spatial encoder}\label{subsec:gat}

The tension at a station depends on the configuration of the line around it and not only on the
motion of the station itself. A grounded station, for instance, does not move at all. The spatial
encoder therefore combines, at each sample, the features of every station with those of its
neighbours along the line. It acts on one sample of the window at a time and returns an embedding
for every station, a learned vector of fixed width, \gatOut{}, that summarises the state of the
station and its surroundings. It is a two-layer graph attention network of the GATv2 form
\citep{brody2022gatv2}, a standard layer that is adopted here unchanged. Graph attention combined
with recurrence has been applied to a floating offshore wind turbine by
\citet{yung2025structural,yung2025omae}, on a graph of the turbine, and the present model follows
that combination but places the graph on the mooring line itself.

Graph attention weights each neighbour's contribution by a learned score
\citep{velickovic2018gat}. In the original form the score that station $i$ assigns to its
neighbour $j$ is given by Eq.~\eqref{eq:gat_score},
\begin{equation}
    e_{ij}^{\mathrm{GAT}} = \mathrm{LeakyReLU}\!\left(
        \mathbf{a}^{\top}\!\left[\mathbf{W}\mathbf{h}_i \,\big\|\,
        \mathbf{W}\mathbf{h}_j\right]\right),
    \label{eq:gat_score}
\end{equation}
with $\mathbf{W}$ a learned weight matrix that projects the embedding of every station into a
common space, shared by all stations, $\mathbf{a}$ a learnable attention vector, $\|$
concatenation and $\mathrm{LeakyReLU}(x) = x$ for $x > 0$, $\alpha x$ otherwise, a rectifier that
keeps a small gradient ($\alpha = 0.2$ here) for negative arguments instead of the zero gradient of
a plain rectified linear unit. Splitting $\mathbf{a}$ over the two halves of the concatenation makes
Eq.~\eqref{eq:gat_score} separable, so one term alone orders all neighbours $j$ irrespective of
the querying station. GATv2 removes this restriction by applying the nonlinearity before the
attention vector rather than after it, and the segment features enter as a third additive term,
as implemented in PyTorch Geometric \citep{fey2019pyg}, which gives the score of
Eq.~\eqref{eq:gatv2_score},
\begin{equation}
    e_{ij} = \mathbf{a}^{\top}\,\mathrm{LeakyReLU}\!\left(
        \boldsymbol{\Theta}_{l}\mathbf{h}_i + \boldsymbol{\Theta}_{r}\mathbf{h}_j
        + \boldsymbol{\Theta}_{e}\mathbf{s}_{ij}\right),
    \label{eq:gatv2_score}
\end{equation}
where $\mathbf{h}_i$ is the embedding that station $i$ carries into the layer, $\mathbf{s}_{ij}$
the concatenated static and dynamic edge features of the segment between $i$ and $j$, and
$\boldsymbol{\Theta}_{l}$, $\boldsymbol{\Theta}_{r}$ and $\boldsymbol{\Theta}_{e}$ independently
learned projections of the querying station, the neighbour and the segment. The biases of the node
projections and of the layer output are omitted. For the first layer $\mathbf{h}_i$ is the
concatenated node feature vector of Fig.~\ref{fig:pipeline}, and for the second it is the output
$\mathbf{h}_i^{(1)}$ of the first after normalisation, activation and dropout, with the line-global
vector of Eq.~\eqref{eq:gc_broadcast} added. The scores are normalised over the neighbourhood
$\mathcal{N}(i)$ of each station, its one or two physical neighbours without the station itself,
unlike in the original GAT \citep{velickovic2018gat}, and weight the projected neighbour states
in the aggregate of Eq.~\eqref{eq:gatv2_aggregate},
\begin{equation}
    \alpha_{ij} = \frac{\exp(e_{ij})}{\sum_{k \in \mathcal{N}(i)} \exp(e_{ik})},
    \qquad
    \mathbf{h}_i' = \sum_{j \in \mathcal{N}(i)} \alpha_{ij}\,
        \boldsymbol{\Theta}_{r}\mathbf{h}_j .
    \label{eq:gatv2_aggregate}
\end{equation}
Because the left projection shifts the operating point of the nonlinearity through which the
right one passes, the ranking of neighbours depends on the querying station. The line needs this,
since the relative importance of the anchor-side and fairlead-side neighbours changes as the
touchdown point migrates. Segment features enter the scores but not the aggregated values, so they
decide how much weight a neighbour receives while the transmitted information remains a projection
of the neighbour's own state. The first layer uses \gatHeads{} attention heads, independent copies
of Eqs.~\eqref{eq:gatv2_score} and~\eqref{eq:gatv2_aggregate} with their own projections, so that
each head can weight the neighbours by a different aspect of their state. Their outputs are
concatenated to width \gatConcat{}, and the second layer uses a single head projecting to \gatOut{}.
Each layer is followed
by layer normalisation \citep{ba2016layernorm}, an exponential linear unit
($\mathrm{ELU}(x) = x$ for $x > 0$, $\exp(x) - 1$ otherwise) and dropout, and the
attention coefficients are also subject to dropout, both at a rate of \gatDropout{}. A learned
linear skip from the raw concatenated input, added to the normalised output of the second layer,
restores the station's own information, so a purely local map remains available while the
attention weights are still uninformative.

Two message-passing layers reach two hops, and the temporal encoder that follows acts on each
station separately, so without further coupling the receptive field of the model is the two-hop
neighbourhood. This matters here. A grounded station does not move, so its own motion channels
carry no information about the load it transmits, and that information sits at the touchdown
point, which on a finely discretised line lies many hops away. A line-global readout removes this
bound. Between the two layers, the station embeddings are averaged over the line,
Eq.~\eqref{eq:gc_pool}, passed through a small network, Eq.~\eqref{eq:gc_mlp}, and added back to
every station, Eq.~\eqref{eq:gc_broadcast},
\begin{align}
    \bar{\mathbf{h}} &= \frac{1}{N_{\mathrm{in}}}\sum_{j} \mathbf{h}_j^{(1)},
        \label{eq:gc_pool}\\[2pt]
    \mathbf{g} &= \mathrm{ELU}\!\left(\mathbf{W}_{g}\,\bar{\mathbf{h}}
        + \mathbf{b}_{g}\right),
        \label{eq:gc_mlp}\\[2pt]
    \tilde{\mathbf{h}}_i^{(1)} &= \mathbf{h}_i^{(1)} + \mathbf{g}
        \qquad \text{for every station } i .
        \label{eq:gc_broadcast}
\end{align}
The mean runs over the whole line, so $\mathbf{g}$ depends on every station at any
discretisation. Equivalently, this is message passing on the graph augmented with one virtual
station joined to all others \citep[a master node, as in][]{gilmer2017mpnn}, which reduces the
effective diameter of the graph from $N_{\mathrm{in}} - 1$ to two. The same vector is added
everywhere, so it summarises the line rather than addressing one station, and localisation comes
from the receiving station, which knows its normalised arc length from its static features.
Equations~\eqref{eq:gc_pool}--\eqref{eq:gc_broadcast} are evaluated at every sample and per line,
so lines processed together do not mix. A third and a fourth message-passing layer would instead
reach only four hops, still short of the line, at the cost of an extra attention computation at
every sample and a risk of oversmoothing.

\subsection{Query decoder}\label{subsec:decoders}

In Stage~1 the tension is required only at the sensed stations, so their embeddings pass directly
to the temporal encoder. Stage~2 requires it at all \nStationsMax{} native stations, most of which
carry no sensor, and therefore needs a component that turns the embeddings of the sensed stations
into embeddings of the output stations. This is the query decoder. It decodes the encoded sensed
stations into a representation at each requested output position, and so decouples the output
positions from the input ones. It does this by cross-attention \citep{vaswani2017attention}, in
which every output position forms a query that reads from the encoded stations, as in the output
queries of Perceiver IO \citep{jaegle2022perceiverio} and the sparse-sensor field reconstruction of
the Senseiver \citep{santos2023senseiver}. Here it is placed between the spatial and temporal
encoders rather than after both (Fig.~\ref{fig:architecture}). Attending at the width \gatOut{} of
the station embeddings rather than at the width \lstmOut{} of the temporal encoder keeps the
attention cost small and leaves the shapes of the temporal encoder and head unchanged, so the
trained Stage-1 model can initialise Stage~2 directly.

Positions along the line enter through a sinusoidal positional encoding
\citep{mildenhall2020nerf} of the normalised arc length $s \in [0,1]$, Eq.~\eqref{eq:fourier},
\begin{equation}
    \gamma(s) = \left[\sin(2^{k}\pi s),\; \cos(2^{k}\pi s)\right]_{k=0}^{\decBands{}-1}.
    \label{eq:fourier}
\end{equation}
The decoder acts on each sample of the window separately, so the sample index is left out.
Queries are formed at the \nStationsMax{} native output stations from their own static features
$\boldsymbol{\xi}_m$ (Table~\ref{tab:app_static_node}) and their arc length, while keys and values
are the output $\mathbf{h}_j^{\mathrm{in}}$ of the spatial encoder at the sensed stations, carrying
an embedding of their arc length, as in Eq.~\eqref{eq:query},
\begin{equation}
    \begin{aligned}
        \mathbf{q}_m &= \mathrm{LN}\!\left(\mathbf{W}_{q}\!\left[\boldsymbol{\xi}_m \,\big\|\,
            \gamma(s_m^{\mathrm{out}})\right] + \mathbf{b}_{q}\right),\\
        \mathbf{k}_j = \mathbf{v}_j &= \mathbf{h}_j^{\mathrm{in}}
            + \mathbf{W}_{k}\,\gamma(s_j^{\mathrm{in}}) + \mathbf{b}_{k},
    \end{aligned}
    \label{eq:query}
\end{equation}
with $\mathrm{LN}$ layer normalisation \citep{ba2016layernorm}. Attention is thus addressed by
position along the line rather than by index, so the same weights serve every sensor layout, and
the queries are always the \nStationsMax{} native stations. Multi-head attention with \decHeads{}
heads, in which the queries do not attend to one another, is followed by a feed-forward block of
width \decFFN{} with an exponential linear unit, each added to its own input and layer-normalised
\citep{vaswani2017attention}, which gives $\mathbf{u}_m$. Dropout at a rate of \decDropout{} acts on
the attention weights and inside the feed-forward block. The decoder returns the embedding of
Eq.~\eqref{eq:decoder_out},
\begin{equation}
    \mathbf{h}_m^{\mathrm{out}} = \bigl(\Pi\,\mathbf{h}^{\mathrm{in}}\bigr)_m
        + \mathbf{W}_{u}\,\mathbf{u}_m + \mathbf{b}_{u},
    \label{eq:decoder_out}
\end{equation}
where $\Pi$ interpolates the station embeddings linearly in arc length from the sensed to the
output positions, and the temporal encoder and head of Sections~\ref{subsec:lstm}
and~\ref{subsec:head} then act on $\mathbf{h}_m^{\mathrm{out}}$ at the output stations. The decoder
is thus a learned correction to an interpolated starting point rather than a map built from
scratch \citep{kim2016vdsr}, and $\mathbf{W}_{u}$ and $\mathbf{b}_{u}$ are initialised to zero, so
the correction starts at exactly zero. At an output station that coincides with a sensed one the
interpolation returns that station's embedding, and with all \nStationsMax{} stations sensed it is
the identity. An untrained Stage-2 model warm-started from Stage~1 therefore reproduces it exactly
there, as in the function-preserving initialisations of \citet{chen2016net2net}, which makes
Stage~2 a strict generalisation of Stage~1 rather than a different model. At the other output
stations the starting point is an interpolation of the station embeddings, which differs from
interpolating the predicted tension because the LSTM and head that follow are nonlinear.

Conditioning the decoder on position needs justification, because the closest study reaches the
opposite conclusion. \citet{xie2026sparse} compare three decoders for a shallow recurrent
architecture \citep{williams2024shred}, find the one they call spatial the weakest and a plain
fully connected decoder the strongest, and conclude that spatial conditioning of the decoder is
unnecessary for their problem. The two findings are consistent, because their task, forecasting
the tension of fixed line segments from platform motion, and the present one, reconstructing
tension from a sensor layout that changes, differ in what is fixed. Their model maps platform
motion of fixed size to a vector of fixed length whose entries are segments known at training
time, so the spatial basis is absorbed into the weights of the output layer. Here the output is
also fixed, the \nStationsMax{} native stations, but the input is not. The number and positions of
the sensed stations change with the layout, and with them which output stations coincide with a
sensor. A fully connected decoder over the station embeddings would have one input slot per sensed
station and would need one model per layout, which this work sets out to avoid, whereas attention
keyed on arc length has no such slots. Position conditioning is therefore what this task requires,
not an improvement on a decoder built for a different one.

\subsection{Temporal encoder}\label{subsec:lstm}

\begin{figure}[pos=t]
\centering
\resizebox{\columnwidth}{!}{%
\begin{tikzpicture}[
    >=Latex,
    font=\normalsize,
  ]
  \colorlet{cLine}{blue!55!black}
  \colorlet{cFill}{blue!7!white}

  \tikzset{
    gate/.style={rounded corners=1.2pt, draw=cLine!75!black, fill=cFill,
                 line width=0.5pt, minimum width=0.62cm, minimum height=0.42cm,
                 inner sep=0pt, font=\scriptsize},
    op/.style={circle, draw=cLine!75!black, fill=white, line width=0.5pt,
               inner sep=0pt, minimum size=0.36cm, font=\scriptsize},
    cellflow/.style={-{Latex[length=1.6mm,width=1.2mm]}, line width=0.6pt,
                     color=cLine!75!black},
  }

  \node[rounded corners=2.5pt, draw=cLine!45!black, fill=cFill!50, line width=0.6pt,
        minimum width=6.65cm, minimum height=4.05cm, inner sep=0pt] (cellbox) at (0,0) {};

  \begin{scope}[shift={($(cellbox.south west)+(1.12,0.48)$)}, x=1cm, y=1cm]
    \draw[cellflow] (-0.30,2.60) -- (5.02,2.60);
    \node[font=\scriptsize, anchor=east] at (-0.32,2.60) {$\mathbf{c}_{t-1}$};
    \node[font=\scriptsize, anchor=west] at (5.04,2.60) {$\mathbf{c}_{t}$};

    \draw[color=cLine!75!black, line width=0.6pt] (-0.30,0.40) -- (4.35,0.40);
    \node[font=\scriptsize, anchor=east] at (-0.32,0.40) {$\mathbf{h}_{t-1}$};
    \draw[color=cLine!75!black, line width=0.6pt] (0.10,-0.02) -- (0.10,0.40);
    \node[font=\scriptsize, anchor=north] at (0.10,-0.04) {$\mathbf{x}_{t}$};

    \node[gate] (gf) at (0.80,1.15) {$\sigma$};
    \node[gate] (gi) at (1.70,1.15) {$\sigma$};
    \node[gate, minimum width=0.72cm] (gc) at (2.62,1.15) {$\tanh$};
    \node[gate] (go) at (4.35,1.15) {$\sigma$};
    \foreach \g/\lab in {gf/{\mathbf{f}_t}, gi/{\mathbf{i}_t},
                         gc/{\tilde{\mathbf{c}}_t}, go/{\mathbf{o}_t}}
      \node[font=\scriptsize, anchor=south west] at ($(\g.north)+(0.05,0.03)$)
        {$\lab$};
    \foreach \g in {gf,gi,gc,go}
      \draw[cellflow] (\g.south |- 0,0.40) -- (\g.south);

    \node[op] (mf) at (0.80,2.60) {$\odot$};
    \node[op] (pl) at (2.20,2.60) {$+$};
    \node[op] (mi) at (2.20,1.85) {$\odot$};
    \node[op, rectangle, rounded corners=0.18cm, minimum width=0.74cm, minimum height=0.36cm] (th) at (3.55,2.12) {$\tanh$};
    \node[op] (mo) at (4.35,1.85) {$\odot$};

    \draw[cellflow] (gf) -- (mf);
    \draw[cellflow] (gi) |- (mi.west);
    \draw[cellflow] (gc) |- (mi.east);
    \draw[cellflow] (mi.north) -- (pl.south);
    \draw[color=cLine!75!black, line width=0.6pt] (3.55,2.60) -- (th.north);
    \draw[cellflow] (th.south) |- (mo.west);
    \draw[cellflow] (go) -- (mo);
    \draw[cellflow] (mo.east) -- (4.80,1.85) -- (4.80,0.40) -- (5.02,0.40);
    \node[font=\scriptsize, anchor=west] at (5.04,0.40) {$\mathbf{h}_{t}$};
  \end{scope}
\end{tikzpicture}%
}
\caption{One LSTM cell (Eqs.~\eqref{eq:lstm_forget}--\eqref{eq:lstm_hidden}), with the cell
state $\mathbf{c}_t$ running along the top as an additive highway.}
\label{fig:lstm_cell}
\end{figure}
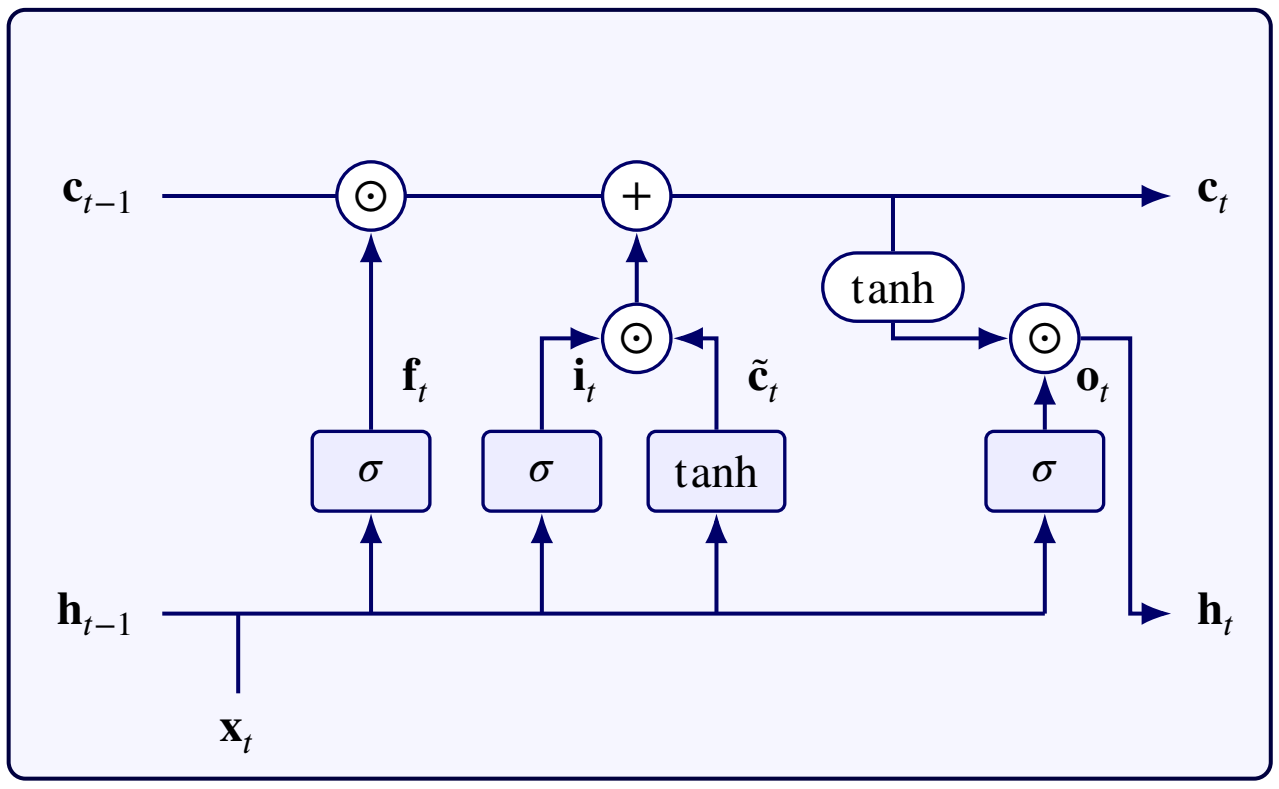

At every station, the sequence of embeddings over the window, from the spatial encoder in Stage~1
and from the query decoder in Stage~2, is passed to a long short-term memory (LSTM) network
\citep{hochreiter1997lstm}, whose gating retains information across a window spanning several wave
cycles. A cell with the forget gate of \citet{gers2000forget} maps an input $\mathbf{x}_t$ together
with the previous hidden and cell states through
\begin{align}
    \mathbf{f}_t &= \sigma\!\left(\mathbf{W}_{f}\mathbf{x}_t
        + \mathbf{U}_{f}\mathbf{h}_{t-1} + \mathbf{b}_{f}\right),
        \label{eq:lstm_forget}\\
    \mathbf{i}_t &= \sigma\!\left(\mathbf{W}_{i}\mathbf{x}_t
        + \mathbf{U}_{i}\mathbf{h}_{t-1} + \mathbf{b}_{i}\right),
        \label{eq:lstm_input}\\
    \tilde{\mathbf{c}}_t &= \tanh\!\left(\mathbf{W}_{c}\mathbf{x}_t
        + \mathbf{U}_{c}\mathbf{h}_{t-1} + \mathbf{b}_{c}\right),
        \label{eq:lstm_candidate}\\
    \mathbf{c}_t &= \mathbf{f}_t \odot \mathbf{c}_{t-1}
        + \mathbf{i}_t \odot \tilde{\mathbf{c}}_t,
        \label{eq:lstm_cell}\\
    \mathbf{o}_t &= \sigma\!\left(\mathbf{W}_{o}\mathbf{x}_t
        + \mathbf{U}_{o}\mathbf{h}_{t-1} + \mathbf{b}_{o}\right),
        \label{eq:lstm_output}\\
    \mathbf{h}_t &= \mathbf{o}_t \odot \tanh(\mathbf{c}_t),
        \label{eq:lstm_hidden}
\end{align}
where $\sigma$ is the logistic sigmoid and $\odot$ the elementwise product. The forget gate,
Eq.~\eqref{eq:lstm_forget}, decides what is discarded from the cell state, the input gate and the
candidate, Eqs.~\eqref{eq:lstm_input} and~\eqref{eq:lstm_candidate}, decide what is written, and the
output gate, Eq.~\eqref{eq:lstm_output}, selects what reaches the hidden state,
Eq.~\eqref{eq:lstm_hidden}. The additive update of Eq.~\eqref{eq:lstm_cell} keeps the gradient
path through time open.
Figure~\ref{fig:lstm_cell} shows one cell.

Two departures from the usual arrangement follow from the task. First, the encoder is
bidirectional \citep{graves2005blstm}, as in earlier mooring-tension networks
\citep{payenda2024rnn,xu2024wec}. One pass runs forward and one backward in time, so the
representation at a sample, Eq.~\eqref{eq:lstm_readout}, draws on the motion both before and after
it,
\begin{equation}
    \mathbf{z}_{i,t} = \mathrm{LN}\!\left(\left[\overrightarrow{\mathbf{h}}_{i,t} \,\big\|\,
        \overleftarrow{\mathbf{h}}_{i,t}\right]\right) \in \mathbb{R}^{\lstmOut{}},
    \label{eq:lstm_readout}
\end{equation}
where $\mathrm{LN}$ is layer normalisation over the concatenation of a forward and a backward
pass of width \lstmHidden{} each. The \lstmLayers{} layers are stacked, each after the first
receiving the concatenated states of the one below, with dropout at a rate of \lstmDropout{}
between them and after the normalisation. Second, the full sequence is kept rather than the state
at the last sample, since an estimate is required at every sample of the window. Each station is
processed independently, because the spatial encoder has already propagated neighbourhood
information into every embedding. The graph encoder thus handles the state of the neighbourhood at
an instant, and the LSTM the evolution of each station.

Since the estimate at an instant draws on motion recorded after it, the model suits digital twins,
fatigue back-analysis and the interpretation of recorded campaigns rather than instantaneous
alarms. Each window is processed on its own, from zero initial
states in both directions, so a sample near either end of a window has context from one side
only.

\subsection{Prediction head}\label{subsec:head}

The head maps the sequence of Eq.~\eqref{eq:lstm_readout} to tension with a two-layer perceptron
applied at every station and sample, Eq.~\eqref{eq:head_output},
\begin{equation}
    \hat{T}_{i,t} = \mathbf{W}_{2}\,\mathrm{ELU}\!\left(\mathbf{W}_{1}\mathbf{z}_{i,t}
        + \mathbf{b}_{1}\right) + \mathbf{b}_{2},
    \label{eq:head_output}
\end{equation}
with a hidden width of \headHidden{} and dropout at a rate of \headDropout{} between the layers.
Since linear layers act on the last axis, one evaluation produces the whole field
$\widehat{\mathbf{T}} \in \mathbb{R}^{W \times N_{\mathrm{out}}}$ at once.

No weight matrix or width in Eqs.~\eqref{eq:gatv2_score}--\eqref{eq:head_output} depends on the
number of stations, which enters only as the normaliser of the line mean in Eq.~\eqref{eq:gc_pool}.
Message passing on a path graph, attention addressed by arc length, a per-station LSTM and a shared
head are all defined station by station, so the parameter count is independent of
$N_{\mathrm{in}}$ and one set of weights serves every layout of Eq.~\eqref{eq:layouts}. All windows
in a batch share one station count, so the graphs of a batch are processed as one block-diagonal
supergraph in a single call, with no message crossing between them.

\subsection{Physics-guided multi-term loss}\label{subsec:loss}

The training loss combines \nLossTerms{} terms through learned weights, two for fit, two for
temporal structure, three for physics and one for safety.

Two terms measure fit, the mean squared and the mean absolute error of Eq.~\eqref{eq:loss_fit},
\begin{equation}
    \begin{aligned}
    L_{\mathrm{mse}} &= \big\langle (\hat{T}_{i,t} - T_{i,t})^{2} \big\rangle, \\
    L_{\mathrm{l1}}  &= \big\langle \lvert \hat{T}_{i,t} - T_{i,t} \rvert \big\rangle,
    \end{aligned}
    \label{eq:loss_fit}
\end{equation}
where $\langle \cdot \rangle$ denotes the mean over the windows of a batch and over every
station $i$ and sample $t$, taken on standardised tension unless stated otherwise, that is on
$(T - \mu_T)/\sigma_T$ with $\mu_T = \SI{\tensionMean}{\kilo\newton}$ and
$\sigma_T = \SI{\tensionStd}{\kilo\newton}$ the mean and standard deviation of the training
tension. Two further terms measure temporal structure, since a small pointwise error can coexist
with a predicted series that has lost its temporal detail, the errors in the step-to-step change
and in the amplitude spectrum of Eq.~\eqref{eq:loss_temporal},
\begin{equation}
    \begin{aligned}
    L_{\mathrm{grad}} &= \big\langle (\Delta \hat{T}_{i,t} - \Delta T_{i,t})^{2} \big\rangle, \\
    L_{\mathrm{spec}} &= \big\langle \big( \lvert \mathcal{F}(\hat{T}_i)_f \rvert
        - \lvert \mathcal{F}(T_i)_f \rvert \big)^{2} \big\rangle, \\
    \mathcal{F}(T_i)_f &= \frac{1}{W} \sum_{t=0}^{W-1} T_{i,t}\, e^{-2\pi\mathrm{i} f t / W},
    \end{aligned}
    \label{eq:loss_temporal}
\end{equation}
with $\Delta$ the first difference along the sample axis and $\mathcal{F}$ the discrete Fourier
transform at the $W/2 + 1$ non-negative frequencies $f$, over which the mean of the second term
runs in place of $t$. The factor $1/W$ makes the amplitude of a component independent of the
window length.

Three terms are residuals of governing relations, forming a hierarchy of material law, static
equilibrium and dynamic equilibrium. Each is zero when its relation holds exactly, which the
simulated tension satisfies only approximately, and each is built only from what the model already
receives, the input positions, the contact flag and the static properties of the line, so no term
adds information. For Stage~2 they are evaluated at the \nStationsMax{} output stations, from the
input positions and contact flag interpolated onto them by the map $\Pi$ of
Eq.~\eqref{eq:decoder_out} and the static properties of the output stations. With
$\mathbf{p}_i = (x_i, z_i)$ the physical input position of station $i$ at sample $t$, the
constitutive term, Eq.~\eqref{eq:loss_constitutive}, compares the prediction with the tension
implied by the pretension and the strain of the segment chord,
\begin{equation}
    L_{\mathrm{const}} = \Bigl\langle \Bigl(
        \frac{\hat{T}_{i+1/2} - \mathrm{relu}(T_{0,i+1/2} + EA\,\varepsilon_{i+1/2})}{\sigma_T}
        \Bigr)^{2} \Bigr\rangle,
    \label{eq:loss_constitutive}
\end{equation}
in units of $\sigma_T$, with $\langle \cdot \rangle$ now the mean over every segment and sample
rather than every station, $\hat{T}_{i+1/2} = \tfrac{1}{2}(\hat{T}_i + \hat{T}_{i+1})$ the
prediction at the segment midpoint in newtons, $EA$ the axial stiffness, $T_{0,i+1/2}$ the
pretension of Table~\ref{tab:app_static_node} averaged onto the segment and
$\varepsilon_{i+1/2} = (\lVert \mathbf{p}_{i+1} - \mathbf{p}_i \rVert - \ell_{i+1/2}) /
\ell_{i+1/2}$ the strain of the chord between input stations $i$ and $i{+}1$ relative to its
initial length $\ell_{i+1/2}$ (Table~\ref{tab:app_static_edge}), at which the tension is the
pretension. The relu keeps the implied tension non-negative, since a slack segment carries
none. The catenary term, Eq.~\eqref{eq:loss_catenary}, is the residual
of the tangential balance $\mathrm{d}T/\mathrm{d}s = w_{s}\,\mathrm{d}z/\mathrm{d}s$ of a
cable under its own submerged weight per unit length $w_{s}$,
\begin{equation}
    L_{\mathrm{cat}} = \Bigl\langle \Bigl(
        \frac{(\hat{T}_{i+1} - \hat{T}_i) - w_{s}(z_{i+1} - z_i)}{\sigma_T} \Bigr)^{2}
        \Bigr\rangle,
    \label{eq:loss_catenary}
\end{equation}
in units of $\sigma_T$. The momentum term is
Newton's second law at an interior station, Eq.~\eqref{eq:loss_momentum},
\begin{equation}
    \begin{aligned}
    m_i\,\ddot{\mathbf{p}}_i &= \mathbf{F}_i^{\mathrm{net}}, \\
    \mathbf{F}_i^{\mathrm{net}} &= \hat{T}_{i+1/2}\,\mathbf{t}_{i+1/2}
        - \hat{T}_{i-1/2}\,\mathbf{t}_{i-1/2} \\
        &\quad + (0, -w_i) + \mathbf{F}_{\mathrm{drag},i},
    \end{aligned}
    \label{eq:loss_momentum}
\end{equation}
with $m_i$ and $w_i$ the mass and submerged weight of Table~\ref{tab:app_static_node},
$\ddot{\mathbf{p}}_i$ the acceleration of the input position, its second central difference in
time divided by $\Delta t^{2}$ for the sampling interval $\Delta t = \SI{\solverStep}{\second}$,
$\mathbf{t}_{i\pm1/2}$ the current (not the static reference) unit chord tangent, and
$\mathbf{F}_{\mathrm{drag},i}$ the drag of Table~\ref{tab:app_dynamic_node}, recomputed from the
input positions. The term, Eq.~\eqref{eq:loss_momentum_residual}, is the mean squared residual of
this balance per force component, in units of $\sigma_T$, over the admissible station-samples
$\mathcal{M}$,
\begin{equation}
    L_{\mathrm{mom}} = \frac{1}{2\lvert\mathcal{M}\rvert} \sum_{(t,i)\,\in\,\mathcal{M}}
        \frac{\big\lVert m_i\,\ddot{\mathbf{p}}_i - \mathbf{F}_i^{\mathrm{net}} \big\rVert^{2}}
             {\sigma_T^{2}},
    \label{eq:loss_momentum_residual}
\end{equation}
where $\mathcal{M}$ excludes the anchor and the fairlead, whose motion is a prescribed
boundary condition, stations in seabed contact, whose reaction this residual does not model,
and the first and last sample of the window, where the second difference is undefined. Every
quantity but $\hat{T}$ is fixed by the inputs, so the gradient reaches the
prediction only through the elastic-transport term, which ties the kinematics to the tension
field through inertia and drag.

The eighth term encodes safety, a quantile, or pinball, loss \citep{koenker1978regression}
applied to every entry whose true tension exceeds \SI{\pinballThr}{\kilo\newton}, the
\pinballPct{}th percentile of training tension, so that it acts on the highest
\SI{15}{\percent} of the training loads, where the peaks that govern design lie,
Eq.~\eqref{eq:loss_pinball},
\begin{equation}
    L_{\mathrm{pin}} = \big\langle \rho_{\tau}(T_{i,t} - \hat{T}_{i,t}) \big\rangle, \qquad
    \rho_{\tau}(u) = \max\!\big(\tau u,\ (\tau - 1) u\big),
    \label{eq:loss_pinball}
\end{equation}
where the averaging brackets $\langle \cdot \rangle$ here denote the mean over those entries only,
and $\tau = \pinballTau{}$, which penalises underprediction nine times as heavily as
overprediction. The loss is two-sided: it penalises overprediction too, with the weight
$1 - \tau$, rather than underprediction alone, which keeps its learned weight in
Eq.~\eqref{eq:kendall} stable. Under this loss the optimal point estimate is the $\tau$
quantile of the conditional distribution rather than its mean \citep{gneiting2011quantiles}, so
the estimate is deliberately and quantifiably biased towards conservatism where the loads are
largest.

Fixed weights for \nLossTerms{} terms whose natural scales differ by orders of magnitude would
have to be found by trial and error, and the balance they strike would be sensitive to small
changes in any term. Each term therefore carries a learnable log-variance $s_k$, and the terms are
combined by the homoscedastic-uncertainty formulation of \citet{kendall2018multitask},
Eq.~\eqref{eq:kendall},
\begin{equation}
    L = \sum_{k} \left[ \tfrac{1}{2}\,e^{-s_k} L_k + \tfrac{1}{2}\,s_k \right].
    \label{eq:kendall}
\end{equation}
The first part down-weights a term the model cannot satisfy, and the second prevents the trivial
solution of switching every term off. The log-variances are excluded from weight decay, which would
otherwise pull them towards zero, that is, towards unit variance. Before the first update they are
set to $s_k = \ln L_k$, at which the $k$-th bracket of Eq.~\eqref{eq:kendall} is stationary, with
$L_k$ the mean of the term over a few validation batches of the initial model (untrained in
Stage~1, warm-started from Stage~1 in Stage~2), so that a term of large numerical scale cannot
dominate the first epochs. The converged weights also rank the terms by how well each describes
the data (Appendix~\ref{app:kendall}).

\subsection{Evaluation metrics, training and model selection}\label{subsec:selection}

Model quality is judged by six metrics on physical-unit tension, supplemented below by peak
statistics at the fairlead, by a per-station $R^{2}$ and by the capture of snap events. The
metrics answer different questions. $R^{2}$ is the share of the variance of the tension that the
reconstruction explains, MAE and MAPE give the size of the error in newtons and relative to the
load, and $S$ measures how well the change from one sample to the next is followed, which $R^{2}$
and MAE can miss when an estimate follows the level of the tension but smooths its waves. The tail
statistics measure the error at the peaks that govern design. $R^{2}$
is pooled over every station, sample and window of the evaluation split, and the tail statistics
of Eq.~\eqref{eq:metrics_tail} over the peak of every station and window. The mean absolute error
(MAE), mean absolute percentage error (MAPE) and temporal skill score ($S$) are instead computed
within each window and then averaged unweighted over
windows, so that a window is not down-weighted for covering fewer stations. With $T_{i,t}$ the
ground truth and $\hat{T}_{i,t}$ the prediction at station $i$ and sample $t$, $\mathcal{W}$ the set
of evaluation windows, a window scored at several sensor layouts counting once per layout, and
$\langle \cdot \rangle_{w}$ the mean over the stations and samples of window $w$,
\begin{equation}
\begin{gathered}
    \mathrm{MAE} = \frac{1}{\lvert\mathcal{W}\rvert} \sum_{w \in \mathcal{W}}
        \big\langle \lvert \hat{T}_{i,t} - T_{i,t} \rvert \big\rangle_{w}, \\[4pt]
    R^{2} = 1 - \frac{\sum_{w,i,t} (\hat{T}_{i,t} - T_{i,t})^{2}}
                     {\sum_{w,i,t} (T_{i,t} - \bar{T})^{2}},
\end{gathered}
\label{eq:metrics_mae_r2}
\end{equation}
with the sums over every window, station and sample of the split, $\bar{T}$ the mean of $T$ over
the same entries and the ratio taken once, so $R^{2}$ is sensitive to offsets between windows and
not only to within-window shape, while MAE weights every window equally regardless of its station
count. The percentage error of Eq.~\eqref{eq:metrics_mape} is likewise averaged per window and
gated against near-slack entries,
\begin{equation}
    \mathrm{MAPE} = \frac{1}{\lvert\mathcal{W}\rvert} \sum_{w \in \mathcal{W}}
    \frac{100\%}{n_{w}} \sum_{(i,t) \in w^{+}}
    \frac{\lvert \hat{T}_{i,t} - T_{i,t}\rvert}{\lvert T_{i,t}\rvert},
    \label{eq:metrics_mape}
\end{equation}
where $w^{+}$ holds the $n_{w}$ entries of window $w$ with
$\lvert T_{i,t}\rvert > T_{\min} = \SI{\mapeThrN}{\newton}$, and a window with none is
excluded from the outer average. Temporal fidelity is a per-window skill score against a
temporally flat predictor, whose step difference is identically zero so its error equals
$\lvert \Delta T_{i,t}\rvert$, again averaged unweighted over windows,
\begin{equation}
\begin{gathered}
    S_{w} = \frac{\big\langle \lvert \Delta \hat{T}_{i,t} - \Delta T_{i,t} \rvert \big\rangle_{w}}
             {\big\langle \lvert \Delta T_{i,t} \rvert \big\rangle_{w}}, \\[4pt]
    S = \frac{1}{\lvert\mathcal{W}\rvert} \sum_{w \in \mathcal{W}} S_{w}, \qquad
    \Delta T_{i,t} = T_{i,t+1} - T_{i,t},
\end{gathered}
\label{eq:metrics_skill}
\end{equation}
with $S < 1$ indicating that the model tracks step-to-step change more accurately, on average
over windows, than a constant prediction. Two further metrics characterise the extreme upper
tail of the distribution of peak tension, the target of the peak-pinball loss and of the selection
gate of Table~\ref{tab:training}.
For each station $i$ and window $w$ let $T^{\mathrm{peak}}_{i,w} = \max_{t \in w} T_{i,t}$ be the
true peak tension, $\hat{T}^{\mathrm{peak}}_{i,w} = \max_{t \in w} \hat{T}_{i,t}$ the peak of the
prediction, irrespective of when in the window either occurs, and
$e_{i,w} = \hat{T}^{\mathrm{peak}}_{i,w} - T^{\mathrm{peak}}_{i,w}$ the signed peak error,
positive when the peak is overpredicted. The $p_q$ tail $\mathcal{T}_q$ is the set of
station-windows whose true peak is at or above the $q$-th percentile $T_q$ of the pooled true
peaks, that is, the top $(100-q)\%$ up to ties, the pool being the split or, in a per-layout table, that
layout. The upper-tail underprediction rate and the under-mass, the mean shortfall
per tail entry, are
\begin{equation}
\begin{gathered}
    u_q = \frac{100\%}{\lvert \mathcal{T}_q \rvert} \sum_{(i,w) \in \mathcal{T}_q}
    \mathbf{1}[e_{i,w} < 0], \\[4pt]
    \text{under-mass}_q = \big\langle \max(-e_{i,w},\, 0) \big\rangle_{\mathcal{T}_q}
    = \frac{\mathrm{MAE}_q - \mathrm{bias}_q}{2},
\end{gathered}
\label{eq:metrics_tail}
\end{equation}
with $\mathrm{bias}_q = \big\langle e_{i,w} \big\rangle_{\mathcal{T}_q}$ and
$\mathrm{MAE}_q = \big\langle \lvert e_{i,w} \rvert \big\rangle_{\mathcal{T}_q}$ the mean
signed and absolute error over the tail, reported at $q = 90$ and $q = 99$. A positive
$\mathrm{bias}_q$ marks a conservative tail. The peak MAE is the mean of
$\lvert e_{i,w} \rvert$ over the stations of a window, averaged over windows as the MAE is.

The fairlead governs the design load and is sensed in every layout, so the peak statistics are
also evaluated at the fairlead alone, with $\mathcal{T}_q$ then the top $(100-q)\%$ of the
fairlead peaks. With $T^{\mathrm{peak}}_{\mathrm{f},w}$ the true and
$\hat{T}^{\mathrm{peak}}_{\mathrm{f},w}$ the predicted fairlead peak of window $w$,
$e_{\mathrm{f},w}$ their difference, and $\mathcal{W}_s$ the windows of a subset such as one
site, three statistics are computed without tail selection, namely the median peak ratio, the
peak percentage error and the peak underprediction share,
\begin{equation}
\begin{aligned}
    \tilde{\rho} &= \operatorname*{median}_{w \in \mathcal{W}_s}
        \frac{\hat{T}^{\mathrm{peak}}_{\mathrm{f},w}}{T^{\mathrm{peak}}_{\mathrm{f},w}}, \\
    \mathrm{MAPE}^{\mathrm{peak}} &= \frac{100\%}{\lvert\mathcal{W}_s\rvert}
        \sum_{w \in \mathcal{W}_s} \frac{\lvert e_{\mathrm{f},w}\rvert}{T^{\mathrm{peak}}_{\mathrm{f},w}}, \\
    u^{\mathrm{peak}} &= \frac{100\%}{\lvert\mathcal{W}_s\rvert}
        \sum_{w \in \mathcal{W}_s} \mathbf{1}[e_{\mathrm{f},w} < 0],
\end{aligned}
\label{eq:metrics_fairlead}
\end{equation}
where $\tilde{\rho} > 1$ marks a conservative peak. $\mathrm{MAPE}^{\mathrm{peak}}$ has one entry
per window and needs no gate against near-slack tension, since every fairlead peak of the test
split exceeds $T_{\min}$.

Accuracy along the line is judged by the coefficient of determination of each station within
each window,
\begin{equation}
    R^{2}_{i,w} = 1 - \frac{\sum_{t \in w} (\hat{T}_{i,t} - T_{i,t})^{2}}
        {\sum_{t \in w} (T_{i,t} - \bar{T}_{i,w})^{2}}, \qquad
    \bar{T}_{i,w} = \big\langle T_{i,t} \big\rangle_{t \in w},
    \label{eq:metrics_station}
\end{equation}
which, unlike the pooled $R^{2}$ of Eq.~\eqref{eq:metrics_mae_r2}, is normalised by the variance
of one station over one window alone. A level offset therefore weighs in proportion to how little
the station oscillates, and $R^{2}_{i,w}$ is negative whenever the error exceeds that
oscillation. Station-windows whose true tension is constant are excluded from these statistics,
since $R^{2}_{i,w}$ is undefined for them. Each station-window is assigned to a
contact region by the fraction $c_{i,w}$ of the window in which the station rests on the seabed,
read from the contact flag of the input features and, where the outputs lie at other stations
than the inputs, interpolated to them in unstretched arc length. A station-window is grounded for
$c_{i,w} \ge \SI{98}{\percent}$, suspended for $c_{i,w} \le \SI{2}{\percent}$ and in the touchdown
region in between. The median of $R^{2}_{i,w}$ and the share of station-windows with $R^{2}_{i,w} < 0$ are
reported per region. Figures~\ref{fig:alongline} and~\ref{fig:foursensor} quote beside
$R^{2}_{i,w}$ the mean absolute error of the station-window,
$\mathrm{MAE}_{i,w} = \langle \lvert \hat{T}_{i,t} - T_{i,t} \rvert \rangle_{t \in w}$.

Snap events are those catalogued in Section~\ref{subsec:snaps}. Snap samples within
\SI{2}{\second} of the first of a group form one event, placed at the sample $t_e$ of largest
tension in the group and at the station $n_e$ carrying it. An event falls inside several
overlapping test windows, and inside the same window at several layouts, so snap loads are scored
per distinct event. With $\mathcal{E}$ the distinct events of a test set and $\mathcal{W}_e$ the
scored windows containing $t_e$, restricted to one layout where a layout is stated, capture is
measured in two ways. The first concerns the line maximum, the largest tension anywhere on the line
at the instant of the snap. With $T^{\mathrm{line}}_{w,t} = \max_i T_{i,t}$ the true and
$\hat{T}^{\mathrm{line}}_{w,t}$ the predicted line maximum over the output stations of window $w$
at sample $t$, the capture of an event is the median over its windows of the ratio of predicted to
true line maximum, Eq.~\eqref{eq:snap_capture},
\begin{equation}
    c_{e} = \operatorname*{median}_{w \in \mathcal{W}_e}
        \frac{\hat{T}^{\mathrm{line}}_{w,t_e}}{T^{\mathrm{line}}_{w,t_e}}.
    \label{eq:snap_capture}
\end{equation}
For a set of events $\mathcal{E}_s \subseteq \mathcal{E}$, such as those of one site, the median
capture $\tilde{c}$ and the share of events captured in full $F$ are
\begin{equation}
\begin{aligned}
    \tilde{c} &= \operatorname*{median}_{e \in \mathcal{E}_s} c_e, \\
    F &= \frac{100\%}{\lvert\mathcal{E}_s\rvert} \sum_{e \in \mathcal{E}_s} \mathbf{1}[c_e \ge 1].
\end{aligned}
\label{eq:snap_summary}
\end{equation}
A capture below one marks an underscaled maximum. The second measure is taken at the station
$n_e$ itself, as the median over $\mathcal{W}_e$ of $\hat{T}_{n_e,t_e} / T_{n_e,t_e}$, is
summarised by the same $\tilde{c}$ and $F$, and exists only where the output layout contains that
station. A variant replaces both
tensions by their maxima over $t_e$ and the two samples either side.

The cost of sensing fewer stations is quoted as the ratio of the MAE with $N_{\mathrm{in}}$
sensed stations to the MAE of the same model with all $N_{\mathrm{out}} = \nStationsMax{}$
stations sensed, on the same test windows, and likewise for $S$.

Table~\ref{tab:training} lists the optimisation and selection hyperparameters, shared by the
two stages unless a stage is given. The models are implemented in PyTorch
with PyTorch Geometric \citep{paszke2019pytorch,fey2019pyg} and trained with the AdamW
optimiser \citep{loshchilov2019adamw}, which adapts the step of every parameter to the history of
its gradients, on a single NVIDIA A100 \citep{DHPC2024}. The learning rate itself is also adapted.
It starts at $\optLr{}$ and is multiplied by \schedFactor{} whenever the validation $R^{2}$ has not
improved by more than $\schedThreshold{}$ for \schedPatience{} epochs, down to a floor of
$\schedFloor{}$, which each stage reached after \schedDecays{} reductions. Training took about
\sOneTrainHours{}~h for Stage~1 and \sTwoTrainHours{}~h for Stage~2, over \sOneEpochsRun{} and
\sTwoEpochs{} epochs, and Appendix~\ref{app:training} shows how the training and validation metrics
evolved. Reconstructing the test windows, about \num{50000} per stage, took \sOneTestMin{} and
\sTwoTestMin{}~min, or \sOneTestMs{} and \sTwoTestMs{}~ms per \windowSeconds{}-s window including
data loading and metric computation. A \simMinutes{}-minute record is thus reconstructed in about
\recordReconSec{}~s, against a median of \solverHoursMed{}~h of solver time for the
finite-element model (Section~\ref{subsec:fe_data}). A window is processed far faster than it is
recorded, so the model can follow a monitoring campaign continuously, with a delay of one
\windowSeconds{}-s window, which the bidirectional encoder needs in full. It therefore serves
monitoring and hindcast analysis rather than instantaneous alarms.

\begin{table}[t]
\centering
\caption{Optimisation and selection hyperparameters.}
\label{tab:training}
\footnotesize
\setlength{\tabcolsep}{3pt}
\begin{tabular}{@{}l l@{}}
\toprule
{Hyperparameter} & {Value} \\
\midrule
\multicolumn{2}{@{}l}{\itshape Optimisation} \\
Optimiser                                       & AdamW \\
Initial learning rate                           & $\optLr{}$ \\
Learning-rate reduction factor                  & \schedFactor{} \\
Plateau threshold on validation $R^{2}$         & $\schedThreshold{}$ \\
Plateau patience                                & \schedPatience{} epochs \\
Learning-rate floor                             & $\schedFloor{}$ \\
Weight decay                                    & $\optWd{}$ \\
Gradient-norm clip                              & \optClip{} \\
Effective batch size                            & \optBatch{} windows \\
Dropout                                         & \gatDropout{} \\
Maximum epochs, Stage~1 / Stage~2               & \sOneEpochsMax{} / \sTwoEpochs{} \\
Early-stopping patience                         & \selEarlyStopPatience{} epochs \\
\addlinespace
\multicolumn{2}{@{}l}{\itshape Model selection} \\
$R^{2}$ tolerance below the best so far, $\Delta R^{2}$ & \selRsqDelta{} \\
Upper-tail cap, $u_{90}$ (Stage~1 only)         & \selUpCap{}\,\si{\percent} \\
Selected epoch, Stage~1 / Stage~2               & \sOneSelEpoch{} / \sTwoSelEpoch{} \\
\bottomrule
\end{tabular}
\end{table}

Selection cannot rest on accuracy alone. Validation $R^{2}$ saturates early and then varies only
in the fifth decimal place, while tail behaviour over the same epochs does not, so ranking epochs
by accuracy is close to ranking them at random on safety. An epoch is therefore admissible when its
validation $R^{2}$ lies within \selRsqDelta{} of the best so far and, for Stage~1, when fewer than
\selUpCap{}\,\si{\percent} of the upper-tail entries are underpredicted, and the admissible epoch
with the lowest $S$ is selected. The underprediction rate is used to reject and never to rank,
because it is a sign statistic on a distribution centred near zero and fluctuates far more than
the behaviour underlying it. The accuracy floor is a rule for choosing which epoch to keep, not a
condition on training, which proceeds whatever the epoch's admissibility. Stage~1 was trained for
up to \sOneEpochsMax{} epochs with early stopping on the selection score (patience
\selEarlyStopPatience{} epochs), which halted it at epoch~\sOneEpochsRun{}, and
epoch~\sOneSelEpoch{} was selected. The patience is long because the selection score is noisy and
admissible epochs are rare, so a short patience could stop training before the next one. The
\sOnePostSelEpochs{} epochs after the selected one, about \sOnePostSelHours{}~h of training,
lowered the validation MAE from \sOneValMAESel{} to as little as \sOneValMAEBest{}~N, but none of
them was admissible. Stage~2 was warm-started from it
and trained for up to \sTwoEpochs{} epochs under the same early-stopping rule, which did not fire,
and epoch~\sTwoSelEpoch{} was selected on the accuracy floor and the skill objective alone. The
tail condition was dropped for Stage~2 after a first training run and before any test evaluation,
because annealing lowers the conservative margin of the $\tau = \pinballTau{}$ term as the model
converges on the conditional mean, so the condition excluded the converged epochs rather than the
unsafe ones. Section~\ref{sec:results} reports what this costs.

The architecture and the optimisation settings were fixed beforehand in shorter runs, scored on
validation windows only, with a pool of \num{\searchPool} training windows and \searchLayouts{} of
the sensor layouts. These runs were made on earlier versions of the data curation of
Section~\ref{subsec:snaps}, so their numbers are comparable only within one comparison. Over the
same epochs, a bidirectional temporal encoder lowered the validation $S$ from \searchSUni{} to
\searchSBi{} against a unidirectional one. Reducing the hidden width from \lstmHidden{} to
\searchNarrow{} per direction, or the number of LSTM layers from two to one, changed $S$ by at most
\searchArchPct{}\,\si{\percent}, less than the \searchSeedPct{}\,\si{\percent} spread between
\searchSeeds{} training seeds of the same configuration. Without the line-global readout of
Section~\ref{subsec:gat}, the median station $R^{2}_{i,w}$ in the grounded region was
$\gcOffGrounded{}$, against \gcOnGrounded{} with it. An absolute rather than a relative threshold in
the plateau test of the learning-rate schedule lowered $S$ by \searchLRPct{}\,\si{\percent}. The
learning rate, the dropout rates and the patience were not searched.

\section{Results and discussion}\label{sec:results}

Both stages are evaluated on test windows built as in Table~\ref{tab:splits},
\oodTestShare{}\,\si{\percent} of them from the two withheld sites, but on different draws.
Stage~1 (Section~\ref{subsec:res_matched}) is scored on its own draw at the stations it senses.
Stage~2 (Section~\ref{subsec:res_sparse}) is scored on an independent draw, paired across sensor
layouts, and compared with an interpolation baseline, in which Stage~1 is run unchanged on the
$N_{\mathrm{in}}$ sensed stations and its predicted tension is interpolated linearly in unstretched
arc length onto the 21 stations. The baseline is thus what Stage~1 delivers without a decoder, and
its difference from Stage~2 is what the query decoder and its training add. The
\sOneFullGapPct{}\,\si{\percent} difference between the baseline MAE with all 21 stations sensed,
\sOneFullMAE{}~N (Section~\ref{subsec:res_sparse}), and the Stage-1 MAE for the same layout in
Table~\ref{tab:res_layouts}, \sOneFullMAETable{}~N, is sampling variation between the two draws,
not a difference in the model. The metrics are those of Section~\ref{subsec:selection}.

\subsection{Reconstruction at matched resolution}\label{subsec:res_matched}

\subsubsection{Accuracy across sensor layouts and sites}\label{subsec:res_layouts}

Table~\ref{tab:res_layouts} lists the Stage-1 test metrics per sensor layout. One set of weights
serves all ten with nearly the same fidelity. The pooled $R^{2}$ lies between \resRsqMin{} and
\resRsqMax{} and $S$ between \resSkillMin{} and \resSkillMax{}, and the MAE falls from
\resMAEFour{}~N at four stations to \resMAEFine{}--\resMAEFineHi{}~N from ten stations upwards.
Because each layout is evaluated at its own stations, differences between rows reflect both the
layout and the stations averaged over.

Stage~1 was also compared with a series of linear regression baselines that use the same inputs
and are scored on the same test windows (Appendix~\ref{app:linear}). The most advanced of them
adds the motion of the station and of the fairlead at lags of up to $\pm$\lrLagLong{}~s, yet its
MAE is \lrMAERatio{} times that of Stage~1, \lrMAEFour{} against \sOneMAE{}~N, and its $S$ is
\lrSFour{} against \lrSOne{}. At the two shallowest sites it follows the change between samples
worse than a constant prediction in most windows.

\begin{table*}[t]
\centering
\caption{Stage-1 test metrics per sensor layout. Tension-field metrics over all stations, peak
metrics over the window maximum of each station, with the tail thresholds $T_{90}$ and
$T_{99}$ taken within each layout.}
\label{tab:res_layouts}
\footnotesize
\setlength{\tabcolsep}{4pt}
\begin{tabular}{@{}S[table-format=2.0]S[table-format=1.4]S[table-format=2.1]S[table-format=1.2]S[table-format=1.3]S[table-format=3.1]S[table-format=2.2]S[table-format=2.1]S[table-format=3.0]S[table-format=2.2]S[table-format=2.1]S[table-format=4.0]@{}}
\toprule
 & \multicolumn{4}{c}{Tension field} & {Peaks} & \multicolumn{3}{c}{$p_{90}$ peak tail} & \multicolumn{3}{c}{$p_{99}$ peak tail} \\
\cmidrule(lr){2-5}\cmidrule(lr){6-6}\cmidrule(lr){7-9}\cmidrule(lr){10-12}
{$N$} & {$R^2$} & {MAE} & {MAPE} & {$S$} & {MAE} & {$T_{90}$} & {$u_{90}$} & {Under-mass} & {$T_{99}$} & {$u_{99}$} & {Under-mass} \\
 &  & {[\si{\newton}]} & {[\si{\percent}]} &  & {[\si{\newton}]} & {[\si{\kilo\newton}]} & {[\si{\percent}]} & {[\si{\newton}]} & {[\si{\kilo\newton}]} & {[\si{\percent}]} & {[\si{\newton}]} \\
\midrule
4 & 0.9992 & 57.9 & 2.12 & 0.389 & 66.1 & 10.58 & 28.8 & 24 & 13.89 & 41.1 & 110 \\
5 & 0.9993 & 48.2 & 1.84 & 0.372 & 76.4 & 9.98 & 6.4 & 74 & 13.43 & 14.1 & 172 \\
6 & 0.9993 & 45.5 & 1.78 & 0.374 & 67.3 & 9.01 & 4.7 & 72 & 13.22 & 11.9 & 248 \\
7 & 0.9994 & 43.2 & 1.74 & 0.360 & 63.8 & 8.45 & 4.5 & 19 & 13.15 & 10.7 & 66 \\
8 & 0.9994 & 42.2 & 1.70 & 0.360 & 61.2 & 8.36 & 5.0 & 76 & 13.03 & 12.6 & 296 \\
10 & 0.9994 & 39.1 & 1.53 & 0.362 & 56.6 & 8.16 & 2.6 & 20 & 12.71 & 7.9 & 41 \\
12 & 0.9994 & 38.4 & 1.48 & 0.364 & 54.0 & 8.06 & 2.2 & 6 & 12.52 & 4.3 & 12 \\
15 & 0.9994 & 38.3 & 1.48 & 0.361 & 55.9 & 8.01 & 2.3 & 16 & 12.60 & 9.8 & 116 \\
18 & 0.9993 & 38.1 & 1.47 & 0.364 & 58.0 & 7.77 & 2.3 & 13 & 12.32 & 6.7 & 50 \\
21 & 0.9993 & 38.4 & 1.51 & 0.363 & 56.8 & 7.78 & 2.5 & 10 & 12.22 & 6.6 & 26 \\
\midrule
{All} & 0.9993 & 42.9 & 1.66 & 0.367 & 61.6 & 8.17 & 4.3 & 26 & 12.74 & 12.6 & 94 \\
\bottomrule
\end{tabular}

\end{table*}

In the upper tail the underprediction rate and the under-mass need not agree. At four stations
$u_{99}$ is \resUpNNFour{}\,\si{\percent}, against \resUpNNFifteen{}\,\si{\percent} at fifteen, yet
the $p_{99}$ under-mass is lower, \resUMNNFour{} against \resUMNNFifteen{}~N, so the misses of the
four-station layout are more frequent but smaller.

\begin{table*}[t]
\centering
\caption{Fairlead-peak metrics per site, pooled over the ten sensor layouts: median peak ratio
$\tilde{\rho}$, peak percentage error $\mathrm{MAPE}^{\mathrm{peak}}$ and peak underprediction
share $u^{\mathrm{peak}}$ (Eq.~\eqref{eq:metrics_fairlead}). $h_0/L_0$: water depth over
unstretched line length.}
\label{tab:res_sites}
\footnotesize
\setlength{\tabcolsep}{4pt}
\begin{tabular}{@{}l S[table-format=1.3] S[table-format=1.3] S[table-format=1.2] S[table-format=2.1] S[table-format=1.3] S[table-format=1.2] S[table-format=2.1]@{}}
\toprule
 & & \multicolumn{3}{c}{Stage~1} & \multicolumn{3}{c}{Stage~2} \\
\cmidrule(lr){3-5}\cmidrule(lr){6-8}
{Site} & {$h_0/L_0$} & {$\tilde{\rho}$} & {$\mathrm{MAPE}^{\mathrm{peak}}$} & {$u^{\mathrm{peak}}$} & {$\tilde{\rho}$} & {$\mathrm{MAPE}^{\mathrm{peak}}$} & {$u^{\mathrm{peak}}$} \\
 &  &  & {[\si{\percent}]} & {[\si{\percent}]} &  & {[\si{\percent}]} & {[\si{\percent}]} \\
\midrule
loc01 & 0.401 & 1.010 & 1.01 & 2.3 & 1.002 & 0.26 & 13.0 \\
loc03 & 0.475 & 1.008 & 0.80 & 3.7 & 1.000 & 0.17 & 54.4 \\
loc04 & 0.425 & 1.008 & 0.87 & 5.2 & 0.997 & 0.35 & 92.6 \\
loc05 & 0.373 & 1.009 & 0.88 & 5.0 & 1.000 & 0.18 & 56.5 \\
loc06 & 0.193 & 1.004 & 0.74 & 16.0 & 0.993 & 0.94 & 96.0 \\
loc07 & 0.388 & 1.008 & 0.81 & 6.9 & 0.999 & 0.17 & 74.0 \\
loc08 & 0.402 & 1.008 & 0.85 & 3.2 & 1.000 & 0.18 & 42.5 \\
loc09 & 0.448 & 1.009 & 0.94 & 1.7 & 1.000 & 0.21 & 53.8 \\
\addlinespace
Training, pooled & {--} & 1.008 & 0.86 & 5.5 & 0.999 & 0.31 & 60.2 \\
\midrule
loc10 & 0.200 & 0.974 & 2.70 & 94.0 & 0.956 & 4.62 & 99.9 \\
loc11 & 0.368 & 1.013 & 1.34 & 2.8 & 1.010 & 1.01 & 3.5 \\
\bottomrule
\end{tabular}

\end{table*}

Table~\ref{tab:res_sites} breaks the fairlead-peak metrics down by site for both stages, each on
its own test draw. At seven of the eight training sites Stage~1 has a $u^{\mathrm{peak}}$ of
\siteDevUpLo{}--\siteDevUpHi{}\,\si{\percent}, and over all eight the median peak ratio is
$\tilde{\rho} = \siteDevRatioSone{}$, slightly above unity, as the $\tau = \pinballTau{}$ term
intends. The exception is loc06, at \siteSixUpSone{}\,\si{\percent}, the only shallow training
site ($h_0/L_0 = \siteSixHL{}$, against \siteNextHL{} or more elsewhere).

The two withheld sites behave differently. The first, loc11, with $h_0/L_0 = \siteElevenHL{}$ and
close to several training sites, transfers with a small offset on the conservative side, with
$\tilde{\rho} = \siteElevenRatioSone{}$ and $u^{\mathrm{peak}} = \siteElevenUpSone{}$\,\si{\percent},
although its $\mathrm{MAPE}^{\mathrm{peak}}$ of \siteElevenErrSone{}\,\si{\percent} exceeds the
\siteDevErrSone{}\,\si{\percent} of the training sites. At loc10, $u^{\mathrm{peak}}$ is
\siteTenUpSone{}\,\si{\percent}, with $\tilde{\rho} = \siteTenRatioSone{}$ and
$\mathrm{MAPE}^{\mathrm{peak}} = \siteTenErrSone{}$\,\si{\percent}. At
$h_0/L_0 = \siteTenHL{}$, loc10 has a single neighbour among the training sites, loc06, itself the
weakest of them. Stage~2 shows the same ordering, with a $\mathrm{MAPE}^{\mathrm{peak}}$ of
\siteTenErrStwo{}\,\si{\percent} at loc10, \siteElevenErrStwo{}\,\si{\percent} at loc11 and
\siteDevErrStwo{}\,\si{\percent} at the training sites. The peak term of the loss
(Eq.~\eqref{eq:loss_pinball}), which moves the prediction to the conservative side, also plays
almost no role at the two shallow sites. Only \pinShareSix{}\,\si{\percent} of the entries at loc06
and \pinShareTen{}\,\si{\percent} at loc10 exceed its threshold of \SI{\pinballThr}{\kilo\newton},
against \pinShareTrainLo{}--\pinShareTrainHi{}\,\si{\percent} at the other training sites. These
findings suggest that more than one shallow training site is needed to remove this limitation.

\subsubsection{Reconstruction along the line}\label{subsec:res_spatial}

Table~\ref{tab:res_regions} gives the median station $R^{2}_{i,w}$ (Eq.~\eqref{eq:metrics_station})
and the share of negative values in each contact region (Section~\ref{subsec:selection}). For
Stage~1 the medians are \sOneGroundedRsq{}, \sOneTouchdownRsq{} and \sOneSuspendedRsq{} in the
grounded, touchdown and suspended regions, with \sOneGroundedNeg{}, \sOneTouchdownNeg{} and
\sOneSuspendedNeg{}\,\si{\percent} of station-windows negative.

\begin{table}[t]
\centering
\caption{Station $R^{2}_{i,w}$ by contact region: median, and share of station-windows below
zero. Stage~2 and interpolation: $N_{\mathrm{out}} = 21$, pooled over $N_{\mathrm{in}}$, with
regions labelled from the input contact flag interpolated to the output stations.}
\label{tab:res_regions}
\footnotesize
\setlength{\tabcolsep}{3pt}
\begin{tabular}{@{}l S[table-format=1.3] S[table-format=1.1] S[table-format=1.3] S[table-format=1.1] S[table-format=1.3] S[table-format=2.1]@{}}
\toprule
 & \multicolumn{2}{c}{Stage~1} & \multicolumn{2}{c}{Stage~2} & \multicolumn{2}{c}{Interpolation} \\
\cmidrule(lr){2-3}\cmidrule(lr){4-5}\cmidrule(lr){6-7}
{Region} & {Median} & {$<0$ [\si{\percent}]} & {Median} & {$<0$ [\si{\percent}]} & {Median} & {$<0$ [\si{\percent}]} \\
\midrule
Grounded & 0.990 & 7.9 & 0.995 & 1.9 & 0.988 & 9.7 \\
Touchdown & 0.991 & 0.0 & 0.991 & 6.6 & 0.172 & 47.8 \\
Suspended & 0.945 & 4.4 & 0.989 & 9.5 & 0.898 & 8.6 \\
\bottomrule
\end{tabular}

\end{table}

\begin{figure*}[pos=p]
\centering
\includegraphics[width=\textwidth]{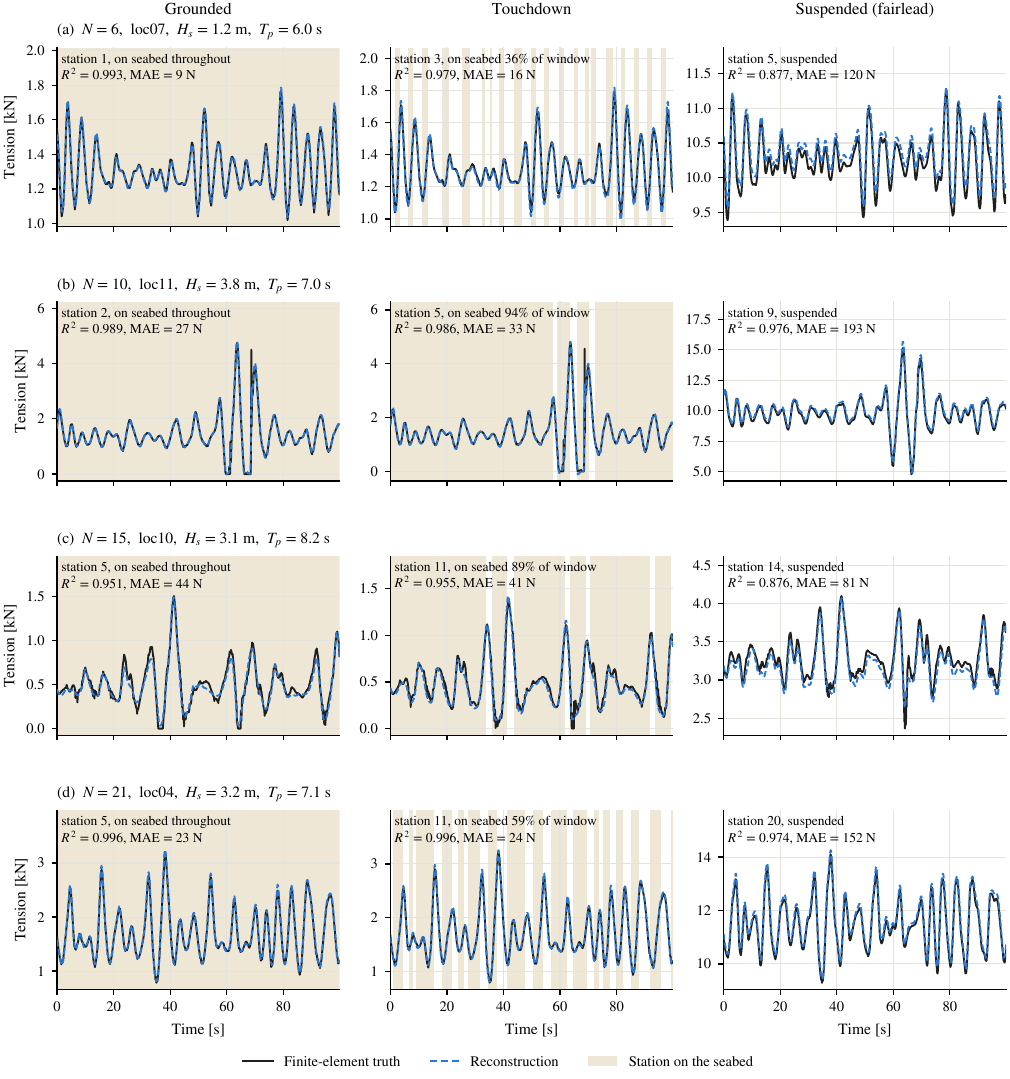}
\caption{Stage-1 reconstruction at a grounded, a touchdown and a suspended station for four
sensor layouts, one test window each, with the $R^{2}_{i,w}$ and MAE of the
station over the window. The touchdown panel shows the station that rests on the seabed for
closest to half of the window. The window of row (b) contains a snap event.}
\label{fig:alongline}
\end{figure*}

Figure~\ref{fig:alongline} shows one test window for each of four layouts, selected on the ground
truth alone and including both withheld sites. The reconstruction follows the grounded stations,
which never leave the seabed, as closely as the suspended ones, and it follows the touchdown
station through its repeated lift-offs in every row and, in rows (b) and (c), through the dips in
which its tension falls to near zero.

The accuracy at the grounded stations is attributable to the line-global readout
(Section~\ref{subsec:gat}). A grounded station does not move, so its load must be inferred from
the touchdown region, which on a fine layout lies beyond the two-hop reach of message passing. In
the shorter runs of Section~\ref{subsec:selection}, removing the readout lowered the median grounded
$R^{2}_{i,w}$ on the validation windows from \gcOnGrounded{} to $\gcOffGrounded{}$.

For Stage~1 the suspended median is the lowest of the three, because of how the loss and the metric
interact. The peak term of the loss (Eq.~\eqref{eq:loss_pinball}) acts on entries whose true
tension exceeds \SI{\pinballThr}{\kilo\newton}, almost all of them at suspended stations, and at
$\tau = \pinballTau{}$ moves the prediction slightly above the truth there, whereas the grounded and
touchdown stations carry far less tension and are not offset. Because $R^{2}_{i,w}$ is normalised by
the variance within the window, and a suspended station in a calm sea oscillates little about a
high mean, even a small offset takes a large share of that variance. The suspended station in
Fig.~\ref{fig:alongline}a illustrates this. Its reconstruction follows every wave but lies above
the truth throughout, which gives it the lowest $R^{2}_{i,w}$ of its row. At loc10, in
Fig.~\ref{fig:alongline}c, the suspended station is reconstructed below the truth instead, by
\alongTenBias{}~N or \alongTenBiasPct{}\,\si{\percent} of its mean tension. The offset is nearly
constant, between \alongTenBiasLo{} and \alongTenBiasHi{}~N in each quarter of the window, with a
trend of only \alongTenDrift{}~N over the \windowSeconds{}~s, so it is a bias in the level rather
than a drift. It is the offset of this shallow withheld site found in
Section~\ref{subsec:res_layouts}.

\subsubsection{Extremes and snap loads}\label{subsec:res_extremes}

\begin{figure*}[pos=t]
\centering
\includegraphics[width=\textwidth]{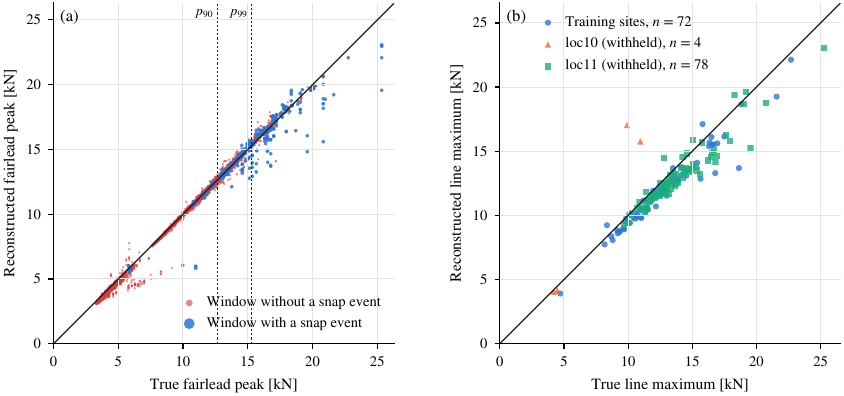}
\caption{Stage~1, predicted against true peaks. (a) Fairlead peak, every test window, with the
$p_{90}$ and $p_{99}$ thresholds. (b) Line maximum at each snap event, one point per distinct
event.}
\label{fig:peaks}
\end{figure*}

The Stage-1 rows of Table~\ref{tab:res_tails} split the fairlead peaks of the upper tail by
whether the window contains a catalogued snap event. Snap windows make up \sOneSnapShareNinety{}\,\si{\percent} of
the tail above $T_{90} = \sOneThrNinety{}$~N and \sOneSnapShareNinetyNine{}\,\si{\percent} of the
tail above $T_{99} = \sOneThrNinetyNine{}$~N. Outside them the peaks are calibrated,
with $u_{90} = \sOneNonSnapUpNinety{}$\,\si{\percent} above $T_{90}$, at a mean bias of
$\sOneNonSnapBiasNinety{}$~N, and $u_{99} = \sOneNonSnapUpNinetyNine{}$\,\si{\percent} above
$T_{99}$, at $\sOneNonSnapBiasNinetyNine{}$~N. In snap windows $u_{90}$ and $u_{99}$ are
\sOneSnapUpNinety{} and
\sOneSnapUpNinetyNine{}\,\si{\percent}, and the under-mass rises from \sOneNonSnapUMNinety{} to
\sOneSnapUMNinety{}~N above $T_{90}$ and from \sOneNonSnapUMNinetyNine{} to
\sOneSnapUMNinetyNine{}~N above $T_{99}$. The tail risk is therefore concentrated in the snap
population, and a pooled $u_q$ such as those of Table~\ref{tab:res_layouts} mixes a calibrated
population with a hard one. Figure~\ref{fig:peaks}a shows the same split, with the scatter below
the diagonal in the tail coming mostly from snap windows.

\begin{table}[t]
\centering
\caption{Fairlead peaks in the upper tail, split by whether the window contains a catalogued snap
event, for Stage~1 and Stage~2. Each stage has its own thresholds.}
\label{tab:res_tails}
\footnotesize
\setlength{\tabcolsep}{3pt}
\begin{tabular}{@{}l l S[table-format=4.0] S[table-format=2.1] S[table-format=2.1] S[table-format=+4.0] S[table-format=4.0]@{}}
\toprule
{Tail} & {Window} & {$n$} & {Share} & {$u$} & {Bias} & {Under-mass} \\
 &  &  & {[\si{\percent}]} & {[\si{\percent}]} & {[\si{\newton}]} & {[\si{\newton}]} \\
\midrule
\multicolumn{7}{@{}l}{\textit{Stage 1}, $T_{90} = \SI{12630}{\newton}$, $T_{99} = \SI{15291}{\newton}$} \\
\addlinespace
$p_{90}$ & no snap & 4584 & 92.5 & 7.9 & +132 & 10 \\
 & snap & 370 & 7.5 & 32.2 & -105 & 296 \\
\addlinespace
$p_{99}$ & no snap & 336 & 67.6 & 10.7 & +227 & 36 \\
 & snap & 161 & 32.4 & 49.7 & -345 & 537 \\
\midrule
\multicolumn{7}{@{}l}{\textit{Stage 2}, $T_{90} = \SI{12622}{\newton}$, $T_{99} = \SI{15278}{\newton}$} \\
\addlinespace
$p_{90}$ & no snap & 4630 & 93.3 & 64.5 & -8 & 40 \\
 & snap & 330 & 6.7 & 65.2 & -526 & 570 \\
\addlinespace
$p_{99}$ & no snap & 340 & 68.0 & 74.1 & -34 & 77 \\
 & snap & 160 & 32.0 & 76.9 & -990 & 1043 \\
\bottomrule
\end{tabular}

\end{table}

Snap loads are assessed per distinct event with the two capture measures of
Section~\ref{subsec:selection}. The first, the capture $c_e$ of the line maximum
(Fig.~\ref{fig:peaks}b), is summarised by site in Table~\ref{tab:res_snaps} through its median
$\tilde{c}$ and the share $F$ of events captured in full.

\begin{table}[t]
\centering
\caption{Stage-1 capture of the line maximum per distinct snap event, by site
(Eq.~\eqref{eq:snap_summary}).}
\label{tab:res_snaps}
\footnotesize
\setlength{\tabcolsep}{3pt}
\begin{tabular}{@{}l S[table-format=3.0] S[table-format=1.3] S[table-format=2.1]@{}}
\toprule
 & {Events} & {Median capture} & {In full [\si{\percent}]} \\
\midrule
Training sites & 72 & 0.938 & 4.2 \\
loc10 & 4 & 1.204 & 50.0 \\
loc11 & 78 & 0.943 & 10.3 \\
\addlinespace
All & 154 & 0.943 & 8.4 \\
\bottomrule
\end{tabular}

\end{table}

The line maximum is not set by the snap itself. The snap rule requires the fairlead condition
(Section~\ref{subsec:snaps}), so no snap peak exceeds the concurrent fairlead tension, and the line
maximum at the event is the fairlead's. Its capture therefore measures the fairlead reconstruction
at the instant of the snap, with a median of \snapCapDev{} at the training sites and
\snapCapEleven{} at loc11 (Table~\ref{tab:res_snaps}). Site loc10 contributes only \snapNTen{}
events.

\begin{figure*}[pos=t]
\centering
\includegraphics[width=\textwidth]{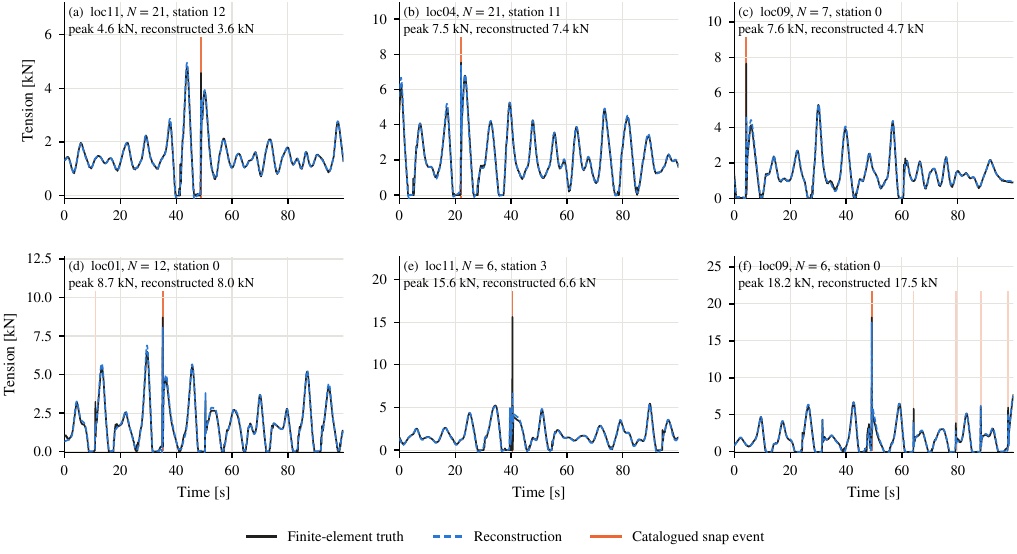}
\caption{Stage-1 tension at the station where a snap event occurs, in \snapStationN{} simulations
ordered by true peak. Vertical lines mark catalogued snap events; the solid one is the event the
panel label quantifies.}
\label{fig:snaptraces}
\end{figure*}

The second measure, the ratio of predicted to true tension at the station where the snap occurs, at
the event sample, is given in Table~\ref{tab:res_snapstation}. For Stage~1 that station is an
output of the sensor layout for \snapStationAvailSone{} of the \snapNEvents{} events, and the other
\snapStationNASone{} are reported as not available. Stage~2 resolves every event, since it emits all
21 stations whatever the input layout. Over the full test sets the median capture is
\snapStationSone{} for Stage~1 and \snapStationStwoFull{} for Stage~2 from 21 sensors, in both cases
below the line maximum of the same events, and \snapStationStwoFour{} from four sensors. The two
stages are scored on different event pools, so their columns are not a controlled comparison. Over
the \snapStationCommonN{} events common to both draws, the paired difference is
\snapStationCommonDiff{}, with Stage~2 higher for \snapStationCommonPct{}\,\si{\percent} of them,
which establishes no difference between the stages. The maxima over the event sample and the two
samples either side barely differ, which places the shortfall in magnitude rather than in timing.
Figure~\ref{fig:snaptraces} shows the reconstruction at the snap station for \snapStationN{}
simulations whose layout includes it. Away from the event the reconstruction follows the station
closely in every panel, whereas at the event it falls short of the truth in all \snapStationN{},
reaching \snapStationLo{} to \snapStationHi{} of the true tension.

\begin{table}[t]
\centering
\caption{Capture at the station where the snap occurs, per distinct event (predicted over true
tension at the event sample). Not available: Stage-1 events whose layout has no station at the
snap. $\pm 2$ samples: maxima over the event sample and the two samples either side. Line maximum:
the capture of Table~\ref{tab:res_snaps} for the same events. The site medians rest on
\snapStationSiteNDevSone{}, \snapStationSiteNTenSone{} and \snapStationSiteNElevenSone{} Stage-1
events and \snapStationSiteNDevStwo{}, \snapStationSiteNTenStwo{} and \snapStationSiteNElevenStwo{}
Stage-2 events (training sites, loc10, loc11).}
\label{tab:res_snapstation}
\footnotesize
\setlength{\tabcolsep}{3pt}
\begin{tabular}{@{}l S[table-format=3.3] S[table-format=3.3] S[table-format=3.3]@{}}
\toprule
 & & \multicolumn{2}{c}{Stage~2} \\
\cmidrule(lr){3-4}
 & {Stage~1} & {$N_{\mathrm{in}}=21$} & {$N_{\mathrm{in}}=4$} \\
\midrule
Events & 97 & 45 & 45 \\
Not available (sensor layout) & 57 & 0 & 0 \\
Median capture & 0.789 & 0.920 & 0.938 \\
Captured in full [\si{\percent}] & 13.4 & 22.2 & 33.3 \\
Median, $\pm 2$ samples & 0.789 & 0.920 & 0.938 \\
Median, line maximum & 0.943 & 0.963 & 0.955 \\
\addlinespace
\multicolumn{4}{@{}l}{Median capture by site} \\
\quad Training sites & 0.826 & 0.938 & 0.926 \\
\quad loc10 & 0.505 & 0.816 & 0.302 \\
\quad loc11 & 0.731 & 0.915 & 0.954 \\
\bottomrule
\end{tabular}

\end{table}

\subsection{Reconstruction from sparse sensing}\label{subsec:res_sparse}

\subsubsection{Sensor-count sweep at the native resolution}\label{subsec:res_sweep}

\begin{figure}[pos=t]
\centering
\includegraphics[width=\columnwidth]{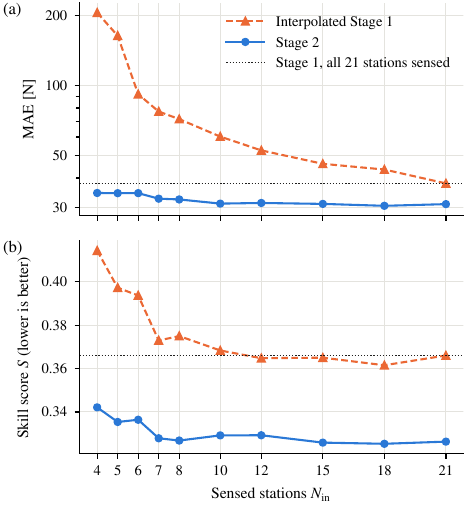}
\caption{Stage~2 and the interpolation baseline at the 21 native stations against the number of
sensed stations. (a) MAE, logarithmic axis. (b) Skill score $S$.}
\label{fig:sweep}
\end{figure}

Figure~\ref{fig:sweep} and Table~\ref{tab:res_sweep} give the accuracy of the 21-station field
for every sensor layout. Stage~2 reconstructs it with an MAE between \srMAELo{} and \srMAEHi{}~N
whatever the number of sensed stations, the largest at four sensors, where the pooled
$R^{2} = \srRsqFour{}$. From four sensors its MAE is thus lower than that of Stage~1 with every
station sensed, \sOneFullMAE{}~N. The interpolation baseline degrades gradually down to six sensors,
to \srPenBaseSix{} times its full-sensing MAE, and then steeply, to \srPenBaseFive{} times at five
sensors and \srPenBase{} times at four, where its $R^{2}$ falls to \srRsqBaseFour{}. For Stage~2 the
same factor is \srPenTest{}, and its $R^{2}$ exceeds that of the baseline at every layout, by the
smallest margin at full sensing.

The baseline fails in space rather than in time, as Fig.~\ref{fig:sweep} shows by comparing its
two panels. With four sensors its $S$ is only \srSkillPenBase{} times its full-sensing value
(Fig.~\ref{fig:sweep}b), whereas its MAE is \srPenBase{} times larger (Fig.~\ref{fig:sweep}a). The
interpolated field follows the wave-by-wave variation of the sensed stations but places it at the
wrong tension level between them. This error in the tension profile along the line is what the
query decoder is built to correct, and Section~\ref{subsec:res_touchdown} locates it. Stage~2 also
follows the dynamics more closely, with $S$ of \srSkillFour{} against \srSkillBaseFour{} at four
sensors.

\begin{table*}[t]
\centering
\caption{Test metrics at the 21 native stations per number of sensed stations, for Stage~2
(St.~2) and the interpolation baseline (Interp.), with the final row pooled over every sensed-station
count.}
\label{tab:res_sweep}
\footnotesize
\setlength{\tabcolsep}{4pt}
\begin{tabular}{@{}S[table-format=2.0] S[table-format=2.1] S[table-format=3.1] S[table-format=1.4] S[table-format=1.4] S[table-format=1.2] S[table-format=2.2] S[table-format=1.3] S[table-format=1.3] S[table-format=2.1] S[table-format=3.1]@{}}
\toprule
 & \multicolumn{2}{c}{MAE [\si{\newton}]} & \multicolumn{2}{c}{$R^2$} & \multicolumn{2}{c}{MAPE [\si{\percent}]} & \multicolumn{2}{c}{$S$} & \multicolumn{2}{c}{Peak MAE [\si{\newton}]} \\
\cmidrule(lr){2-3}\cmidrule(lr){4-5}\cmidrule(lr){6-7}\cmidrule(lr){8-9}\cmidrule(lr){10-11}
{$N_{\mathrm{in}}$} & {St.~2} & {Interp.} & {St.~2} & {Interp.} & {St.~2} & {Interp.} & {St.~2} & {Interp.} & {St.~2} & {Interp.} \\
\midrule
4 & 34.5 & 205.6 & 0.9994 & 0.9803 & 2.56 & 15.53 & 0.342 & 0.414 & 53.1 & 231.4 \\
5 & 34.5 & 163.6 & 0.9994 & 0.9830 & 2.64 & 12.59 & 0.335 & 0.397 & 58.4 & 193.4 \\
6 & 34.5 & 91.8 & 0.9994 & 0.9951 & 2.56 & 6.42 & 0.336 & 0.394 & 56.1 & 119.5 \\
7 & 32.6 & 77.3 & 0.9994 & 0.9973 & 2.37 & 4.58 & 0.328 & 0.373 & 55.5 & 103.4 \\
8 & 32.4 & 71.7 & 0.9994 & 0.9974 & 2.43 & 3.52 & 0.327 & 0.375 & 52.9 & 96.1 \\
10 & 31.1 & 60.3 & 0.9994 & 0.9984 & 2.36 & 2.59 & 0.329 & 0.368 & 43.6 & 82.0 \\
12 & 31.3 & 52.6 & 0.9994 & 0.9989 & 2.34 & 2.42 & 0.329 & 0.365 & 48.9 & 70.4 \\
15 & 31.0 & 46.1 & 0.9994 & 0.9992 & 2.31 & 1.98 & 0.326 & 0.365 & 47.3 & 65.5 \\
18 & 30.4 & 43.5 & 0.9994 & 0.9992 & 2.27 & 1.78 & 0.325 & 0.362 & 49.0 & 64.5 \\
21 & 30.9 & 38.0 & 0.9994 & 0.9993 & 2.29 & 1.49 & 0.326 & 0.366 & 51.0 & 57.5 \\
\midrule
{Pooled} & 32.3 & 85.0 & 0.9994 & 0.9948 & 2.41 & 5.29 & 0.330 & 0.378 & 51.6 & 108.4 \\
\bottomrule
\end{tabular}

\end{table*}

Part of the advantage of Stage~2 over Stage~1 with all 21 stations sensed does not come from the
decoder. The \srFullGainPct{}\,\si{\percent} lower MAE of Stage~2 there, \srMAEfull{} against
\sOneFullMAE{}~N, with $S$ of \srSkillFull{} against \sOneFullSkill{}, results from its further
training at the learning-rate floor and its selection on accuracy alone
(Section~\ref{subsec:selection}).

MAPE is the only field metric on which the baseline is more accurate, from \srMAPECross{} sensors
upwards. MAPE weights errors in low-tension windows heavily, and such windows are common at the
shallow withheld site loc10. The same asymmetry appears in the $\mathrm{MAPE}^{\mathrm{peak}}$ of
Table~\ref{tab:res_sites}, where, pooled over the sensor layouts, Stage~2 is more accurate than
Stage~1 at the training sites (\siteDevErrStwo{} against \siteDevErrSone{}\,\si{\percent}) but less
accurate at loc10 (\siteTenErrStwo{} against \siteTenErrSone{}\,\si{\percent}).

\subsubsection{The four-sensor field and the touchdown region}\label{subsec:res_touchdown}

The Stage-2 and interpolation columns of Table~\ref{tab:res_regions} locate the difference
along the line. Pooled over the sensor layouts, the interpolation baseline keeps the grounded
region, with a median $R^{2}_{i,w}$ of \srGroundedRsqBase{}, loses part of the suspended
region, at \srSuspendedRsqBase{}, and loses the touchdown region, at \srTouchdownRsqBase{} with
\srTouchdownNegBase{}\,\si{\percent} of station-windows below zero. Stage~2 reaches
\srGroundedRsq{}, \srTouchdownRsq{} and \srSuspendedRsq{}.

\begin{figure*}[pos=t]
\centering
\includegraphics[width=\textwidth]{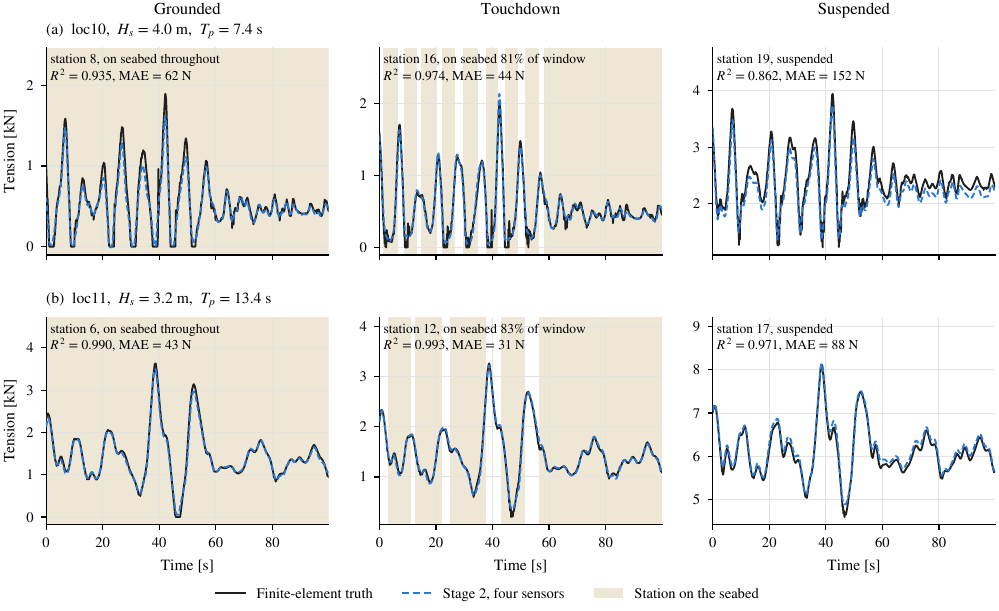}
\caption{Stage~2 from four sensors, at $s = 0$, $1/3$, $2/3$ and $1$, in one test window at each
withheld site. Each row shows three of the 21 stations, none of them sensed, one per region of the
line.}
\label{fig:foursensor}
\end{figure*}

Figure~\ref{fig:foursensor} shows the mechanism at the two withheld sites, one window each, at a
grounded, a touchdown and a suspended station. Only unsensed stations are drawn. The four sensors
sit at $s = 0$, $1/3$, $2/3$ and $1$, so only the anchor and the fairlead fall on a native station.
At \figFourSite{}, the shallow withheld site, with $H_s = \figFourHs{}$~m, the touchdown point lies
at $s = \figFourTdArc{}$, beyond the sensor at two thirds of the line, which is itself still on the
seabed. The tension ramp therefore spans the last third of the line, and even interpolating the
true tension at the four sensed stations gives an MAE of \figFourTdMAEInterp{}~N, with
$R^{2}_{i,w} = \figFourTdRsqInterp{}$, whereas Stage~2 follows the station through its lift-offs
with an MAE of \figFourTdMAE{}~N and $R^{2}_{i,w} = \figFourTdRsq{}$. At \figFourBSite{}, a
longer-period sea, the touchdown point lies at $s = \figFourBTdArc{}$, between the sensors at one
and two thirds, and interpolation errs by \figFourBTdMAEInterp{}~N against \figFourBTdMAE{}~N for
Stage~2.

\subsubsection{Extremes and snap loads from four sensors}\label{subsec:res_sparse_tail}

Stage~2 gives up the calibrated upper tail of Stage~1 (Table~\ref{tab:res_tails}).
Its fairlead peaks above $T_{90} = \sTwoThrNinety{}$~N that come from windows without a snap
have $u_{90} = \sTwoNonSnapUpNinety{}$\,\si{\percent}, at a mean bias of
$\sTwoNonSnapBiasNinety{}$~N and an under-mass of \sTwoNonSnapUMNinety{}~N. Snap windows remain the harder population, with under-masses of
\sTwoSnapUMNinety{}~N above $T_{90}$ and \sTwoSnapUMNinetyNine{}~N above
$T_{99} = \sTwoThrNinetyNine{}$~N.

At the snap's own station, over the \snapNStwo{} Stage-2
events, capture from four sensors against 21 is \snapStationSiteDevFour{} against
\snapStationSiteDevFull{} at the training sites and \snapStationSiteElevenFour{} against
\snapStationSiteElevenFull{} at loc11, and the single loc10 event is captured at
\snapStationSiteTenFour{} against \snapStationSiteTenFull{} (Table~\ref{tab:res_snapstation}).

\begin{figure*}[pos=t]
\centering
\includegraphics[width=\textwidth]{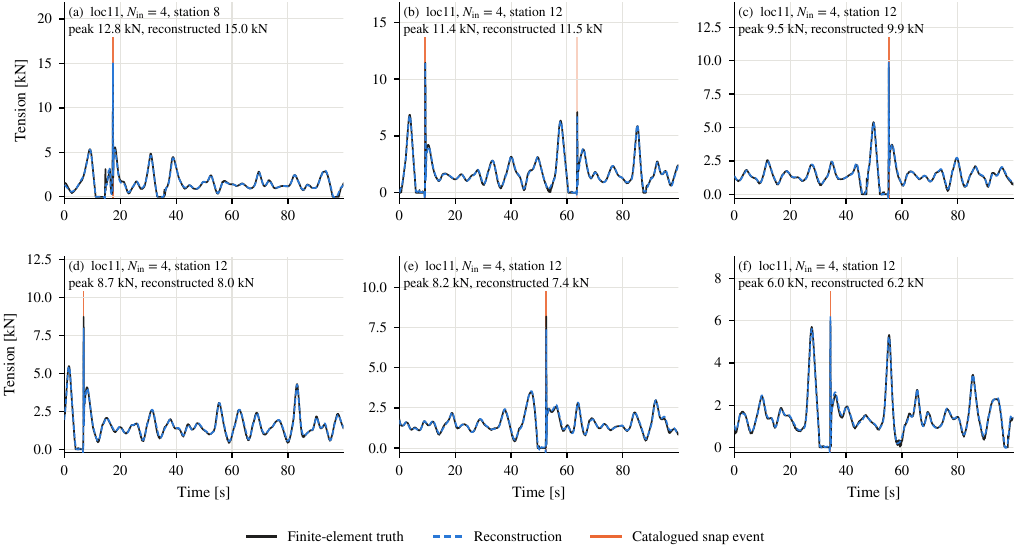}
\caption{Stage-2 tension from four sensors at the station where a snap event occurs, in
\snapStationTwoN{} simulations ordered by true peak. No sensor sits at the station. Vertical lines
mark catalogued snap events; the solid one is the event the panel label quantifies.}
\label{fig:snaptraces_s2}
\end{figure*}

Figure~\ref{fig:snaptraces_s2} shows the four-sensor reconstruction at the snap's own station for
\snapStationTwoN{} Stage-2 simulations, as Fig.~\ref{fig:snaptraces} does for Stage~1, restricted to
stations at which the layout places no sensor. All \snapStationTwoN{} are from the withheld site
loc11, the only site with a qualifying four-sensor window, so the panels are illustrations rather
than a sample. Away from the event the reconstruction follows the wave-frequency response closely
in every panel, and at the event the panels span \snapStationTwoLo{} to \snapStationTwoHi{} of the
true tension, with a median of \snapStationTwoMedian{} against \snapStationStwoFour{} over all
events from four sensors.

\subsection{Sensitivity to motion-measurement noise}\label{subsec:res_noise}

\begin{table*}[t]
\centering
\caption{Stage~2 (St.~2) and the interpolation baseline (Interp.) from four sensed stations, at
the 21 native stations, per level of motion-measurement error. $\mathrm{SNR}_f$ is relative to the
fairlead motion, the ratio is Stage~2 over interpolation, and snap capture is the median
$\tilde{c}$ of the Stage-2 line-maximum capture.}
\label{tab:res_noise}
\footnotesize
\setlength{\tabcolsep}{4pt}
\begin{tabular}{@{}l l S[table-format=2.1] S[table-format=3.1] S[table-format=3.1] S[table-format=1.2] S[table-format=2.1] S[table-format=1.4] S[table-format=1.4] S[table-format=2.2] S[table-format=2.2] S[table-format=1.3]@{}}
\toprule
 & & & \multicolumn{3}{c}{MAE [\si{\newton}]} & {MAPE [\si{\percent}]} & \multicolumn{2}{c}{$R^2$} & \multicolumn{2}{c}{$S$} & {Snap} \\
\cmidrule(lr){4-6}\cmidrule(lr){7-7}\cmidrule(lr){8-9}\cmidrule(lr){10-11}\cmidrule(lr){12-12}
{Error} & {Level} & {$\mathrm{SNR}_f$ [\si{\deci\bel}]} & {St.~2} & {Interp.} & {Ratio} & {St.~2} & {St.~2} & {Interp.} & {St.~2} & {Interp.} & {capture} \\
\midrule
None & {--} & {--} & 34.5 & 205.6 & 0.17 & 2.6 & 0.9994 & 0.9803 & 0.34 & 0.41 & 0.955 \\
\addlinespace
White & \SI{0.01}{\metre} & 30.6 & 91.3 & 225.5 & 0.40 & 6.7 & 0.9957 & 0.9793 & 1.43 & 1.47 & 0.960 \\
 & \SI{0.05}{\metre} & 16.6 & 228.3 & 297.0 & 0.77 & 14.6 & 0.9834 & 0.9671 & 6.42 & 4.65 & 0.979 \\
 & \SI{0.10}{\metre} & 10.6 & 323.5 & 341.3 & 0.95 & 19.5 & 0.9690 & 0.9548 & 9.01 & 5.88 & 0.958 \\
 & \SI{0.25}{\metre} & 2.6 & 491.3 & 448.4 & 1.10 & 26.1 & 0.9295 & 0.9278 & 12.04 & 9.06 & 0.989 \\
\addlinespace
White, velocity differenced & \SI{0.05}{\metre} & 16.6 & 428.4 & 480.9 & 0.89 & 23.2 & 0.9342 & 0.9211 & 15.46 & 17.20 & 0.919 \\
\addlinespace
Fixed offset & \SI{0.05}{\metre} & {--} & 247.8 & 304.2 & 0.81 & 19.4 & 0.9436 & 0.9603 & 0.76 & 0.68 & 0.980 \\
\bottomrule
\end{tabular}

\end{table*}

A deployed model receives motion measured by GNSS receivers and inertial measurement units, which
carry measurement error. The selected Stage-2 weights were therefore scored again with the positions
of the sensed stations perturbed. The error is introduced before any derived input is computed, so
velocity, contact state, seabed reaction and drag all reflect the noisy measurements, while the
target tension is unchanged. The noise is zero-mean Gaussian and independent between samples,
stations and channels. For an inertial solution that receives position fixes more often than once a
second, the white components of the inertial sensor noise outweigh its time-correlated components
\citep{blum2020colored}, which makes white noise a sensible first-order model. Velocity is taken as reported by the sensor, with an error of its own, except at one level where it
is differenced from the noisy positions, which multiplies the error by
$1/(\sqrt{2}\,\Delta t) \approx \noiseVelFactor{}$~\si{\per\second}. A further level replaces the
noise by an offset per station, fixed over each window. The interpolation baseline sees the same
error on the same test windows.

Table~\ref{tab:res_noise} lists the levels. White noise has a standard deviation $\sigma$ in metres
for position and the same value in \si{\metre\per\second} for velocity. \citet{grafe2024virtual}
assumed \SI{0.05}{\metre} in heave and \SI{0.05}{\metre\per\second} in velocity for a commercial
GNSS-aided inertial unit. To relate each level to the measured motion, the table also gives
$\mathrm{SNR}_f$ (in \si{\deci\bel}), the ratio of the variance of the fairlead position, whose
median standard deviation is \noiseFairX{}~m horizontally and \noiseFairZ{}~m vertically, to that
of the noise, both coordinates summed.

At \SI{0.05}{\metre}, the heave accuracy of such a unit ($\mathrm{SNR}_f = \noiseSnrInstr{}$~\si{\deci\bel}),
the four-sensor reconstruction of Stage~2 keeps an $R^{2}$ of \noiseRsqFourInstr{} and an MAE of
\noiseMAEFourInstr{}~N, against \noiseMAEFourInstrB{}~N for interpolation. Relative to the load,
its MAPE rises from \noiseMAPEFourClean{}\,\si{\percent} without noise to
\noiseMAPEFourInstr{}\,\si{\percent} (Table~\ref{tab:res_noise}), and its median capture
of the line maximum moves only from \noiseSnapFourClean{} to \noiseSnapFourInstr{}. Its advantage
over interpolation persists up to \noiseKeepSigma{}~m and is lost by \noiseLoseSigma{}~m. The
\SI{1}{\metre} horizontal error that \citet{grafe2024virtual} assumed lies outside the tested range.

\begin{figure}[pos=t]
\centering
\includegraphics[width=\columnwidth]{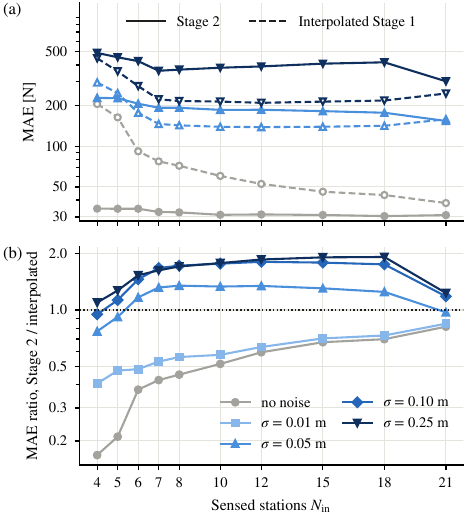}
\caption{Sensitivity to white motion-measurement noise. (a) MAE at the 21 native
stations against the number of sensed stations, colours as in (b). (b) Ratio of the Stage-2 to
the interpolated MAE.}
\label{fig:noise}
\end{figure}

Figure~\ref{fig:noise} shows how sensor density interacts with the noise. The advantage of Stage~2
is largest in sparse layouts, whereas from \SI{0.05}{\metre} upwards the baseline is more accurate
at intermediate densities, where interpolation averages the independent errors of neighbouring
sensors.

Noise degrades the step-to-step dynamics before the tension level. Already at \SI{0.01}{\metre},
$S$ exceeds one for both models with four sensors (Table~\ref{tab:res_noise}), so neither tracks
the step-to-step change better than a temporally flat predictor.

How velocity is obtained matters for the accuracy of the reconstruction. When it is differenced
from the noisy positions rather than
reported, the four-sensor Stage-2 MAE at \SI{0.05}{\metre} rises from \noiseMAEFourInstr{} to
\noiseMAEFourDiff{}~N and $R^{2}$ falls from \noiseRsqFourInstr{} to \noiseRsqFourDiff{}
(Table~\ref{tab:res_noise}), so velocity should be measured, not derived. A fixed offset of
\SI{0.05}{\metre} instead shifts the level, raising the MAE from \noiseMAEFourClean{} to
\noiseMAEFourOff{}~N, while $S$ reaches only \noiseSkillFourOff{}, against \noiseSkillFourClean{}
without noise and \noiseSkillFourInstr{} with white noise of the same size. Regional $R^{2}_{i,w}$
is not reported under noise, because the regions are labelled from the contact flag, which is
itself computed from the perturbed positions.

\subsection{Practical implications and limitations}\label{subsec:res_limits}

Section~\ref{sec:intro} named three limitations of existing surrogates. The first, a tension at
one point of the line only, is removed. The model returns the tension at all 21 stations,
including the grounded and touchdown regions, from as few as four sensors
(Sections~\ref{subsec:res_spatial} and~\ref{subsec:res_touchdown}). The second, a loss of accuracy
outside the training conditions, is reduced but not removed. One withheld site transfers with a
small conservative offset, whereas the shallow site loc10, whose depth-to-length ratio the training
set barely covers, carries a level offset (Section~\ref{subsec:res_layouts}). The third, the largest
errors at the peaks, is resolved for Stage~1 outside snap transients, where the upper tail is
calibrated (Section~\ref{subsec:res_extremes}), a calibration that Stage~2 gives up
(Section~\ref{subsec:res_sparse_tail}). Snap loads remain the harder population.

The models were trained and tested on finite-element simulations alone, with no laboratory or
field record used, and the fairlead follows the local wave-particle motion rather than a floater,
so the coupling between the line and a vessel is not represented. The ground truth also carries
the discretisation of the solver, its element size, time step and the tolerance of its nonlinear
iterations. These were verified by convergence studies (Section~\ref{subsec:fe_data}), but they
bound how finely a transient such as a snap is resolved, and no surrogate trained on the solver can
be more accurate than the solver itself. The results are further limited to one line type in two
dimensions, a single training seed, the 21 native output stations and sites whose depth-to-length
ratio the training set covers, which loc10 does not (Section~\ref{subsec:res_layouts}). The
bidirectional encoder needs the whole window, so the model trails the measurements by one
\windowSeconds{}-s window. It supports monitoring and hindcast analysis, not instantaneous
alarms, and the velocity of the sensed stations should be measured rather than differenced
(Section~\ref{subsec:res_noise}).

The sea states follow the joint distribution observed at each site (Section~\ref{subsec:sites}),
so the training data concentrate on the conditions the lines experience, and rare combinations of
wave height, period and current are represented by few cases. Within the sampled conditions this
does not show in the error. At the training sites the per-case MAPE is uncorrelated with how
sparsely the neighbourhood of its sea state is sampled, with rank correlations of
\seaErrRhoSone{} for Stage~1 and \seaErrRhoStwo{} for Stage~2, and $S$ is higher by less than
\seaErrSRise{} in the sparsest fifth of the cases than in the densest. At the withheld sites the
error is likewise unrelated to the distance to the nearest training sea state, with rank
correlations no larger than \seaErrRhoWithheld{} in magnitude. Conditions that the joint
distribution of these sites does not produce were not tested, and a surrogate meant for a design
study that must cover them would call for a space-filling sample.

The training settings were chosen by hand rather than by a systematic search. The shorter runs of
Section~\ref{subsec:selection} varied the architecture and the plateau test of the schedule, but
not the learning rate, the dropout rates or the patience, and training was not as efficient as it
could have been. The long early-stopping patience made Stage~1 run \sOnePostSelEpochs{} epochs,
about \sOnePostSelHours{}~h, beyond the selected one, and Stage~2 spent its last \sTwoLateEpochs{}
epochs, about \sTwoLateHours{}~h, lowering the validation $S$ by less than \sTwoLateSkillDrop{}
(Appendix~\ref{app:training}). An automated search over these settings, and a schedule matched
to the convergence seen there, could reach the same accuracy at a lower cost.

The synthesised wave train repeats within each record, and \repeatShare{}\,\si{\percent} of the
temporal test windows at the training sites cover forcing that also occurs in training. The
comparison of Section~\ref{subsec:task} bounds any benefit at \repeatEquivBound{}\,\si{\percent},
and the withheld sites are unaffected, but a campaign with non-repeating forcing would remove the
question.

Snap loads and solver artifacts are separated by rules whose thresholds were set from the data
(Section~\ref{subsec:snaps}), without an independent reference. A numerical event that shows every
signature of a snap therefore cannot be ruled out, since the rules test the consistency of an event
with the axial balance of the simulated line, not its physical origin. The fairlead condition was applied
to the test windows after training. The windows it removes are under \SI{1}{\percent} of each test
draw, yet on the unscreened draws they carried \resSSShareOne{} and
\resSSShareTwo{}\,\si{\percent} of the squared error of Stage~1 and Stage~2, and
\catPoolShare{}\,\si{\percent} of the training and validation windows contain
such a sample. A dedicated study of
the exclusion policy, covering the sensitivity of each threshold, retraining with the screen and a
solver run at a finer time step, would characterise both populations more fully.

\section{Conclusions}\label{sec:conclusions}

This study reconstructed the tension along a catenary mooring line from motion measured on the
line. The main findings are as follows.
\begin{itemize}
    \item One set of weights reconstructs the tension field across \nLayouts{} sensor layouts and
    \nSites{} sites, including the grounded stations, which do not move, with $R^{2} = \sTwoRsq{}$
    and an MAE of \sTwoMAE{}~N pooled over every layout.
    \item From four sensors the query decoder reconstructs all \nStationsMax{} stations at
    \srMAEfour{}~N, only \srPenTest{} times its error with every station sensed, whereas
    interpolation raises the error \srPenBase{}-fold and loses the touchdown region.
    \item Linear regressions on the same inputs, even with the motion of the preceding and
    following \lrLagLong{}~s, have \lrMAERatio{} times the MAE of Stage~1 and follow the change
    between samples far less closely, with $S$ of \lrSFour{} against \lrSOne{}, so the map from
    motion to tension is markedly nonlinear.
    \item Outside snap transients the extreme peaks are calibrated at both the $p_{90}$ and the
    $p_{99}$ tail. Snap loads concentrate the tail risk, and at the station where a snap occurs a
    median \snapStationSonePct{}\,\si{\percent} of its tension is recovered across
    \snapStationAvailSone{} Stage-1 events, and \snapStationStwoFourPct{}\,\si{\percent} across
    \snapNStwo{} Stage-2 events from four sensors.
    \item Excluding solver artifacts per window rather than per simulation keeps
    \snapReachAfter{}\,\si{\percent} of the snap events usable, against
    \snapReachBefore{}\,\si{\percent}.
    \item With position errors of \SI{0.05}{\metre}, as assumed for a commercial GNSS-aided
    inertial unit, the reconstruction from four sensors keeps an $R^{2}$ of
    \noiseRsqFourInstr{} and stays ahead of interpolation. Its step-to-step dynamics degrade
    first, and velocity should be measured rather than differenced from position.
    \item A \simMinutes{}-minute record is reconstructed in about \recordReconSec{}~s, against a
    median of \solverHoursMed{}~h for the solver, so the model can follow a monitoring campaign
    continuously, one \windowSeconds{}-s window behind the measurements. It supports digital twins
    and hindcast fatigue assessment of floating structures rather than instantaneous alarms.
\end{itemize}

\appendix
\makeatletter
\@addtoreset{table}{section}
\@addtoreset{figure}{section}
\makeatother
\renewcommand{\thetable}{\Alph{section}.\arabic{table}}
\renewcommand{\thefigure}{\Alph{section}.\arabic{figure}}

\section{Wave-train repeat: a leakage check}\label{app:repeat}

Section~\ref{subsec:task} notes that the synthesised wave train repeats within each record
(Section~\ref{subsec:campaign}), so \repeatShare{}\,\si{\percent} of the temporal test windows at
the training sites cover forcing that the same site's training windows have already seen. This
appendix tests, in three steps, whether either stage benefits from the repeat, using the
per-window MAE, MAPE and $S$ of Section~\ref{subsec:selection} on the temporal test windows of the
training sites. MAE and MAPE would reveal an advantage in the level of the tension and $S$ one in
its dynamics, the two ways in which a remembered forcing could help. First, windows that cover
repeated forcing are compared with those that do not.
The affected windows score better, with an
MAE \repeatSoneMAEDrop{}\,\si{\percent} lower for Stage~1 and \repeatStwoMAEDrop{}\,\si{\percent}
lower for Stage~2, and a lower MAPE and $S$. This alone does not show that the repeat is exploited,
because the affected windows are by construction the last part of each record, and the forcing
need not be equally severe along it.

Second, the end of the record is tested directly for being calmer. The measure is the activity of
a window, the mean absolute change of the true tension from one sample to the next. It is the
denominator of $S_w$ in Eq.~\eqref{eq:metrics_skill} and depends on the simulation alone, not on
any model. The comparison is made separately at each training site, between the Stage-1 test
windows that cover repeated forcing (the affected windows) and those that do not, among the test
windows starting at or after sample \num{\repeatSpanStart}, the final \SI{30}{\percent} of a full
record. Each site's result is the percentage by which the mean activity of the affected windows
differs from that of the unaffected ones. To see whether such a difference could arise from the
draw alone, the test windows are resampled with replacement \num{2000} times, each window counted
once however many sensor layouts it is evaluated at, and the \SI{95}{\percent} confidence
interval is the central \SI{95}{\percent} of the resampled differences. At every site the affected
windows are less active, by \repeatActSiteMin{} to \repeatActSiteMax{}\,\si{\percent}. Pooled over
the test windows of all eight sites, the reduction is \repeatActPooled{}\,\si{\percent}
(95\,\% CI \repeatActPooledLo{} to \repeatActPooledHi{}\,\si{\percent}). The interval of every
site, like the pooled one, lies entirely below zero, so the lower activity is not a chance effect
of the draw. The affected windows are therefore calmer, and a calmer window is easier to
reconstruct.

Third, windows of equal activity are compared. Within each site and sensor layout, the windows are
sorted by activity and split into ten groups of equal size, and each affected window is then
compared only with the unaffected windows of its own group. Compared in this way, the differences
in MAE, MAPE and $S$ fall to at most
\repeatMatchMax{}\,\si{\percent} at either stage. Even the most favourable end of each
\SI{95}{\percent} confidence interval shows the affected windows better by no more than
\repeatEquivBound{}\,\si{\percent}. The advantage of the affected windows thus comes from their
calmer forcing.

\section{Justification of threshold values}\label{app:thresholds}

This appendix justifies the bounds of the shape rule, the snap rule and the magnitude ceiling of
Section~\ref{subsec:snaps}. Where the axial balance, Eq.~\eqref{eq:axial_balance}, determines a
bound, the bound is derived from it. Every other bound is tested against the archive, by moving it
and recording what changes. Two groups of simulations serve as reference. The \num{\catCleanSims} unflagged simulations are those in which neither the shape rule
nor the non-finite check fires. The \num{\ceilingRefSims} reference simulations are the unflagged
ones in which the fairlead condition is also never violated.

\subsection{Fairlead condition}

The fairlead condition rejects a snap when a station other than the fairlead carries more tension
than the fairlead at the same sample. To test its bound of 1, take the samples that meet every
other snap condition, whether or not they meet this one, and call them candidates.
Equation~\eqref{eq:axial_balance} makes the fairlead the most heavily loaded station of the
simulated line, and two observations support the bound. First, the candidates stop sharply at it.
In the unflagged simulations, \catCandBelow{} have a largest station-to-fairlead ratio between 0.9
and 1, which the condition accepts, and only \catCandAbove{} between 1 and 1.1, which it rejects.

Second, the recorded motion cannot account for a station above the fairlead. A genuine snap could
lift one only through inertia, which would appear in the archived positions.
Equation~\eqref{eq:axial_balance} was therefore evaluated between the anchor and the fairlead at
every candidate in the archive, with accelerations from the archived positions and drag omitted,
since inertia dominates during these events. The residual, the part of the tension difference that
the motion does not explain, has a median of \catMomBelow{}~kN at ratios below 1 and
\catMomAbove{}~kN above 1.5, rising steadily in between.

\subsection{Coherence bound}

The coherence ratio of Eq.~\eqref{eq:coh} compares a station with its more heavily loaded
neighbour, and both rules use the bound 3. The shape rule flags a spike that exceeds both
neighbours by more than this factor, and the snap rule accepts one only when it does not.

The bound rests on Eq.~\eqref{eq:axial_balance} applied to one segment. The tensions of two
adjacent stations differ by the submerged weight of the segment, at most
$w_{s}L_{0}/20 = \thrSegWeight$~kN, plus the inertia and drag of the two lumped masses. This was
checked on the station-samples above \SI{10}{\kilo\newton} of the reference simulations, with
accelerations from the archived positions. The difference to the more heavily loaded neighbour has
a median of \thrSegDiffMed~N, and the balance reproduces it to a median residual of
\thrSegResOrd~N. Because the neighbour above a station carries more by the weight of the segment,
only the fairlead, whose single neighbour lies below it, can exceed its neighbour appreciably, and
the archive agrees. Among the \thrCleanSamplesM{} million station-samples above \SI{10}{\kilo\newton}
in the reference simulations the largest $\mathrm{coh}$ is \thrCohCleanMax{}, and every value above
1.1 occurs at the fairlead.

Larger ratios occur only in flagged simulations. The \thrCohHiN{} station-samples above
\SI{10}{\kilo\newton} with $\mathrm{coh}$ above 1.25 all lie there, and for them the balance fails
by a median of at least \thrCohResLo{}~kN in every band of the ratio, against \thrSegResSnap{}~N at
the catalogued snap events, whose median $\mathrm{coh}$ is \thrCohEvMed{}. The bound of 3 therefore
lies \thrCohMargin{} times above the largest genuine value. Any value between 1.5 and 8 changes
little. In the shape rule it
changes the \num{\maskSamples} flagged samples by at most \thrCohMaskMax{} and flags no further
simulation, and in the snap rule it changes the catalogue by at most \thrCohCatMax{}\,\si{\percent}
and leaves the non-snap tail statistics of Section~\ref{subsec:res_extremes} unchanged at the
precision reported there.

The shape rule has a second spatial test, $\mathrm{fair} > 3$, for spikes that extend over two
adjacent stations. Each station of such a pair has a neighbour of similar tension, so
$\mathrm{coh}$ stays near 1 and cannot detect the spike. The test compares the station instead with
the largest tension that the fairlead reaches in the record, Eq.~\eqref{eq:cat}. Since
$\mathrm{fair} \le \mathrm{cat}$, and a record that meets the fairlead condition has
$\mathrm{cat} \le 1$ above \SI{10}{\kilo\newton}, the bound of 3 again lies a factor of 3 above
every genuine sample. The test flags \thrFairOnlySamples{} samples in \thrFairOnlySims{}
simulations that no other rule flags.

\subsection{Isolation bounds}

The isolation ratio of Eq.~\eqref{eq:iso} compares a sample with the mean of its two neighbours in
time. A smooth wave crest has a value near 1 and a one-sample peak a large one.

In the snap rule, the bound of 3 follows from three observations. First, wave crests lie well
below it. Of the \thrCrestM{} million crests above \SI{3}{\kilo\newton} that no rule flags and
that are not preceded by slack, \SI{99.99}{\percent} have $\mathrm{iso}$ below \thrCrestPct{} and
only \thrCrestAbove{} reach 3. Second, a slack line that is pulled taut and then stays taut gives
$\mathrm{iso} \le 2$ however abruptly it rises, because the sample after the peak is at least as
large as the peak, and such a rise persists beyond one sample. A value above 3 therefore requires
the tension to fall by more than a third of the peak within one sample. Third,
the catalogued events have this shape. At the event station the sample before the peak carries a
median of \thrPrevMed{}\,\si{\percent} of it and the sample after it \thrNextMed{}\,\si{\percent},
and over the following second the tension settles at a median of \thrSettleMed{}\,\si{\percent}
of the peak, so the line stays taut at a level well below the overshoot.

Unlike the coherence ratio, $\mathrm{iso}$ shows no gap among the peaks that follow slack that
would fix the bound, so the size of the catalogue depends on it. Table~\ref{tab:app_iso} therefore
repeats the analyses that use the catalogue for bounds from 2 to 10. The non-snap p90 tail of Section~\ref{subsec:res_extremes} stays
calibrated throughout, with $u_{90}$ between \thrTabUpLo{} and \thrTabUpHi{}\,\si{\percent} and a
positive bias, and for bounds up to 5 the share of events reachable under per-window exclusion
stays between \thrTabReachLo{} and \thrTabReachHi{}\,\si{\percent}. The p99 tail is stable up to a
bound of 4, with $u_{99}$ between \thrTabNinetyNineLo{} and \thrTabNinetyNineHi{}\,\si{\percent}.
Beyond that, the weaker spikes that a stricter bound no longer catalogues fall into the non-snap
group, and $u_{99}$ rises to \thrTabNinetyNineTen{}\,\si{\percent} at a bound of 10. The capture of
Section~\ref{subsec:res_extremes} also depends on the bound, since sharper spikes are captured
less well. At their own station the median in Stage~1 is \thrCapSoneLo{} for the
\thrCapSoneLoN{} events with $\mathrm{iso}$ up to 4 and \thrCapSoneHi{} for the \thrCapSoneHiN{}
above 4, and from four sensors Stage~2 gives \thrCapStwoLo{} over \thrCapStwoLoN{} events and
\thrCapStwoHi{} over \thrCapStwoHiN{}. The main text uses the bound of 3, where the calibration
and the reachable share are stable.

\begin{table}[t]
\centering
\caption{Snap catalogue and dependent results for different isolation bounds of the snap rule,
Stage~1. Reachable: share of events reachable under per-window exclusion. $u_{q}$ and bias:
underprediction rate and mean signed error of the fairlead peaks above $T_{q}$, in windows without
a catalogued snap event. The adopted bound is 3.}
\label{tab:app_iso}
\footnotesize
\setlength{\tabcolsep}{3pt}
\begin{tabular}{@{}r S[table-format=4.0] S[table-format=2.1] S[table-format=2.1] S[table-format=+3.0] S[table-format=2.1] S[table-format=+3.0]@{}}
\toprule
{Bound} & {Events} & {Reachable} & {$u_{90}$} & {Bias$_{90}$} & {$u_{99}$} & {Bias$_{99}$} \\
 &  & {[\si{\percent}]} & {[\si{\percent}]} & {[\si{\newton}]} & {[\si{\percent}]} & {[\si{\newton}]} \\
\midrule
2 & 1157 & 72.5 & 7.7 & +133 & 9.5 & +247 \\
3 & 902 & 72.5 & 7.9 & +132 & 10.7 & +227 \\
4 & 706 & 71.5 & 7.9 & +131 & 11.7 & +200 \\
5 & 557 & 71.5 & 8.2 & +130 & 14.4 & +178 \\
7 & 376 & 67.8 & 8.6 & +128 & 16.4 & +160 \\
10 & 263 & 63.9 & 9.2 & +121 & 20.1 & +107 \\
\bottomrule
\end{tabular}

\end{table}

In the shape rule, the bound of 10 means that the two samples around a spike together carry less
than a fifth of its value. It is not what distinguishes artifacts from snaps, since
\thrEvIsoTenPct{}\,\si{\percent} of the catalogued snaps exceed it, and coherence does that
instead. The bound matters only for the station-samples above \SI{10}{\kilo\newton} that already
fail a spatial test, with $\mathrm{coh} > 3$ or $\mathrm{fair} > 3$. Their isolation falls into two
groups. \thrNlSust{} have $\mathrm{iso} \le 1.05$, so the neighbouring samples are as large as the
sample itself and the divergence is sustained, which the magnitude ceiling handles, and
\num{\thrNlIso} have $\mathrm{iso} > 10$ and are single-sample spikes. Only \thrNlMid{} lie
between. Any value of the bound between 7 and 15
therefore changes the flagged samples by at most \thrIsoMaskMax{} and flags no further simulation.

\subsection{Magnitude ceiling, floors and slack fraction}

The magnitude ceiling of \SI{50}{\kilo\newton} is set from simulations it does not filter. In the
\num{\ceilingRefSims} reference simulations the largest tension recorded anywhere is
\ceilingCleanMax{}~kN, and the ceiling is \ceilingRatio{} times that value. It fires in
\ceilingCases{} simulations, all already flagged by the other rules, so it implicates none of its
own.

The \SI{3}{\kilo\newton} floor of the snap rule lies
just above the mean tension of the archive, \tensionMean{}~kN, so a snap must at least exceed the
typical load. Lowering the \SI{10}{\kilo\newton} floor of the shape rule to \SI{7.5}{\kilo\newton}
would flag only \thrFloorAdd{} further samples, \thrFloorEval{} of which the fairlead condition
already screens from the test windows. The slack fraction of 0.25 has little effect, because the
tension at the station falls to zero in the second before \thrSlackZeroPct{}\,\si{\percent} of the
catalogued events, and moving the fraction between 0.1 and 0.75 changes the catalogue by at most
\thrSlackCatMax{} events.

\section{Feature matrices}\label{app:features}

Tables~\ref{tab:app_static_node}--\ref{tab:app_dynamic_edge} define the four feature matrices of
Section~\ref{subsec:graph} (Fig.~\ref{fig:pipeline}) channel by channel, in the column order of the
implementation, with index 0 the first channel. Channels computed directly from other channels are
given as equations, and station indices $i,j$ range over $0,\dots,N_{\mathrm{in}}-1$. The reference
length $L_i$ of station $i$, the line length lumped at it ($L/(N_{\mathrm{in}}-1)$), is halved at the
two end stations, as for the lumped mass (channel 6 of Table~\ref{tab:app_static_node}). The constants $k_b$ and $c_b$ are the seabed spring stiffness and
damping, $\rho$ is the sea-water density, $C_{d,t}$ and $C_{d,n}$ are the axial and normal drag
coefficients, and $d$ is the line diameter (channel 7 of Table~\ref{tab:app_static_node}), all
fixed properties of the line and seabed. The velocity components $v_{t,i}$ and
$v_{n,i}$ are $(\dot x_i,\dot z_i)$ projected on the instantaneous unit tangent and normal at
station $i$, recomputed at every sample from the instantaneous position, unlike the static
reference tangent of Table~\ref{tab:app_static_edge}. Both drag channels use the velocity of the
line itself rather than its velocity relative to the water,
$\mathbf{u}_{\mathrm{wave}}+\mathbf{u}_{\mathrm{current}}-\dot{\mathbf{u}}$ in the solver
\citep[Eq.~24]{agarwal2026finitestrain}, because the instantaneous wave-particle velocity at a
station is not an input, so $F_{t,i}$ and $F_{n,i}$ approximate the drag of the solver; the
momentum term built on them still keeps a finite learned weight in
Tables~\ref{tab:app_kendall_s1} and~\ref{tab:app_kendall_s2}, unlike the constitutive term.

\begin{table}[t]
\centering
\caption{Static node features, $[N_{\mathrm{in}} \times \nStatNodeFeat{}]$,
fixed for a simulation.}
\label{tab:app_static_node}
\footnotesize
\setlength{\tabcolsep}{3pt}
\begin{tabular}{@{}c l >{\raggedright\arraybackslash}p{0.50\linewidth}@{}}
\toprule
{\#} & {Channel} & {Description} \\
\midrule
0  & \texttt{s\_over\_L\_line}  & normalised arc length, $s_i/L \in [0,1]$ \\
1  & \texttt{is\_anchor}        & anchor indicator, 1 at $i=0$ \\
2  & \texttt{is\_fairlead}      & fairlead indicator, 1 at $i=N_{\mathrm{in}}-1$ \\
3  & \texttt{is\_intermediate}  & interior-station indicator \\
4  & \texttt{x0}                & initial horizontal position [m] \\
5  & \texttt{z0}                & initial vertical position [m] \\
6  & \texttt{nodal\_mass}       & lumped mass [kg], halved at the two end stations \\
7  & \texttt{diameter}          & line diameter [m] \\
8  & \texttt{area}              & cross-sectional area [m$^2$] \\
9  & \texttt{submerged\_weight} & lumped submerged weight [N], halved at the two end
                                   stations \\
10 & \texttt{z\_bed}            & seabed elevation [m] \\
11 & \texttt{tension0}          & static pretension [N] \\
\bottomrule
\end{tabular}
\end{table}

\begin{table}[t]
\centering
\caption{Dynamic node features,
$[W \times N_{\mathrm{in}} \times \nDynNodeFeat{}]$, one row per sample of
the window.}
\label{tab:app_dynamic_node}
\footnotesize
\setlength{\tabcolsep}{3pt}
\begin{tabular}{@{}c l >{\raggedright\arraybackslash}p{0.50\linewidth}@{}}
\toprule
{\#} & {Channel} & {Description} \\
\midrule
0  & \texttt{x\_abs}             & horizontal position [m] \\
1  & \texttt{z\_abs}             & vertical position [m] \\
2  & \texttt{x\_vel}             & horizontal velocity [m/s], finite difference \\
3  & \texttt{z\_vel}             & vertical velocity [m/s], finite difference \\
4  & \texttt{contact\_flag}      & $c_i = \mathbb{1}[z_i \le z_{\mathrm{bed},i}]$ \\
5  & \texttt{penetration\_depth} & $p_i = \max(z_{\mathrm{bed},i} - z_i,\ 0)$ [m] \\
6  & \texttt{bed\_reaction\_z}   & $R_i = c_i\big[w_i + (k_b p_i - c_b \dot z_i)\,d\,L_i\big]$ [N] \\
7  & \texttt{F\_drag\_t}         & $F_{t,i} = -\tfrac{1}{2}\rho C_{d,t}(\pi d L_i)\,
                                    v_{t,i}|v_{t,i}|$ [N] \\
8  & \texttt{F\_drag\_n}         & $F_{n,i} = -\tfrac{1}{2}\rho C_{d,n}(d L_i)\,
                                    v_{n,i}|v_{n,i}|$ [N] \\
9  & \texttt{h0}                 & mean water depth [m] \\
10 & \texttt{Hs}                 & significant wave height [m] \\
11 & \texttt{Tp}                 & peak spectral period [s] \\
12--20 (even) & \texttt{d1}--\texttt{d5} & current-profile depths, five
                                   levels [m] \\
13--21 (odd) & \texttt{v1}--\texttt{v5} & current velocity at each
                                   depth, five levels [m/s] \\
\bottomrule
\end{tabular}
\end{table}

\begin{table}[t]
\centering
\caption{Static edge features, $[E \times \nStatEdgeFeat{}]$, fixed for a
simulation and shared by the forward and backward directed edge of a
segment.}
\label{tab:app_static_edge}
\footnotesize
\setlength{\tabcolsep}{3pt}
\begin{tabular}{@{}c l >{\raggedright\arraybackslash}p{0.50\linewidth}@{}}
\toprule
{\#} & {Channel} & {Description} \\
\midrule
0 & \texttt{segment\_length}                & chord length between the initial positions of the two
                                               stations [m] \\
1 & \texttt{segment\_length\_over\_L\_line} & normalised segment length \\
2 & \texttt{EA}                             & axial stiffness [N] \\
3 & \texttt{mass\_per\_length}              & linear density [kg/m] \\
4 & \texttt{diameter}                       & line diameter [m] \\
5 & \texttt{tangent\_x}                     & $t_x = \Delta x/\ell$,\ \
                                               $\Delta x = x_{0,i+1}-x_{0,i}$ \\
6 & \texttt{tangent\_z}                     & $t_z = \Delta z/\ell$,\ \
                                               $\Delta z = z_{0,i+1}-z_{0,i}$,\ \
                                               $\ell = $~\texttt{segment\_length} \\
7 & \texttt{edge\_is\_anchor}               & anchor-side segment indicator \\
8 & \texttt{edge\_is\_intermediate}         & interior-segment indicator \\
9 & \texttt{edge\_is\_fairlead}             & fairlead-side segment indicator \\
\bottomrule
\end{tabular}
\end{table}

\begin{table}[t]
\centering
\caption{Dynamic edge features, $[W \times E \times \nDynEdgeFeat{}]$, one
row per sample of the window. For a segment with endpoint stations $i,j$,
each channel is built from the corresponding node channel of
Table~\ref{tab:app_dynamic_node}.}
\label{tab:app_dynamic_edge}
\footnotesize
\setlength{\tabcolsep}{3pt}
\begin{tabular}{@{}c l >{\raggedright\arraybackslash}p{0.50\linewidth}@{}}
\toprule
{\#} & {Channel} & {Description} \\
\midrule
0 & \texttt{edge\_contact\_fraction}   & $\tfrac{1}{2}(c_i + c_j)$ \\
1 & \texttt{edge\_penetration\_mean}   & $\tfrac{1}{2}(p_i + p_j)$ [m] \\
2 & \texttt{edge\_bed\_reaction\_mean} & $\tfrac{1}{2}(R_i + R_j)$ [N] \\
3 & \texttt{edge\_touchdown\_flag}     & $\mathbb{1}[c_i \ne c_j]$ (marks the
                                          touchdown segment) \\
\bottomrule
\end{tabular}
\end{table}

\section{Learned uncertainty weights as a physics diagnostic}\label{app:kendall}

Tables~\ref{tab:app_kendall_s1} and~\ref{tab:app_kendall_s2} rank the \nLossTerms{} loss terms of
Section~\ref{subsec:loss}, from smallest to largest learned scale $\sigma_k = e^{s_k/2}$ of the
log-variance $s_k$ in Eq.~\eqref{eq:kendall}, on the unrounded values. Values are for the selected
model of each stage, epoch~\sOneSelEpoch{} for Stage~1 and epoch~\sTwoSelEpoch{} for Stage~2. A
term with a small $\sigma_k$ is satisfied tightly and carries a large weight, $1/(2\sigma_k^{2})$,
whereas a term the model cannot satisfy takes a large $\sigma_k$ and is down-weighted. Since the
terms differ in units, $\sigma_k^{2}$ carries the units of its own term, and $\sigma_k$ is
unrelated to the tension scale $\sigma_T$ of Section~\ref{subsec:loss}.

The two stages order the terms alike, except that the L1 and pinball terms exchange places, and in
Stage~2 these two agree to the digits shown. The constitutive term comes last in both by a wide
margin and so receives almost no weight. This is physically interpretable: at these strain levels
the chord-based strain estimate is swamped by discretisation error, so an $EA\,\varepsilon$ law
computed from resampled node positions carries almost no usable information about tension.

\begin{table}[t]
\centering
\caption{Loss terms of the Stage-1 model (epoch~\sOneSelEpoch{}), ranked by learned scale.}
\label{tab:app_kendall_s1}
\footnotesize
\setlength{\tabcolsep}{4pt}
\begin{tabular}{@{}c l l r r@{}}
\toprule
{Rank} & {Loss term} & {Symbol} & {$s_k$} & {$\sigma_k$} \\
\midrule
1 & Spectral & $L_{\mathrm{spec}}$ & $-13.76$ & 0.001026 \\
2 & Mean squared error & $L_{\mathrm{mse}}$ & $-6.62$ & 0.03653 \\
3 & Temporal gradient & $L_{\mathrm{grad}}$ & $-6.15$ & 0.04625 \\
4 & Catenary & $L_{\mathrm{cat}}$ & $-5.81$ & 0.05483 \\
5 & Pinball & $L_{\mathrm{pin}}$ & $-3.85$ & 0.1458 \\
6 & Mean absolute error & $L_{\mathrm{l1}}$ & $-3.74$ & 0.1542 \\
7 & Momentum & $L_{\mathrm{mom}}$ & $-2.75$ & 0.2524 \\
8 & Constitutive & $L_{\mathrm{const}}$ & $12.21$ & 448.0 \\
\bottomrule
\end{tabular}

\end{table}

\begin{table}[t]
\centering
\caption{Loss terms of the Stage-2 model (epoch~\sTwoSelEpoch{}), ranked by learned scale.}
\label{tab:app_kendall_s2}
\footnotesize
\setlength{\tabcolsep}{4pt}
\begin{tabular}{@{}c l l r r@{}}
\toprule
{Rank} & {Loss term} & {Symbol} & {$s_k$} & {$\sigma_k$} \\
\midrule
1 & Spectral & $L_{\mathrm{spec}}$ & $-14.16$ & 0.0008428 \\
2 & Mean squared error & $L_{\mathrm{mse}}$ & $-6.98$ & 0.03048 \\
3 & Temporal gradient & $L_{\mathrm{grad}}$ & $-6.38$ & 0.04125 \\
4 & Catenary & $L_{\mathrm{cat}}$ & $-5.89$ & 0.05254 \\
5 & Mean absolute error & $L_{\mathrm{l1}}$ & $-3.92$ & 0.1409 \\
6 & Pinball & $L_{\mathrm{pin}}$ & $-3.92$ & 0.1409 \\
7 & Momentum & $L_{\mathrm{mom}}$ & $-3.78$ & 0.1514 \\
8 & Constitutive & $L_{\mathrm{const}}$ & $8.32$ & 64.18 \\
\bottomrule
\end{tabular}

\end{table}

\section{Training history}\label{app:training}

\begin{figure*}[t]
\centering
\includegraphics[width=\textwidth]{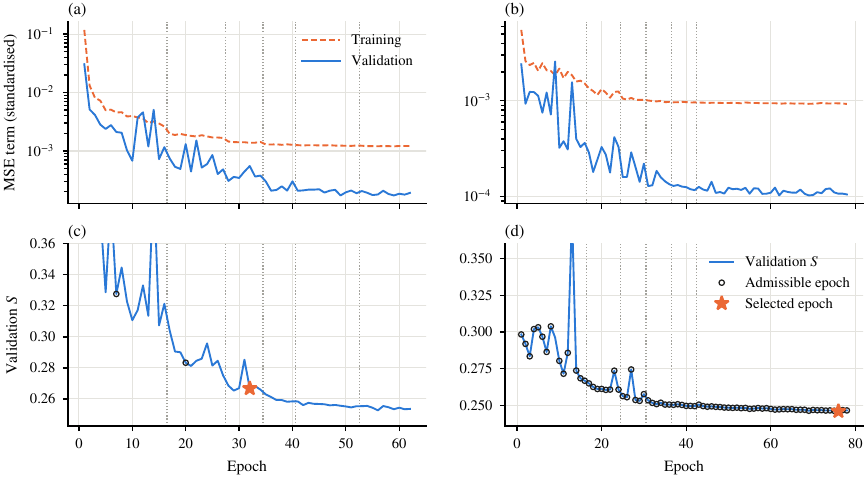}
\caption{Training history of Stage~1 (a, c) and Stage~2 (b, d). (a, b) Data-fit term of the loss,
the mean squared error of standardised tension. (c, d) Validation $S$, with the epochs that pass
the selection rule of Section~\ref{subsec:selection} and the one selected. Dotted lines mark the
learning-rate reductions.}
\label{fig:app_training}
\end{figure*}

Figure~\ref{fig:app_training} shows how the mean squared error of Eq.~\eqref{eq:loss_fit} and the
validation $S$ evolved in both stages. The training value is averaged over each epoch in training
mode, with dropout active and the weights still changing, whereas the validation value is computed
at the end of the epoch without dropout, which is why the training curve lies above the validation
one. The two fall and then level off together, and the validation error shows no sustained rise
late in training, so neither stage shows the divergence between the curves that marks
overfitting. The withheld sites of Section~\ref{sec:results} test generalisation more directly.
Only \sOneAdmissibleN{} of the \sOneEpochsRun{} Stage-1 epochs are admissible, because the
conservative margin set by the peak term of the loss shrinks as the model converges, and the long
early-stopping patience of Section~\ref{subsec:selection} allows for this. In Stage~2, where the
tail condition was dropped, \sTwoAdmissibleN{} of the \sTwoEpochs{} epochs are admissible.

\section{Linear regression models versus the Stage~1 model}\label{app:linear}

\begin{figure*}[t]
\centering
\includegraphics[width=\textwidth]{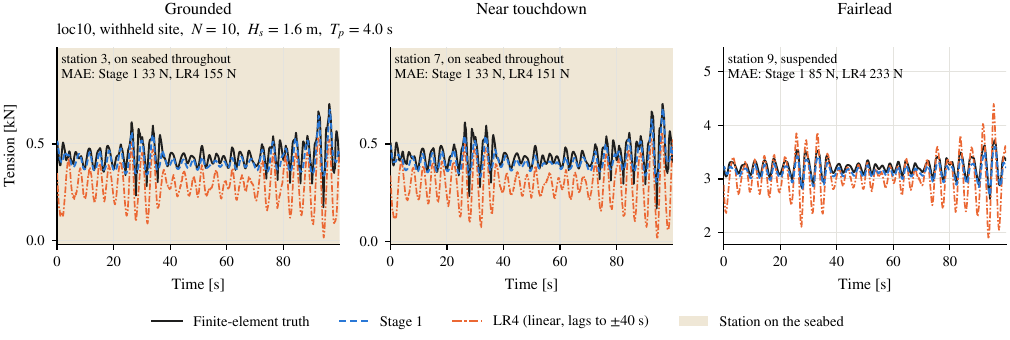}
\caption{Tension at three stations in one Stage-1 test window at the withheld site loc10, with
the mean absolute error of each station-window for Stage~1 and LR4.}
\label{fig:app_linear}
\end{figure*}

\begin{table}[t]
\centering
\caption{Stage~1 and the linear models on the screened Stage-1 test windows. LR0 is constant, so
$S_{w} = 1$ in every window.}
\label{tab:app_linear}
\footnotesize
\setlength{\tabcolsep}{3pt}
\begin{tabular}{@{}l r r r r r r@{}}
\toprule
 & Stage~1 & LR0 & LR1 & LR2 & LR3 & LR4 \\
\midrule
Pooled $R^{2}$ & 0.999 & 0.973 & 0.988 & 0.995 & 0.996 & 0.996 \\
MAE [\si{\newton}] & 42.9 & 279.7 & 186.3 & 128.8 & 107.0 & 106.5 \\
$S$ & 0.37 & 1.00 & 0.94 & 0.71 & 0.85 & 0.81 \\
\addlinespace
\multicolumn{7}{@{}l}{Median station $R^{2}_{i,w}$} \\
\quad Grounded & 0.99 & $-$1.57 & $-$0.03 & 0.68 & 0.79 & 0.80 \\
\quad Touchdown & 0.99 & $-$1.07 & 0.12 & 0.77 & 0.84 & 0.85 \\
\quad Suspended & 0.95 & $-$0.56 & 0.52 & 0.80 & 0.88 & 0.88 \\
\addlinespace
\multicolumn{7}{@{}l}{Windows with $S_{w} > 1$ [\si{\percent}]} \\
\quad loc06 & 0 & -- & 79 & 84 & 83 & 78 \\
\quad loc10 & 0 & -- & 97 & 82 & 97 & 96 \\
\bottomrule
\end{tabular}

\end{table}

This appendix tests whether a linear map from the same inputs could replace Stage~1. Five linear
models predict the tension of one station at one sample, each from the inputs of the previous
one plus a further block. LR0 holds the static pretension of the station constant (channel 11 of
Table~\ref{tab:app_static_node}). LR1 uses the \lrInOne{} input channels of the station itself,
the static and dynamic node features of Tables~\ref{tab:app_static_node}
and~\ref{tab:app_dynamic_node}. LR2 adds the motion channels (dynamic channels 0--8) of the
fairlead, of the two neighbouring stations and their mean over the line, for \lrInTwo{} inputs.
LR3 adds the motion channels of the station and of the fairlead at lags from \lrLagMin{} to
\lrLagShort{}~s before and after the sample (\lrInThree{} inputs), and LR4 adds lags at every
whole second up to \lrLagLong{}~s (\lrInFour{} inputs), most of the \windowSeconds{}-s window that
Stage~1 sees. The four fitted models are ordinary least-squares fits on the Stage-1 training
windows of Section~\ref{subsec:task}, LR3 and LR4 on one sample in \lrStride{} of a random
\lrSubsetPct{}\,\% of them; a ridge penalty did not lower the validation error. All six models are
scored on the \lrTestWindows{} screened Stage-1 test windows with the metrics of
Section~\ref{subsec:selection}.

The pooled $R^{2}$ is high for every model in Table~\ref{tab:app_linear}, \lrRsqConst{} even for
LR0, because the static tension level dominates it; the station $R^{2}_{i,w}$ and $S$ remove that
level and separate the models. Against LR4, Stage~1 has the lower MAE in \lrWinMAE{}\,\% of the
test windows and the lower $S$ in every one. Lengthening the lags from \lrLagShort{} to
\lrLagLong{}~s lowers the validation RMSE of the linear model by only \lrMemGain{}\,\%, so the
shortfall is not a matter of how much of the window the model sees. At the shallow sites loc06 and
loc10, every fitted linear model follows the change between samples worse than a constant
prediction ($S_{w} > 1$) in most windows. Figure~\ref{fig:app_linear} shows a loc10 window chosen
on the ground truth alone: among the randomly drawn stored loc10 windows with at least ten
stations, in a simulation that no artifact rule flags and with no masked sample, the one with the
largest standard deviation of the true fairlead tension. The full comparison,
with every metric per site group and further test windows, is part of the code and evaluation
outputs listed under Data availability.

\printcredits
\section*{Declaration of competing interest}
The authors declare that they have no known competing financial interests or personal
relationships that could have appeared to influence the work reported in this paper.

\section*{Funding}
This research did not receive any specific grant from funding agencies in
the public, commercial, or not-for-profit sectors.

\section*{Data availability}
\begin{sloppypar}
The code used to train and evaluate the models reported in this study, and the scripts that
generate the tables and figures of this manuscript, are publicly available under the MIT license at
\url{https://github.com/AthKon/PI-GAT-LSTM/releases/tag/v1.3}. The trained model checkpoints and the
evaluation outputs are archived at 4TU.ResearchData \citep{konstantaras2026dataset} under the CC-BY 4.0
license, \url{https://doi.org/10.4121/e3f167ac-2925-4815-956e-5ef3d883f1a5.v1}. The raw finite-element
simulation dataset, generated with the mooring model of \citet{agarwal2026finitestrain}, is not
shared owing to its size (tens of gigabytes) and because it is jointly held by the model's
developers; it is available from the corresponding author upon reasonable request.
\end{sloppypar}

\section*{Declaration of generative AI and AI-assisted technologies in the manuscript preparation process}
During the preparation of this work, the author(s) used Claude (Anthropic) in order to improve
the language, clarity and style of the manuscript text, and to assist with code debugging and
review during the development of the data-processing and analysis scripts used to produce the
reported results. These tools were used only for language editing and coding assistance; they
were not used to generate results, perform the scientific analysis, or make scientific decisions,
and all technical content, analyses and interpretations remain the author(s)' own. After using
this tool/service, the author(s) reviewed and edited the content as needed and take(s) full
responsibility for the content of the published article.

\section*{Acknowledgements}
\begin{sloppypar}
The authors acknowledge the use of computational resources of the DelftBlue supercomputer,
provided by the \citet{DHPC2024} (\url{https://www.tudelft.nl/dhpc}). Both models were trained on a single NVIDIA A100 80GB PCIe
GPU with 16 CPU cores and \SI{64}{\giga\byte} of host memory per job. Matrix multiplications ran in TF32 precision.
\end{sloppypar}

\bibliographystyle{cas-model2-names}
\bibliography{refs}

\end{document}